\documentclass{emulateapj}
\usepackage{placeins}
\usepackage{graphicx}

\shorttitle{Protostellar disk accretion} \shortauthors{Zhu et al.}

\begin{document}

\title{Challenges in Forming Planets by Gravitational Instability: Disk Irradiation and Clump Migration, Accretion \& Tidal Destruction}

\author{Zhaohuan Zhu\altaffilmark{1,2}, Lee Hartmann\altaffilmark{1}, Richard P. Nelson\altaffilmark{3}, and
Charles F. Gammie \altaffilmark{4,5}}

\altaffiltext{1}{Dept. of Astronomy, University of Michigan, 500
Church St., Ann Arbor, MI 48109}
\altaffiltext{2}{Department of Astrophysical Sciences, Princeton University, 
4 Ivy Lane Peyton Hall, Princeton, NJ 08544 }
\altaffiltext{3}{Astronomy Unit, Queen Mary, University of London,
Mile End Road, London E1 4NS UK}
\altaffiltext{4}{Dept. of
Astronomy, University of Illinois at Urbana-Champaign, 1002 W. Green
St., Urbana, IL 61801}
\altaffiltext{5}{Dept. of Physics, University
of Illinois at Urbana-Champaign}

\email{zhuzh@umich.edu, lhartm@umich.edu, r.p.nelson@qmul.ac.uk, gammie@illinois.edu}

\newcommand\msun{\rm M_{\odot}}
\newcommand\lsun{\rm L_{\odot}}
\newcommand\au{\, \rm AU}
\newcommand\msunyr{\rm M_{\odot}\,yr^{-1}}
\newcommand\mjupiteryr{\rm M_{\rm J}\,yr^{-1}}
\newcommand\be{\begin{equation}}
\newcommand\en{\end{equation}}
\newcommand\cm{\rm cm}
\newcommand\kms{\rm{\, km \, s^{-1}}}
\newcommand\K{\rm K}
\newcommand\etal{{\rm et al}.\ }
\newcommand\sd{\partial}
\newcommand\mdot{\rm \dot{M}}
\newcommand\rsun{\rm R_{\odot}}
\newcommand\yr{\rm yr}

\begin{abstract}
We present two-dimensional hydrodynamic simulations of
self-gravitating protostellar disks subject to axisymmetric, continuing mass loading from
an infalling envelope and irradiation from the central star, 
to explore the growth of gravitational instability (GI) and disk 
fragmentation. We assume that the disk is built gradually and smoothly by the infall,
resulting in good numerical convergence.
We confirm that for disks around solar-mass stars, 
infall at high rates at radii beyond $\sim 50 \au$ leads to disk 
fragmentation. At lower infall rates, however, irradiation suppresses 
fragmentation. We find that, 
once formed, the fragments or clumps migrate inward on typical type I 
time scales of $\sim 2\times 10^3$~yr initially, but with a stochastic
component superimposed due to their interaction with the GI-induced spiral arms.
Migration begins to deviate from the type I time scale when 
the clump becomes more massive than the local disk mass, and/or
when the clump begins to form a gap in the disk.
As they migrate, clumps accrete from the  disk at a rate between
10$^{-3}$ to 10$^{-1} \mjupiteryr$, consistent with analytic estimates 
that assume a 1-2 Hill radii cross section. 
The eventual fates
of these clumps, however, diverges depending on the migration speed:
3 out of 13 clumps in our simulations become massive enough (brown dwarf mass range)
to open gaps in the disk and essentially stop migrating;
4 out of 13 are tidally destroyed during inward migration;  
6 out of 13 migrate across the inner boundary of the simulated disks.  
A simple analytic model for clump evolution explains
the different fates of the clumps and reveals some limitations of
2-D simulations.  Overall, our results indicate that
fast migration, accretion, and tidal destruction of the clumps 
pose challenges to the scenario of giant planet formation by GI in situ,
although we cannot address whether or not remnant solid cores can survive 
after tidal stripping. The models where the massive clumps are not disrupted 
and open gaps may be relevant to the formation of close binary systems 
similar to Kepler 16.  A clump formed by GI-induced fragmentation can be as 
large as 10 AU and as luminous as 2$\times$10$^{-3}$L$_{\odot}$, which will be 
easily detectable with ALMA, but its lifetime before dynamically
collapsing is only $\sim 1000$ years. 
\end{abstract}


\section{Introduction}
Even a small amount of rotation of a protostellar molecular cloud
core will result in most of the core mass falling onto a disk
rather than the central protostar.  Unless non-gravitational 
angular momentum transport is very efficient or the initial
angular momentum is very small, the resulting disk will become quite
massive and subject to gravitational instability (GI) 
(Vorobyov \& Basu 2005, 2006; Rice \etal 2010; Zhu \etal 2010b), 
which can serve as an efficient mechanism to transport material 
inward (e.g., Durisen \etal 2007, and references therein). However,
GI disks have been shown to be subject to
fragmentation if the disk cooling time is shorter than or comparable to the orbital
period (Gammie 2001; Rice \etal 2005; Paardekooper \etal 2011). 
Ignoring irradiation from the central star/accretion disk,
this cooling criterion can be satisfied at radii beyond tens of AU 
(Nelson \etal 2000; Boley \etal 2006; Rafikov 2007; Stamatellos \&
Whitworth 2008; Stamatellos et al. 2010;  Boley 2009;  see Armitage 2010 for a review). 

Gravitational fragmentation has been proposed as a mechanism to form giant planets beyond
tens of AU  (Boss 1997; Boley 2009), explaining recent observations of the presence of
massive planets at large radii (e.g. Fomalhaut, Kalas et al. 2008; HR 8799, Marois et al.
2008).  Alternatively, GI-induced fragmentation has been proposed to explain multiple star
systems (Kratter \etal 2008, 2010a,b; Offner \etal 2010; Hayfield \etal 2011).  On the other
hand, GI fragmentation is complicated by the tendency of fragments to migrate inward; if they
reach the star, they could produce jumps in accretion luminosity like those of FU Orionis
outbursts (Vorobyov \& Basu 2005, 2006; but see Zhu \etal 2010a,b for an alternative model).
If the migration stops at the inner disk, it may provide a formation mechanism for close
binary systems (\S 7.1). 

The variety of clump masses obtained in previous simulations suggests GI-induced
fragmentation depends on the initial condition of the disks, 
which should be determined by both the rate of infall and the mass of the growing central star.
A more realistic treatment of GI requires consideration of both the way in which
mass is loaded onto the disk as infall proceeds, and irradiation heating of the
outer disk by the central star; the latter is known to weaken and/or suppress GI (Cai
\etal 2008). Recent analytic studies (Levin 2007; Rafikov 2009; Cossins \etal
2010; Zhu \etal 2010b) have shown that irradiation may even suppress fragmentation 
beyond 50 AU. However, simulations with a large parameter space have not been 
carried out to confirm these analytic studies.

Investigations of the evolution of clumps have been limited so far, largely because of
numerical difficulties. Most 3D hydrodynamic simulations use arithmetically spaced
computational grids in the radial direction, which can only cover a relatively small range of
radii. Their limited resolution also makes it difficult to resolve the collapsing clump.
Particle-based SPH simulations can also suffer from deficient numbers of particles,
especially in 3D (Nelson 2006). Many initial attempts to treat disk cooling in SPH
simulations have only applied the orbital cooling prescription which leads to a runaway
collapse. Recent studies have more realistic radiative transfer treatment implemented for
both grid based (e.g. Boley 2009) and SPH (e.g. Stamatellos \& Whitworth 2008, Forgan \etal
2009, 2010, Boley 2009, Rogers \& Wadsley 2011) calculations.  But since 3D simulations are
time consuming, long timescale high-resolution simulations that can follow clump evolution
are only feasible in 2D.  While some analytic work has been done (Nayakshin 2010abc) to
explore long-term clump evolution, the migration process is simplified, and further mass
growth through gas accretion is ignored once the clump has formed.  Furthermore, 2D
simulations enable us to study clump formation in a large parameter space of both disk
and infall properties.

In order to study GI-induced disk fragmentation under the influence of both irradiation and 
mass infall, we have carried out high-resolution two-dimensional (2D) hydrodynamic 
simulations with thermal physics included simply.  We show how fragmentation depends upon
the mass infall rate and the mass infall radius. Furthermore, 
with high resolution long-timescale simulations, we are able to trace the clump 
evolution. We find that clumps migrate inward rapidly, unless they become
massive enough to open gaps, and the accretion rate onto the clump is usually
high due to the large cross section. Inward migrating clumps can be subject to 
tidal destruction as they migrate inward because of their shrinking Hill radii.  
But the detailed accretion physics around the clump and the dynamically unstable
second core phase complicates this issue.

In \S 2, we introduce our numerical model. In \S 3 and 4,  we present our 
simulation results regarding disk fragmentation and clump evolution, 
in \S 5 we discuss the fate of the clumps, while in \S 6, we develop analytic 
models for disk fragmentation and clump evolution to compare with numerical 
results. The implications are discussed in \S 7, and conclusions are drawn in \S 8.

\section{Models}

We use the FARGO-ADSG code (Baruteau \& Masset 2008) in a two dimensional ($R, \phi$) fixed
polar grid. FARGO-ADSG is an extended version of FARGO (Masset 2000), which uses a finite
difference scheme with standard source and transport steps that are similar to ZEUS (Stone
\etal 1992).  However, additional orbital advection has been applied which removes the
average azimuthal velocity so that truncation error is reduced and the timestep allowed by
the Courant-Friedrichs-Lewy condition is significantly increased.  Thus FARGO-ADSG enables us
to perform high resolution simulations over long timescales. Furthermore, the effect of the
central star acceleration due to the non-axisymmetric potential of the disk is included
as an indirect potential term in the fluid equations (Baruteau \& Masset 2008).

\subsection{Radiative cooling and infall}

Starting from the publicly available version of FARGO-ADSG, we have implemented 
the hydrodynamic equations with both a simplified radiative cooling treatment 
using detailed opacities (Appendix A, updated from Zhu \etal 2009b) and mass infall.
The hydrodynamic equations are
\begin{equation}
\frac{\partial \Sigma}{\partial t}+ \nabla\cdot (\Sigma
\rm{v})=f_{in}\,,\label{eq:mass}
\end{equation}
\begin{equation}
\Sigma\left(\frac{\partial \rm{v}}{\partial
t}+\rm{v}\cdot\nabla\rm{v} \right)=-\nabla P-
\Sigma\nabla\Phi-\nabla\cdot\Pi+\rm{F_{in}}\,,\label{eq:momen}
\end{equation}
\begin{equation}
\frac{\partial E}{\partial t}+\nabla\cdot(E {\rm v})=-P\nabla\cdot
\rm{v}+Q^{+}-\Lambda_{C}+E_{in}\,,\label{eq:energy}
\end{equation}
where $\Sigma$ is the disk surface density, $E=\int e dz$ is the internal energy per 
unit area, $P=\int p dz$ is the vertically integrated pressure, $\Pi$ is the
viscous stress tensor (which is turned off in this paper), and $\rm{v}$ and 
$\Phi$ are the velocity and gravitational potential (including the self-gravity 
potential). $\rm{v}$ is constant along $z$ and $\Phi$ is derived from the thin disk 
approximation.  An ideal gas equation of state 
with $\gamma$=7/5 has been assumed, so that $P=E(\gamma-1)$. Then the averaged
disk 2D temperature is $T_{2D}=P\mu/\Re \Sigma$, where $\Re$ is the gas constant. It can 
be shown that if $T (Z)^4\propto T_{0}^4 \int \rho dZ$ in the vertical direction of a 
3D disk, which is close to the structure of a viscous heating dominated disk with a 
constant opacity, the disk's central temperature $T_{c}=4/5 T_{2D}$ where
$T_{2D}$ is calculated above with the total pressure and surface density. With irradiation
included, $T_{2D}$ is closer to $T_{c}$, so we set $T_{2D} = T_{c}$.
f$_{\rm in}$, F$_{\rm in}$ and E$_{\rm in}$ model the effects 
of infall on the disk mass, angular momentum, and energy. 
These terms are described below in Eqs. 6-8.

The cooling rate per unit area, $\Lambda_{C}$, is modeled via
\begin{equation}
\Lambda_{C}=\frac{16}{3}\sigma(T_{c}^{4}-T_{\rm ext}^{4})\frac{\tau}{1+\tau^{2}}\,,
\label{eq:cooling}
\end{equation}
where $\tau=(\Sigma/2)\kappa_{R}(\rho_{c},T_{c})$ is the optical depth to the 
disk midplane at radius $R$. The Rosseland mean opacity $\kappa(\rho_{c},T_{c})$
is calculated using the opacity table given in Appendix A with the disk
midplane density $\rho_{c}$=$\Sigma$/2H and the midplane temperature $T_{c}$. The
particular form $\tau/(1+\tau^{2})$ is chosen so that the cooling
term has the correct form in the optically-thick and -thin limits. 
(In the optically thin limit, $\kappa$  should be the Planck mean opacity.  
Because the dominant opacity is from dust, which has a relatively slow variation with 
wavelength, we just use the Rosseland mean opacity for simplicity.)

The term $\sigma T_{\rm ext}^{4}$ represents the heating effect of the irradiation from 
the central star, assumed to vary as
\begin{equation}
T_{\rm ext}^{4}=\frac{f(R)L_{*}}{4\pi R^{2}\sigma}\,,\label{eq:text}
\end{equation}
where $L_{*}$ is the total luminosity of the star and $f(R)$ accounts
for the normal component of the 
irradiation from the central star to the disk. Here
we set $f(R)$ to have the constant value $f(R)=0.1$, which is
roughly what is expected for flared disks (Kenyon \& Hartmann 1987; 
Chiang \& Goldreich 1997; D'Alessio \etal 1998), and $L_{*}=$L$_{\odot}$. For disks 
without internal heating by spiral shocks, the temperature will relax to 
$T_{\rm ext}$ based on Eq. 4.

The effect of infall onto the disk is represented by
f$_{\rm in}$, F$_{\rm in}$ and E$_{\rm in}$ in Eqs. 1-3. 
f$_{\rm in}$ and E$_{\rm in}$ are the rate of disk mass and internal energy
increase per unit area due to the infall, which can be directly
derived from an envelope infall model. The form of F$_{\rm in}$, however,
is less immediately apparent; it is the equivalent shear force induced from merging
infalling mass with the disk mass when they have different velocities.
The relationship between F$_{\rm in}$ and the azimuthal velocity of the infalling 
envelope can be derived if we write the angular momentum equation from 
Eqs. \ref{eq:mass} and \ref{eq:momen}
\begin{eqnarray}
\lefteqn{\frac{\partial J}{\partial t}+\frac{1}{R}\left(\frac{\partial
(v_{\theta}J)}{\partial \theta}+\frac{\partial(Rv_{R}J)}{\partial R}\right)}
\nonumber \\
& & =-\frac{\partial P}{\partial \theta}-\Sigma\frac{\partial
\Phi}{\partial \theta}+R({\rm f_{in}}v_{\theta}+{\rm F_{in,\theta}})\,,
\end{eqnarray}
where $J=\Sigma R v_{\theta}$. Thus
$R({\rm f_{in}} v_{\theta}+{\rm F_{in,\theta}})$ is the additional angular momentum brought
by the envelope falling onto the disk per unit area, which should be
$R {\rm f_{in}} v_{\rm in,\theta}$, where $v_{\rm in,\theta}$ is the azimuthal
velocity of the infalling envelope as it lands on the disk at radius $R$.
If the infalling envelope has the same velocity as the disk
rotational velocity ($v_{\rm in,\theta} = v_{\theta}$), there is no shear 
between the disk and the infalling envelope and F$_{\rm in,\theta}$=0.

In our simulations, we minimize the envelope's impact on the disk by
assuming F$_{\rm in,\theta}=0$ in order to study the disk's response just to
the mass loading from the infalling envelope.
This is actually not a bad approximation based on the rigid rotating envelope 
infall model of \cite{tsc} and Cassen \& Moosman (1981), which suggest most of the
infalling material falls onto the centrifugal radius ($R_{c}$) where it
has just slightly less than Keplerian rotation ($v_{\rm in, \theta}\sim v_{\theta}$). 
This also implies F$_{{\rm in},R}$=0. In our simulation we add mass to the disk within 
a small range in radii ($R_{a}$ to $R_{b}$) according to
\begin{equation}
{\rm f_{in}}(R)=\frac{\dot{M_{\rm in}}}{2\pi
R}\frac{(1-p)R^{-p}}{R_{b}^{1-p}-R_{a}^{1-p}}\label{eq:infallfin}
\end{equation}
so that by integrating over $R_{a}$ to $R_{b}$ the total infall rate is
$\dot{M_{\rm in}}$. The parameter $p$ is chosen to be 0.75 so that the disk within
$R_{a}$ and $R_{b}$ has the same $Q$ during the infall as if the disk is irradiation 
dominated (see Eq. 5). We also assume that the infalling material has a temperature 
equal to $T_{\rm ext}$ such that
\begin{equation}
{\rm E_{in}}(R)={\rm f_{in}}(R)\frac{k_{\rm B}T_{\rm ext}}{(\gamma-1)\mu m_{\rm H}}.
\end{equation}
This is not precisely correct; backwarming from the envelope can
actually heat the disk.  We ignore this effect for simplicity in our simulations, 
but we discuss its possible impact in \S 7.2.

\subsection{Initial conditions and simulation setup}
Currently, most simulations studying GI start with massive and non-irradiated disks.
In such cases the occurence or otherwise of fragmentation depends
upon the initial conditions (Boley 2009). Without external heating (irradiation), 
a disk with a finite initial temperature can cool to a point such that the
Toomre parameter $Q \equiv c _{\rm s}\kappa / \pi G \Sigma \sim 1$, where $\kappa$
is the epicyclic frequency,
and the disk becomes gravitationally unstable. The equivalent
orbital cooling time ($t_{\rm cool}$) sensitively depends on the initial surface
density distribution,
\begin{equation}
t_{\rm cool}\Omega\sim\frac{\Omega E}{\sigma T_{\rm eff}^{4}}\,.
\end{equation}
In the optically thick, viscous heating dominated case 
$T_{\rm c}^{4} = (3/8) T_{\rm eff}^{4}\,\tau$ and the cooling time becomes
\begin{equation}
t_{cool}\Omega\sim\frac{\Omega \Sigma
c_{v}T_{c}3 \Sigma C T_{c}^{a}}{\sigma
16 T_{c}^{4}}\propto\frac{1}{\Sigma^{4-2a}R^{10.5-3a}}\,,\label{eq:tcoo}
\end{equation}
where we have assumed that $Q$ is constant to constrain the temperature , $\kappa_{R}= C
T_{\rm c}^{a}$, where $C$ is a constant, and the disk follows the Keplerian rotation
law (so that $\kappa=\Omega=\Omega_{K}$) in obtaining the final proportionality relation.  
Since $a=1.5$ for our grain opacity (Appendix A), larger initial surface densities at 
larger radii lead to shorter cooling times and fragmentation.

During a realistic star formation process the disk is built up over a finite
time from the infall of the parent molecular cloud core.  Thus the
initial conditions depend on the mass and angular momentum distribution
within the cloud, which can be quite complicated, and in any event are
poorly constrained. Moreover, the thermal history of the disk can affect
disk fragmentation (Clarke et al. 2007). To simulate the disk mass growth
from infall until it becomes gravitationally unstable we start with a
relatively low mass disk and then gradually add mass to the disk at a
constant rate.  When the disk becomes massive enough and gravitationally
unstable, spiral arms generated by the GI will transport the infalling
mass to the central star at a rate matching the infall rate if the system
can achieve a steady-state.  With this procedure, we avoid strong
relaxation at the beginning of our simulations which may cause
fragmentation due only to initial conditions; this turns out to be
important for simulation convergence, discussed in \S 2.3. It also allows
us to control the infall rates and infall radii as two free parameters.

We have performed a variety of simulations with infall at different rates
and radii.  Since the disk undergoing GI tries to accrete to match the
infall rate, $\dot{M}\sim\dot{M_{\rm in}}$ in order to maintain $Q\sim$1,
we can vary $\dot{M_{\rm in}}$ to control the disk accretion rate.  These
simulations are summarized in Table 1.  Different cases are named as
R+infall radii+infall rates. Thus R40\_3e-4 means the typical infall
radii is 40 AU, while the infall rate at $\sim$ 40 AU is
$3\times10^{-4}\msunyr$.  The special cases have notations after the
infall rates which are self-explanatory or explained in the footnote.  In
the standard cases, the azimuthal grid resolution is 512.  We use
logarithmically spaced radial grids and choose the radial resolution such
that every grid cell has the same azimuthal and radial size. In most
of our simulations the radial resolution is $\sim$ 400, with the inner
radius at 5 or 10 AU and the outer radius at 1000 or 1500 AU (see Table
1). We set $f(R)$=0.1 in Eq.~\ref{eq:text}, so that $H/R\sim 0.029(R/{\rm
AU})^{0.25}$ (or T$_{\rm irr}$(1 AU)=215 K with irradiation heating
only).  The initial disk surface density is $\Sigma= 4.16\times 10^{4}
(R/{\rm AU})^{-1.75}$g cm$^{-2}$ with an exponential cut-off beyond
$R_{b}$  (obtained by multiplying the
surface density by the factor $\exp(5-5R/R_{b})$ if $R>R_{b}$, where $R_{b}$ is
defined in Eq.~\ref{eq:infallfin}).  With this setup, the disk is
gravitationally stable ($Q\sim$2) within $R_{b}$, and the disk surface
density is extremely small close to the outer boundary at $\sim 1000$ AU.
With infall, the disk gradually becomes gravitationally unstable near 
infall radii; spiral arms develop and start transporting mass inwards,
resulting in the inner disk gradually becoming gravitationally unstable.

We ran our simulations for up to 5$\times$10$^{4}$ years for
non-fragmenting cases. If the disk fragments, we run the simulation until
a clump is close to the inner boundary and causes negative densities
(details in Appendix B).  (In the R200\_3e-6 case we ran the simulation
for 2$\times$10$^{5}$ years to test whether this disk fragments on a
longer timescale.) A timescale of 5$\times$10$^{4}$ years is not much
smaller than the estimated length of the infall phase for low-mass
protostars, and it is equivalent to 17 orbits at our largest infall
radius (200 AU), long enough to see if the disk will fragment.

We assume the central protostar has a fixed mass (M$_{\odot}$) throughout
our simulations in order to study the disk response to the mass infall
only.  Thus our simulations are approximating the disk evolution near the
end of protostellar collapse, when the central object is quite massive
and emits significant radiation. The expected mass of the GI disk from
$R_{\rm in}$ to $R_{\rm out}$ can be estimated by assuming $Q=1$
with the temperature set by our irradiation prescription
(Eq.~\ref{eq:text}):
\begin{equation}
M_{d}=0.22 \left(\left({R_{\rm out}\over {\rm AU}}\right) ^{0.25}-
\left({R_{\rm in}\over {\rm AU}}\right)^{0.25}\right) {\rm M}_{\odot} \,.
\end{equation}
With R$_{\rm out}$=100 AU and R$_{\rm in}$=10 AU, we have $M_{d}=0.3$
M$_{\odot}$.  Thus the disk is massive compared with the central
star if irradiation is considered. 

To study how irradiation affects fragmentation, we ran three additional
simulations with high, low, and no irradiation, respectively. The high
irradiation case, R200\_3e-5L100, is similar to R200\_3e-5 but with 100
$L_{\odot}$, typical of an A type star.  On the contrary, the decreasing
irradiation case, R100\_1e-5I, is similar to case R100\_1e-5, but after
5$\times$10$^{4}$ years, with the same infall rate, the irradiation
parameter in Eq.~\ref{eq:text} gradually decreases at a rate
\begin{equation}
f(R)=0.1 \left(1-\frac{t}{5\times10^{4} \rm{years}}\right)^{8}\,,\label{eq:frt}
\end{equation}
until the disk fragments. Equation \ref{eq:frt} is chosen so that the
disk scale height due to irradiation decreases linearly with time and
the disk mass decreases linearly with time (assuming $Q=$constant). The
resulting mass accretion rate due to the decreasing irradiation is much
smaller than the infall rate at R$>$85 AU, so the disk accretion
rate is still controlled by the infall rate.  Finally, we performed one
simulation similar to R200\_3e-6 but without irradiation at all, which is
labeled as R200\_3e-6noirr. The entire suite of parameters explored in
our simulations is summarized in Table 1.

\subsection{Numerical tests}

To explore how robust our simulations are, we have carried out
simulations with different settings. The most important issue is the
convergence on the fragmentation criterion raised by Meru \& Bate 2011.
Some other issues include the $\gamma$ value and the gravitational force
smoothing length. 

\subsubsection{Resolution tests}

To study whether the critical fragmentation radius depends on the
numerical resolution, we have re-calculated several marginal
fragmentation and non-fragmentation cases, but with double the resolution
in both radial and azimuthal directions as shown in Table 1.
Non-fragmentation cases remain non-fragmenting at higher resolution and
the fragmentation case still fragments.  A particular concern applies to
the non-fragmentation case, in which numerical dissipation may prevent
disk fragmentation. Thus, for R100\_1e-5 case, we double and even
quadruple the resolution in both radial and azimuthal direction, but the
disk still remains non-fragmenting. We want to point out that the
irradiated disk has larger scale height which is better resolved
numerically. We conclude that an azimuthal resolution of 512 cells is
enough to study disk fragmentation, which is expected considering each
disk scale height is resolved by 7.5 grid cells at 100 AU (considering
$H/R\sim 0.029(R/{\rm AU})^{0.25}$) and the neutral stability critical
wavelength ($L=2\pi H$ for $Q=1$) is resolved by 45 grid cells, far larger
than that required by the condition given in Nelson (2006).

Recent SPH simulations by Meru \& Bate (2011) have questioned the
convergence of the orbital cooling fragmentation criterion for GI disks.
However, in our resolution tests from 512 to 2048 azimuthal grid cells
and the corresponding quadruple resolution increase in the radial
direction, our results converged for both non-fragmenting and fragmenting
disks. We attribute this discrepancy to the fact that we gradually build
the disk from a gravitationally stable state to an unstable state by adding
mass to the disk. This process not only simulates the real astronomical
condition of infall but also allows the disk to regulate itself.  In a
sense this is similar to Clark \etal (2007), who increased disk cooling
gradually.  Simulations by Paardekooper \etal (2011) confirmed that the
treatment in Clark \etal (2007) leads to a convergent cooling criterion,
which is consistent with our finding that gradually built-up disks
converge.

\subsubsection{Adiabatic index}

The two dimensional adiabatic index $\gamma$ is related to the three
dimensional $\Gamma$ in the static limit (e.g., Goldreich \etal 1986). In
a strongly self-gravitating disk, $\gamma=3-2/\Gamma$. If the three
dimensional $\Gamma=1.4$, the two dimensional $\gamma$=11/7$\sim$ 1.57. We
explore both the quasi-steady GI and highly dynamic fragmenting states,
so it is not obvious what $\gamma$ value we should choose.  We have
simulated  two cases, non-fragmenting and fragmenting disks, with both
$\gamma$=1.4 and 1.57 (Table 1).  Our results suggest that the
fragmentation criterion does not depend sensitively on $\gamma$ in the
range from 1.4 to 1.57.  This is consistent with Rice \etal (2005).  On
the other hand, Boley \etal (2007) have found that variation of $\gamma$
due to thermalization of the rotational levels of molecular hydrogen can
affect gravitational instability (Boley \etal 2007).  In our already
simplified 2D simulations, however, we simply assume $\gamma$ is
constant. 

\subsubsection{Gravitational force smoothing length}

To avoid small scale density fluctuations leading to a run-away collapse,
we use a smoothed gravitational force (Baruteau \& Masset 2008).
Considering that the disk vertical structure can smooth the disk's global
gravitational potential, we assume a smoothing length of 0.0264 R, which is
0.6 disk scale height at 5 AU and 0.3 disk scale height at 100 AU. 
 Since a large smoothing length may lead to bigger clumps, we
have carried out one simulation with smoothing length of 10 times smaller. 
The clump mass from this simulation turns out to be close to
the clump mass from the corresponding large smoothing length simulation.
This is consistent with Baruteau \etal's (2011) finding that a small
smoothing length does not change the properties of the gravo-turbulence.

\section{Results: Fragmentation}

\begin{figure}
\includegraphics{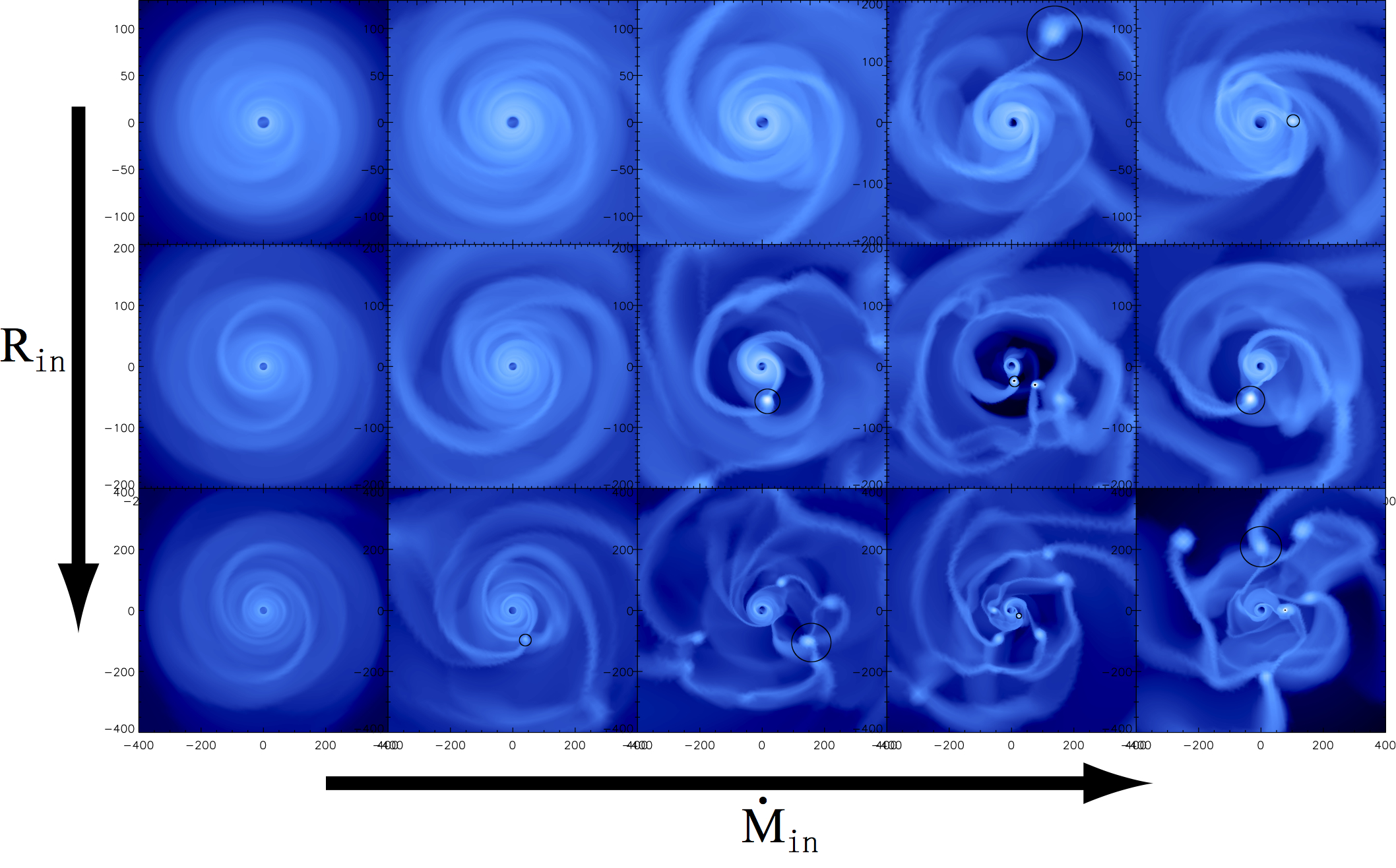} \caption{The disk surface density distribution at the end of the simulations 
(time shown in Table 1) with different infall rates (increasing from 3$\times$10$^{-6}\msunyr$ on the left to 
3$\times$10$^{-4}\msunyr$ on the right) and infall radii (65 AU, 100 AU, 200 AU from the top to 
bottom). The black circle labels the Hill radius of the selected clump if the disk fragments. } \label{fig:fig2map2}
\end{figure}

Images of the disk surface density for models with different infall rates
and infall radii at intermediate times are shown in Figure
\ref{fig:fig2map2}. In this figure, the infall rate increases
horizontally from 3$\times$10$^{-6}\msunyr$ (left) to
3$\times$10$^{-4}\msunyr$ (right).  The infall radius runs vertically
from 65 AU (top) to 200 AU (bottom).  These runs are listed from
R65\_3e-4 to R200\_e3e-6 in Table 1.  At a given infall rate (moving from
top to bottom down a column in Fig.~\ref{fig:fig2map2}), the spiral arms
become narrower and more unstable with increasing infall radius and
eventually break into fragments. At marginal fragmentation only one
clump forms, while for a more unstable case, several clumps form
simultaneously.  Higher infall rates at the same radius (moving from
left to right along a row in Fig.~\ref{fig:fig2map2}) also lead to more
unstable spiral arms.  Similarly, fragmentation becomes more severe (one
clump to multiple clumps) with higher infall rates at the same radius.
The critical radius where the disk starts to fragment increases with
decreasing infall rate (marginal fragmentation cases form a diagonal line
in Fig.~\ref{fig:fig2map2} moving from top right toward bottom left).  

Figure \ref{fig:frag} summarizes our fragmentation results, and they are
in approximate agreement with our previous analytic estimates for the
critical radius for fragmentation as a function of the mass infall rate
(Zhu \etal 2010b, and further details in \S 6). These estimates are shown
as curves in Fig.~\ref{fig:frag}.

The special case R100\_1e-5I (decreasing irradiation) is shown in
Fig.~\ref{fig:figclump2}. This disk fragments at R=400 AU when $f(R)$
decreases to 0.013.  A comparison simulation that maintains full
irradiation (run R100\_1e-5), which is shown in the lower right panel of
Figure \ref{fig:figclump2}, does not fragment.  We thus confirm that
irradiation tends to suppress fragmentation at fixed infall rate. In the
zero irradiation limit model R200\_3e-6noirr fragments, unlike its
counterpart with full irradiation, which again confirms that irradiation
suppresses the fragmentation. 

In the high irradiation limit (run R200\_3e-5L100) the disk did not
fragment by the end of our simulation (5$\times$10$^{4}$ years), unlike
its 1 L$_{\odot}$ counterpart.  Again: irradiation suppresses
fragmentation.  This also suggests that for high-luminosity systems
(where $L \sim 100$ L$_{\odot}$) gravitational instability can transfer
mass inwards smoothly from hundreds of AU; an analytic treatment of this
result is given in \S 6.1. 

\section{Results: Clump evolution}

In a fragmenting disk, clump formation and evolution should resemble
protostar formation and evolution (Larson 1969, Bodenheimer \etal 1980).
During disk/cloud fragmentation, a quasi-hydrostatic core forms (this
stage is called the ``first core'' in protostellar studies). When the
core's central temperature rises from $\sim$1500-2000 K, the dust
sublimates, then hydrogen molecules begin to dissociate.  At this point
the ratio of specific heats $\gamma$ decreases below 4/3, the clump
becomes dynamically unstable and it collapses to form the so-called
``second core'' (Larson 1969).

In our 2D simulations we cannot simulate the dynamically unstable
``second core'', and whenever dust sublimates at the midplane our
assumption that the disk has uniform opacity vertically becomes invalid.
Thus, although we allow the clump to be heated up above 1500 K, our
results at those stages are not accurate.  Whenever the clump is above
2000 K and the center of the clump has collapsed to the ``second core'',
a circum-clump disk will form, and its accretion becomes important for
the clump's further growth. In this paper, we focus on the ``first core''
stage.  Even for this stage, we found there are three distinct fates for
the clumps.  During the clump's inward migration, if the clump grows
massive enough to open a gap, it slows down or even stops migration. On
the other hand, if the clump fails to open a gap, continuous migration
can eventually lead to tidal destruction.  There is a third possibility
that the clump manages to migrate inward all the way to the central star
without being tidally destroyed.  Since our inner boundary is at $\sim$
10 AU, we can not follow the clump all the way to the central star to
confirm this possibility, but provide further analytic estimates to
investigate this in \S 6.2.  Nonetheless, our simulations do exhibit
clumps that manage to reach the inner boundary of the computational
domain without being tidally disrupted.

We want to emphasize that all the clumps studied hereinafter are limited
to the marginally fragmentation cases, in which only one or two clumps
form in the disk. They do not suffer the strong clump-clump scattering
seen in cases of multiple fragmentation, which complicates their
evolution. 

\subsection{Clump structure}

Figure \ref{fig:fig2grid} shows the clump from the run R100\_3e-5 with
the numerical grid overlaid on the surface density distribution.  The
left panel shows the clump when it first forms at 150 AU after
2.2$\times$10$^{4}$ years, and the right panel shows its density 2000
years later when it has migrated in to 50 AU.  The clump is well resolved
at both times.

Figure \ref{fig:relT} shows the temperature distribution of the clump
(corresponding to the right panel of Figure \ref{fig:fig2grid}).
Velocity vectors are superimposed showing the velocity field measured in
a frame rotating at the same rate as the centre of the clump.  Shocked
regions form at the edges of the clump where the infalling material from
the disk collides with the pressure-supported clump (a similar phenomenon
is observed in the 1D simulations by Larson 1969). As the clump accretes
gas with high relative angular momentum, it develops significant rotation
(2 km s$^{-1}$, compared with the Keplerian rotation speed of 4 km
s$^{-1}$ in the clump rotating reference frame, as discussed in detail
below). The prograde clump rotation is similar to that seen in
simulations of circumplanetary disks (Lubow \etal 1999), presumably
because of a similar flow pattern occurring as gas enters the clump Hill
sphere from the surrounding protostellar disk. This is also consistent
with the results of Voroybov \& Basu (2010) and Boley \etal (2010). 

The clump's structure at the different times shown in
Fig.~\ref{fig:fig2grid}  are revealed in detail in Fig.~\ref{fig:f220}.
The clump's central density and temperature increase with time during the
collapse of the clump.  For a hydrostatic core, the clump's self-gravity
should be balanced by the sum of the pressure and centrifugal forces due
to the clump's rotation.  When the clump first collapses, it collapses
radially until it is pressure supported.  Since at this early stage
little mass has been accreted that adds angular momentum to the clump,
the clump rotates slowly. The pressure gradient $\partial P/\partial R$
then balances the clump's self-gravity and is almost one order of
magnitude larger than the centrifugal force (left panel of
Fig.~\ref{fig:fig2grid}).  Later, loss of thermal energy due to radiative
cooling leads to further clump contraction, and the clump spins up. Disk
mass with high specific angular momentum is also accreted, increasing the
rotation of the clump, and centrifugal force becomes comparable to the
pressure gradient as shown in the right panels of Fig.~\ref{fig:f220}. In
order to qualitatively estimate how much angular momentum is brought by
the accreted gas from the disk, an analogy can be drawn between accreting
giant planets in disk gaps and our accreting clumps.  Ayliffe \&
Bate (2009) have shown that the accreted gas from the disk falling onto
the planet forms a circumplanetary disk with a radius 1/3 of the planet's
Hill radius.  Thus if the angular momentum of the accreted gas dominates
our clump's initial angular momentum, our clump is expected to be close
to or even larger than 1/3 of the planet's Hill radius considering the
clump is also partly pressure supported.  This is confirmed in our
simulations, shown below, where our clump radii are close to half their
Hill radii.

To study clump evolution further, we define a clump mass, $M_{c}$, and
radius, $r_{c}$, as follows.  First, we choose a large circle around the
clump and integrate the total mass. If the Hill radius
$r_{H}=R(M_{c}/3M_{*})^{1/3}$ calculated by this mass is larger than the
the initial radius of the circle, we choose a smaller circle by iteration
until the Hill radius is the same as the radius of the circle.
Then we define the clump center as the position of peak surface
density and we compute the average disk surface density $\Sigma_{H}$
around the circumference of the clump's Hill radius.  After this step, we
look for the closed curve $r(\theta)$ around the clump center where its
surface density is equal to $0.5*
(\log_{10}\Sigma_{H}+\log_{10}\Sigma_{max}$), where $\Sigma_{max}$ is the
surface density at the clump center.  We then define the clump radius as
the average distance from the curve to the clump center ($r_{c}$=$\int
r(\theta)d\theta/2 \pi$, where $\theta$ is the angle between the line
connecting each position on the curve and the clump center with respect
to a reference direction, taken to be the line joining the the central
star to the clump).  This clump radius can be used to determine whether
or not the clump can be tidally destroyed, by comparing it with the Hill
radius, which will be discussed in detail in \S 5.2.

Quantitatively, the clump in the left panel of Fig.~\ref{fig:fig2grid}
for R100\_3e-5 case is at 139 AU with mass 0.026 M$_{\odot}$ and Hill
radius 26 AU. The total specific angular momentum J with respect to the
clump center is 4$\times$10$^{17}$ cm$^{2}$ s$^{-1}$ which is similar to
Boley \etal (2010). When the clump accretes mass, both mass and angular
momentum increase.  For the clump in the right panel of Fig.
\ref{fig:fig2grid}, its mass is 0.159 M$_{\odot}$ (the upper panel in
Fig.~\ref{fig:clump}) and the specific angular momentum is 10$^{19}$ cm$^{2}$ s$^{-1}$.
Such high specific angular momentum can limit contraction of the clump
even in the ``first core'' stage.  For an order of magnitude estimate,
the clump in the right panel cannot contract to become smaller than $\sim$AU scale.
This is roughly consistent with the 8 AU clump radius calculated using
the procedure defined in the last paragraph.  Even with the significant
growth in mass, the core radius remains approximately half the
Hill radius at all times, as shown by the right-hand vertical axis in the
middle panel of Fig.~\ref{fig:clump}.  However, the angular momentum
distribution within a clump and how it is transported outwards is
unclear. We do not include explcit viscosity in these simulations,
so evolution of the clump spin angular momentum distribution can arise 
through tides and numerical diffusion.

Considering that the clump's dynamical timescale (free-fall timescale) is
significantly shorter than its Kelvin-Helmholtz timescale (Nayakshin
2010a), the clump can be approximated as quasi-hydrostatic during its
collapse (more timescale estimates are discussed in \S 5.2). This is
confirmed by the bottom panel of Fig.~\ref{fig:clump} where $GM_{c}/\Re
r_{c}T_{c}\sim 1.5$ is almost constant throughout the clump evolution. As
shown in Eq.~\ref{eq:gmort} in Appendix C, for a spherical, hydrostatic,
radiative clump, we expect $GM_{c}/\Re r_{c}T_{c}\sim 0.7$. However,
considering the rather rough definition of $r_{c}$ in our 2D simulation,
the geometric difference between 2D and 3D, and using the disk midplane
opacity to represent the envelope opacity at that disk radius, a factor
of two uncertainty is expected and can be roughly considered as our error
bar.
 
\subsection{Clump accretion}

In the previous subsection we explored clump structure by following an
individual clump.  From this subsection forward we will study general
clump evolution (e.g. accretion and migration). After the disk starts
fragmenting, a clump quickly evolves to a quasi-hydrostatic state. As
shown in the previous section, the core radius is approximately half of
the Hill radius, leaving plenty of space for disk material to accrete
onto the collapsed clump. Due to the stochastic nature of the clump
accretion and migration in GI disks, we will follow the clumps from a
number of disk fragmentation cases to increase our sample.

Clump accretion rates as a function of time are shown in the left panel
of Fig.~\ref{fig:clumpdm}. They vary from 10$^{-5}\msunyr$ at 200 AU to
10$^{-4}\msunyr$ at tens of AU. These accretion rates can be estimated by
assuming that the accretion cross section corresponds to 1-2 clump Hill
radii:
\begin{equation}
 \dot{M_c} = 2 \Sigma \int_{0}^{x_{\rm max}} dx v_{\rm shear}\label{eq:dmclump}
\end{equation} 
where $v_{\rm shear} = 3/2 \Omega x$, $x=R - R_{\rm clump}$ and $R_{\rm
clump}$ is the clump's radial position in the disk. $x_{\rm max}$
corresponds to the impact parameter where the last unbound orbit sits.
This impact parameter has been studied in the context of planetary rings
by Petit \& Henon (1986) and Henon \& Petit (1986), and in giant
molecular cloud scattering by Gammie \etal (1991).  We can understand the
gas flow in a gaseous Keplerian rotating disk by applying the results
from the circular restricted three-body problem. If disk material is
orbiting at a radius very close to the clump (small impact parameter),
the Jacobi integral is large and the disk material is restricted to the
horseshoe region in the vicinity of the corotation radius. If disk
material orbits far from the clump, it will follow its Keplerian orbit
with very little perturbation from the gravity of the clump. Only disk
material with intermediate impact parameters can be captured.  The impact
parameter of the last unbound orbit as in Eq. \ref{eq:dmclump} normally
lies between 1-2 Hill radii (Gammie \etal 1991; Bryden \etal 1999; Lubow
\etal 1999). The planet-disk interaction studies of Bryden \etal (1999)
and Lubow \etal (1999) have clearly demonstrated how the planet accretes
from the gaseous disk. Although spiral shocks complicate the orbits of
the disk material, gas in the circumstellar disk orbiting 1-2
Hill radii away from the planet is deflected into the Hill sphere by the
planet's gravity.  In a disk undergoing GI, the clump forms within a high
density spiral arm, and mass can also flow along this arm onto the clump,
as indicated by Fig.~\ref{fig:relT}. Even with this further complication,
we find that Eq. \ref{eq:dmclump} can describe the accretion rate onto a
clump accurately.

Assuming $x_{\rm max}=f_{c}r_{H}$, we can estimate
\begin{equation}
\dot{M_{c}}=\frac{3}{2}\Sigma\Omega (f_{c}r_{H})^{2}\,,\label{eq:dmclump2}
\end{equation}  
where higher order terms in $x/R$ are neglected.  Because the disk
surface density fluctuates vigorously with time and the migration is
stochastic, the accretion rate onto the clump also fluctuates.  Comparing
Eq.~\ref{eq:dmclump2} with our simulations, we found $f_{c}\sim 1.7$, and
\begin{equation}
\dot{M_{c}}=4\Sigma\Omega r_{H}^{2}\,,\label{eq:dmclump3}
\end{equation}
in basic agreement with the discussion above. With our parameter it is
$\dot{M_{c}}=1.7\times10^{-5}$ $ (R/100 AU)^{-1.25} $ $(M_{c}/10 M_{J})^{2/3}M_{\odot}/yr$.
  Equation \ref{eq:dmclump}
shares similarity with the planet accretion rate from D'Angelo 
\& Lubow(2008), but in our cases the Hill radius is larger than the disk scale height
and the cross section is $\sim r_{H}$ instead of $\sim r_{H}^{2}$. Accretion rates obtained
from the simulations, and estimated using Eq.~\ref{eq:dmclump3}, are
shown in the left panel of Fig.~\ref{fig:clumpdm}.  In our simulations, the
accretion rates are higher than what Boley \etal 2010 assumed. We think this is
due to the effect equivalent to gravitational focusing. When the flow
passes around the planet, it is gravitationally focused to the planet so that
its cross section increases (Fig. ~\ref{fig:relT}).  Clearly this result
is only valid if the core radius is smaller than the Hill radius;
otherwise the clump is tidally limited and accretion is suppressed.
Further reduction of the Hill radius would cause the clump to be tidally
disrupted.

\subsection{Clump migration} 

In most of our fragmentation cases, the clumps migrate inward roughly on the
type I timescale initially, given by the expression (Tanaka \etal 2002)
\begin{equation}
\tau_{mig}=\frac{R}{\dot{R}}=\frac{h^{2}}{4C q \mu}\frac{2\pi}{\Omega}\,,\label{eq:mig}
\end{equation}
where $C=3.2+1.468 \xi$, $\xi$ is the slope of the surface density 
$\Sigma \propto r^{-\xi}$, $h=H/R$, $\mu=\pi\Sigma(R)R^{2}/M_{*}$, and $q=M_{c}/M_{*}$,
where $M_{c}$ is the clump mass. With our disk parameters, this becomes
\begin{equation}
\tau_{mig}=784 \left(\frac{M_{c}}{0.01 {\rm M}_{\odot}}\right)^{-1}
\left(\frac{R}{100 \, {\rm AU}}\right)^{1.75} \, {\rm yrs}. \,\label{eq:mig2}
\end{equation}
Our simulations confirm this timescale estimate early in the simulations
shortly after clump formation, as shown in the right panels of
Fig.~\ref{fig:clumpdm} (the smooth solid curve is from
Eq.~\ref{eq:mig2}). This is consistent with recent papers on the same
subject by Baruteau \etal (2011) and Michael \etal (2011). The upper
solid lines in these panels are the migration trajectories followed by
the clumps in the simulations, and the spikes are a stochastic component
due to the interaction between the clumps and the spiral arms (and
sometimes other clumps).  Our simulations, however, suggest that as the
clump mass increases due to accretion, the migration deviates from the
typical type I migration time scale given by Eq.~\ref{eq:mig2}. This
appears to occur because the mass of the clump begins to approach the
local disk mass, such that the inertia of the clump plays a role in
slowing the migration rate. The type I migration timescale given by
Eq.\ref{eq:mig2} is obtained using a linear theory that assumes that the
underlying disc structure is unperturbed by the clump.  Although the
clumps are not able to clear gaps in the disk because of the large
effective viscosity, the disc response to the clump gravity is non linear
because the clump Hill radii exceed the local disc scale height. We find
that the clump migration rates can be reasonably-well fit by the
following expression that accounts for the slowing of migration because
of the clump inertia 
\begin{equation}
\tau_{mig,fit}=\tau_{mig}\left(1+\left(\frac{M_{c}}{2 R^{2}\Sigma(R)}\right)^{2}\right)\,.
\label{eq:mig3}
\end{equation}
Fits obtained using the above expression are shown by dashed lines
in the right panels of Fig.~\ref{fig:clumpdm}.

The clump represented by the black curve shown in the upper panel of
Fig.~\ref{fig:clumpdm} (run R100\_3e-5 ) slows down its migration when it
reaches $\sim$50 AU since it is massive enough ($\sim$0.3 M$_{\odot}$) to
open a gap in the disk (see the discussion in \S 5.1).
 
\subsection{Clump mass}

The clumps accrete mass from the disk during their migration as discussed
in \S 4.2. There are several characteristic masses which mark transitions
from differing types of evolution, as discussed further in \S 5. 

The mass limits of clumps in disks undergoing GI can be estimated
analytically and compared with our simulation results. The initial clump
mass can be estimated as the mass contained in the disk patch whose
lengthscale is equal to that of the most unstable mode (wavelength
$\lambda$=2c$_{s}^{2}$/G$\Sigma$ with Q=1). Thus
\begin{equation}
M_{c,ini}=\pi (\lambda/2)^{2}\Sigma(R)=\frac{ \pi c_{s}^{4}}{ G^{2}\Sigma(R)}=
\frac{\pi H^{4}}{R^{4}}\frac{M_{*}^{2}}{R^{2}\Sigma}\,. \label{eq:mclump}
\end{equation}
In our disk setup, $M_{c,ini}$=4.7$\times$10$^{-4} M_{*}(R/{\rm
AU})^{0.75}$ (the lower dotted curve in the upper left panel of
Figs.~\ref{fig:clumpgap}, \ref{fig:clumptidal}, \& \ref{fig:clumpmig}).
When the clump first forms from the break-up of a spiral arm, its shape
is elongated due to tidal forces from both the star and the rest of
spiral arm.  This elongation together with the shock structure along the
spiral arms makes the real initial clump mass different from that
estimated above. 

When the clumps are identifiable and become quasi-spherical, separating
from the spiral arms,  we found they are close to that would be
implied by Equation \ref{eq:mclump}.  But some clumps are a factor of 2 less
massive (e.g. the upper left panel of Fig. \ref{fig:clumptidal}) \footnote{We
want to point out that these initial clumps were identified by eye, and
when we are certain they are gravitationally bound; the clumps may be
less massive when they initially become bound objects.}.  This is
slightly higher than more detailed clump mass calculations and simulations by
Boley \etal (2010), suggesting that the initial clump mass can be a
factor of several smaller than that estimated using Equation
\ref{eq:mclump}. And our initial clumps are $\gtrsim$10 M$_{J}$,
larger than that found by Boley \etal (2010) and Hayfield \etal (2011).
We attribute this to the strong irradiation in our disks requiring larger 
disk masses before fragmentation can arise.

After a clump collapses, its size is smaller than its Hill radius,
enabling it to accrete from the disk during its orbit around the central
star (as discussed in \S 4.2). If the clump accretes all the mass along
its orbit, the maximum clump mass can be calculated by considering its
``isolation mass'', defined to be the mass contained in an annulus
centered on the orbital radius with half-width equal to the clump Hill
radius (e.g. Lissauer 1987; Rafikov 2001):
\begin{equation}
r_{H}=\left(\frac{M_{c}}{3M_{*}}\right)^{1/3}R\sim
\left(\frac{4\pi R r_{H}\Sigma(R)}{3M_{*}}\right)^{1/3}R\,,
\end{equation}
and
\begin{equation}
M_{c,iso}=4\pi R r_{H}\Sigma(R)=\frac{(4\pi\Sigma)^{3/2}R^3}{(3M_{*})^{1/2}}=
M_{*}\left(\frac{64 H^{3}}{3R^{3}Q^{3}}\right)^{1/2}\,.\label{eq:miso}
\end{equation}
With our disk parameters, $M_{c,iso}=0.023 M_{*}(R/{\rm AU})^{0.375}$.
This isolation mass is plotted as the upper dotted line in the upper left
panel of  Figs.~\ref{fig:clumpgap}, \ref{fig:clumptidal}, and
\ref{fig:clumpmig}. It should be noted that this mass is derived assuming
that the clump does not migrate during accretion. In reality, the
clump migrates and is therefore able to grow to a mass that is larger
than the isolation mass.

In our simulations, the surviving clumps start at a mass smaller than the
mass estimated by Equation \ref{eq:mclump} but quickly accrete to the
isolation mass.  This high accretion rate, and the resulting formation of
massive clumps, provides a different picture of clump evolution than the
analytic calculations of Nayakshin (2010a), where accretion onto the
clump was neglected.

\section{Clump fates}

Due to the stochastic nature of the migration and accretion, and the
different environments in which they form, the clumps
have different fates in our simulations. In this paper, we have only
traced the clumps in the marginal fragmentation cases, where 1 or 2
clumps coexist in the disk. When multiple clumps form,
clump-clump interaction becomes significant and examination of this
is left for a future study.  A total of 13 clumps are
generated in the marginally fragmenting cases (summarized in the 
last column of Table 1 and Table 2). Of these, 3 clumps open  gaps and 
stop migrating (as shown in Fig.~\ref{fig:clumpgap}), 4 clumps are tidally 
destroyed during their migration (three are shown in Fig.~\ref{fig:clumptidal}), 
and 6 clumps migrate across the inner boundary (three are shown in
Fig.~\ref{fig:clumpmig}). 

\subsection{Gap opening}

Figure \ref{fig:clumpgap} shows the evolution of the clumps that open
gaps and almost stop migration in the disks at the end of our
simulations. Results are taken from the simulations R65\_1e-4, R100\_3e-5
and R200\_1e-5 listed in Table 1.  In the R100\_3e-5 case (black curves),
we can see the trend that the clump's inward migration slows down when it
opens a gap at $\sim$50 AU. The change in surface density during gap
formation is shown in Fig.~\ref{fig:surf}. Two conditions need to be
satisfied for gap opening in a non self-gravitating disk. The first
condition is that the Hill radius of the clump needs to be larger than
the disk scale height. Using  Eq.~\ref{eq:mclump}, it can be shown that
this condition is met when the clump mass $M_{c}$ is smaller than the
initial clump mass calculated from Eq.~\ref{eq:mclump} by a factor of 12
(the lower dashed line in the upper left panel of
Figs.~\ref{fig:clumpgap}, \ref{fig:clumptidal}, and \ref{fig:clumpmig}).
Thus, this condition is satisfied almost all the time considering that the
initial clump is only a few times less massive than the mass given by 
Eq.~\ref{eq:mclump} .
The second condition is that $M_{c}/M_{*}>40/R_{\rm e}$ (Lin \&
Papaloizou 1979) where the effective Reynolds number at the clump
location is $R_{\rm e}= R v_{\rm k}/\alpha H c_{s}$.  Here, $v_{\rm k}$
is the Keplerian velocity.  For gap opening to occur, the clump mass must
satisfy the relation 
\begin{equation}
M_{c,gap} > 40 M_{*}\frac{\alpha H^{2}}{R^{2}}\,.\label{eq:mgap}
\end{equation}
With the irradiation temperature we adopted, $H/R \sim 0.029(R/{\rm
AU})^{0.25}$, and $M_{c,gap} >  0.034\, \alpha \, M_{*}(R/{\rm
AU})^{0.5}$. Since $\alpha$ can be as large as 1 (see 
\S 6.1), the gap opening value of $M_{c}$ is shown as the upper dashed
line in the upper left panels of Figs.~\ref{fig:clumpgap},
\ref{fig:clumptidal}, and \ref{fig:clumpmig} \footnote{The real gap
opening mass can be slightly smaller than that shown in these figures
since we assume the maximum $\alpha$=1 in these plots. }.
Fig.~\ref{fig:clumpgap} shows clearly that when the clump mass is above
this line, the migration slows down significantly. The gap opening is
illustrated further in Fig.~\ref{fig:surf}, and the gap forms when the
clump mass $M_{\rm c} \sim 0.17 M_{\odot}$, corresponding to the point
where the clump is close to the critical mass in Eq.~\ref{eq:mgap} and
stops migrating. Numerous previous studies have showed that even in the
presence of a gap, gas can accrete onto the gap forming object by
diffusion of gas through the gap (e.g. Bryden et al. 1999; Kley 1999;
Nelson et al. 2000).  It is for this reason that some of the clump masses
are able to grow beyond the gap forming mass in Fig.~\ref{fig:clumpgap}. 

For the black curve in the top right panel of Fig.~\ref{fig:clumpgap},
the clump appears to migrate inward sharply at the end of the simulation.
By investigating this inward migration in detail, we found another clump
forms at the edge of the gap, and the interaction between these two
clumps leads to this inward migration \footnote{The interaction between
these two clumps can be seen in the upper right panel of
Fig.~\ref{fig:clumpgap} where there are periodic oscillation for the
clump's position in the disk as soon as the second clump forms in the
disk.  The oscillation in the cyan curve is also due to the interaction
with another clump in the disk.}

\subsection{Tidal destruction}

Out of 13 clumps in our simulations, 4 are tidally destroyed during their
migration, as illustrated by the upper left panel of
Fig.~\ref{fig:clumptidal}). These cases support the clump tidal
destruction picture suggested by  Boley \etal (2010), Nayakshin (2010abc),
and Vorobyov(2011). As pointed out by these papers, since the condition
within the clump can be very different from the disk, different chemistry
and solid processing are expected. If the core can be tidally destroyed,
the products can be released back into the protoplanetary disk, affecting
planet formation.  The process of tidal destruction is shown in
Fig.~\ref{fig:figtidal}.  Starting from the upper left panel, the clump
is initially well within its Hill sphere. When the clump migrates inward,
its Hill radius decreases until it approaches the clump radius (upper
right panel). At this stage, due to the strong tidal force, the clump is
elongated along the radial direction (lower left panel). The clump is
continuously stretched and loses mass when it migrates further inwards,
and finally it dissolves into the spiral arm.

Whether or not the clump is tidally disrupted depends on the competition
between cooling/shrinking and migration.  If the migration is fast, and
the clump radius becomes larger than its Hill radius ($r_{c}>r_{H}$), the
clump will be tidally destroyed.  Assuming $r_{c}\sim r_{H}$ when the
clump first forms, this condition becomes
\begin{equation}
-\dot{r_{c}}<-\dot{r_{H}}=-\frac{r_{H}\dot{M_{c}}}{3M_{c}}-\frac{r_{H}\dot{R}}{R}
\end{equation}   
or
\begin{equation}
\tau_{\rm cool}>\frac{\tau_{\rm mig}}{-3\tau_{\rm mig}/\tau_{\rm acc}+1} \label{eq:tidal}
\end{equation}
where $\tau_{\rm cool}=-r_{c}/\dot{r_{c}}$ due to cooling, $\tau_{\rm
acc}=M_{c}/\dot{M_{c}}$, and $\tau_{\rm mig}=-R/\dot{R}$, if $r_{c}\sim
r_{H}$. The left side is the cooling timescale, while the right side is a
combination of the accretion  and  migration timescales. As shown above,
although the migration and cooling timescales are important, rapid
accretion prevents the clump's tidal destruction since increasing clump
mass increases the Hill radius.  The cooling timescale is given by
Equation \ref{eq:cooltimescale} below as
\begin{equation}
\tau_{\rm cool,r}= 104  \left(\frac{r_{c}}{10 {\rm AU}}\right)^{-2.5}
\left(\frac{M_{c}}{0.01 {\rm M}_{\odot}}\right)^{0.5} {\rm yrs}
\end{equation}
if the clump/core is radiative with the polytrope we assumed (see Appendix C), and
\begin{equation}
\tau_{\rm cool,c}= 122  \left(\frac{r_{c}}{10 {\rm AU}}\right)^{-1.8}
{\rm yrs},
\end{equation}
if the clump is convective.  During the clump accretion and contraction,
this timescale gets significant larger with time.  On the other hand,
with Eq.~\ref{eq:dmclump2}, the accretion timescale is
\begin{eqnarray}
\tau_{acc}& = & \frac{2 M}{3\Sigma \Omega (1.7 R_{H})^2}  \nonumber \\
          & = & 1112 
\left(\frac{M}{0.01 {\rm M}_{\odot}}\right)^{1/3}
\left(\frac{R}{100 {\rm AU}}\right)^{1.25} {\rm yrs}.
\end{eqnarray}
And the migration timescale from Eq.~\ref{eq:mig2} is
\begin{equation}
\tau_{mig}=784 \left(\frac{M_{c}}{0.01 {\rm M}_{\odot}}\right)^{-1}
\left(\frac{R}{100 {\rm AU}}\right)^{1.75} {\rm yrs}.
\end{equation}
Since these three timescales are close to each other, it is natural that
the clumps may have different fates due to the stochastic migration and
accretion. Equation \ref{eq:tidal} above suggests that rapid migration
combined with slow accretion can lead to clump tidal destruction.  As
shown in Fig.~\ref{fig:figtidal2}, two similar clumps can form at the
same time, but the one that undergoes rapid migration is tidally
destroyed, while the one that has a longer time to cool and shrink
survives. Rapid migration also limits a clump's ability to accrete, since
it is unable to cool and contract during the migration, and ends up more
or less filling its Hill (Roche) sphere. In the extreme cases this leads to
tidal disruption as described above.

\subsection{Migration to the inner boundary}

Although 6 out of 13 clumps migrate through the inner boundary without
gap opening and tidal destruction, this is simply because we adopt an
inner boundary that is located at $\sim$ 2.5 to 10 AU, so the ultimate
fate of these clumps is not known.  Some illustrative cases are shown in
Fig.~\ref{fig:clumpmig}, and are described in Table 1.  There is one gap
opening case and one tidal destruction case occuring close to 10 AU, as
shown in the Figs.  \ref{fig:clumpgap} and \ref{fig:clumptidal}, showing
that both gap opening and tidal destruction can occur at small radii.
With a smaller inner boundary, we would clearly expect more clumps to
open gaps or be tidally destroyed as they migrate into the inner disc. It
is unclear if any clump can survive migration all the way to the central
star. The naive timescale arguments given above seem to suggest that the clump
may be subject to tidal destruction at smaller radii due to the fact that
the migration timescale decreases significantly at smaller radii with
bigger clump mass ($M_{c}^{-1}$), while the cooling timescale increases
with bigger clump mass ($M_{c}^{0.5}$ or independent of clump mass). 
We note that when the clump mass exceeds the local disc mass the
migration slows down, and the migration time
increases with clump mass as indicated by Eq.\ref{eq:mig3}. Also,
because the clump central temperature is almost always above 1500 K when
it migrates to the inner disk (see the bottom panels in Figs.
\ref{fig:clumpgap}, \ref{fig:clumptidal}, and \ref{fig:clumpmig}), the
clump becomes dynamically unstable and should collapse to form a second
core (subject to overcoming the angular momentum barrier provided by its
spin). It will shrink considerably, leaving the final fate of the clump 
unclear. Further accretion may lead to gap formation and slowed
migration, increasing the survival probability of the clump.

From a statistical point of view, however, we might expect a fragmenting
disk to form a binary or multiple star system. If a less massive clump is
tidally destroyed or migrates to the central star, a new clump can still
form in the outer disk and migrate inwards. If one clump forms and
happens to become massive enough to open a gap, it will stay there and
survive to form a binary companion. The common occurence of close
binary star systems indicates that inward migration of clumps cannot
always lead to tidal disruption.

\section{Fragmentation and clump evolution: analytic models}

In this section, we will review analytic estimates for fragmentation
radii and develop a simple model for clump evolution. By comparing the
numerical results with the analytic estimates, we can constrain some
fundamental parameters in the analytic model (e.g. we will constrain the
critical $\alpha$ for the disk fragmentation) and study the roles played
by different physical processes.   

\subsection{Analytic estimates for fragmentation}

The parameter space for GI-induced disk fragmentation has been explored
analytically by several authors (Rafikov 2005, 2007, 2010; Levin 2007;
Kratter \etal 2008, 2010a; Cossins \etal 2010; Zhu \etal 2010b). The
derivations are summarized in Appendix D. 

Analytic estimates for the critical radius of fragmentation as a function
of the mass infall rate are shown as curves in Fig.~\ref{fig:frag} with
$\alpha_{c}$=0.06 or 1, where $\alpha_{c}$ is the critical fragmentation
viscosity parameter.  When the equivalent viscosity parameter of the
gravitationally unstable disk is larger than $\alpha_{c}$, the disk will
fragment.  Rice \etal (2005) has shown that for non-irradiated disks and
a variety of values of specific heat ratios, $\alpha_{c}\sim 0.06$.
Kratter \& Murray-Clay(2011) suggested that irradiation may alter the
disk fragmentation criterion.  Recently, Rice \etal (2011) have done 2-D
simulations  showing that the critical  $\alpha_{c}$ decreases with
irradiation.

However, this $\alpha_{c}$ was derived by assuming the disk cools at a
multiple of the orbital timescale throughout fragmentation. With realistic
radiative cooling this may not be the case because of the trapping of
radiative energy, as discussed later.

The trend of these curves can be understood under two limits: the disk
temperature is dominated by viscous heating due to the GI, or it is
irradiation dominated. In the viscous heating dominated limit, the
fragmentation radius $R_{f} \propto \alpha_{c}^{0.3} \dot{M}^{-0.07}
M_{*}^{1/3}\kappa_{R}^{2/9}$ (see Eq.~\ref{eq:fragr}). Thus, $R_{f}$ has
a very weak dependence on $\dot{M}$.  This is shown in
Fig.~\ref{fig:frag} where $R_{f}$ is almost vertical at $\sim$ 50 AU when
the disk accretion rate is high. On the other hand, if the opacity
depletes by a factor of 1000 ($\kappa_{R}$ is 0.1$\%$ of the nominal
value), $R_{f}$ decreases by a factor of 5 to $\sim$10 AU. 

In the irradiation dominated limit,
$R_{f}\propto(\dot{M}/\alpha_{c})^{-4/3}M_{*}^{1.95}$ (see
Eq.~\ref{eq:fragirr}). Thus, $R_{f}$ depends sensitively on $\dot{M}$ as
shown in Figure \ref{fig:frag}, but is independent of the opacity.  On
the other hand, in the envelope irradiation dominated limit (assuming
$T_{\rm irr}$=const) , the disk will fragment everywhere as long as
$\dot{M}>2c_{s,irr}^{3}\alpha_{c}/G$ ($Q \sim 1.5$ is assumed,
corresponding to the horizontal dotted line in Fig.~\ref{fig:frag}),
below which the disk will not fragment at any radius.
 
Furthermore, as shown in Fig.~\ref{fig:frag}, $R_{f}$ is very sensitive
to the value of $\alpha_{c}$ in the irradiation dominated case
($R_{f}\propto \alpha_{c}^{4/3}$) although it is less sensitive in the
viscous heating dominated case (R$_{f}\propto \alpha_{c}^{0.3}$). 

Fig.~\ref{fig:frag} shows that the analytic estimate agrees with
simulations reasonably well.  But it also suggests that $\alpha_{c}$ is
somewhat larger than 0.06, especially at low infall rates. We think this
discrepancy is caused by the following:

1) We assume the infall rates equal the disk accretion rates in
Eq.~\ref{eq:frag} and Fig.~\ref{fig:frag}.  In reality, in order to
conserve angular momentum, some of the infall mass will be transported
outwards, so the net inward mass flux will be smaller than the infall
rate. We estimate that all points in Fig.~\ref{fig:frag} should shift down
by $\lesssim$0.2 dec if the vertical axis is the disk accretion rate. 

2)  With irradiation, the disk is gradually built up from a gravitational
stable state to an unstable state by infall. This gradual build-up toward
GI leads to a transient marginal state which avoids the strong density
fluctuations that arise from an initially massive unstable disk as
considered in most previous simulations.  The gradual build-up of the GI
in our simulations makes the disk less likely to fragment.  This
mechanism is similar to that discussed by Clarke et al. (2007), who
showed that, compared with the cases starting with rapid cooling,
gradually increasing the cooling rate suppresses fragmentation.  

3)  With realistic cooling instead of orbital cooling, radiative trapping
makes gravitationally bound clumps harder to cool during their collapse,
suppressing disk fragmentation.  For example, the cooling time is defined
as the ratio of the thermal energy of the clump, $E$, to the clump
luminosity, $L_{c}$, which is given by
\begin{equation}
L_{c}=\frac{4\pi r_{c}^{2}\sigma T_{c}^{4}}{\kappa\Sigma_{c}}\,.
\end{equation} 
If we assume $\kappa=\kappa_{0}T^{a}$, and $\Sigma_{c}r_{c}^{2}\sim
M_{c}$, we have
\begin{equation}
\tau_{cool}\sim\frac{M_{c}^{2}c_{V} \kappa_{0}}{4\pi r_{c}^{4}\sigma T_{c}^{3-a}}
\end{equation}
where $c_{V}$ is the specific heat capacity. If we assume $T_{c} r_{c}$
is constant ($r_{0}T_{0}$), the above equation is
\begin{equation}
\tau_{cool}=\frac{M_{c}^{2}C_{V}\kappa_{0}T_{c}^{1+a}}{4\pi r_{0}^{4}T_{0}^{4}\sigma}\,.
\end{equation}
Since $a=1.5$, $t_{cool}\propto T^{2.5}$ (a more quantitative estimate is
given in Eq.~\ref{eq:cooltimescale}).  As the clump collapses and heats
up, the cooling time gets longer. Clumps with realistic cooling have
longer cooling times than clumps undergoing constant orbital cooling in
previous simulations used to determine $\alpha_{c}$.  We need a smaller
t$_{cool}$ (than that used in constant orbital prescription) to start
with for the clump to continue to collapse. This implies, because
$\alpha\propto 1/t_{cool}$, a larger equivalent $\alpha_{c}$ in the
analytic fragmentation estimate. 

Finally, notice that even if $\alpha_{c}$ is the same in the irradiation
dominated cases as the viscous dominated cases, irradiation still
suppresses disk fragmentation (as shown by the flattening of the critical
$R$ in the irradiation dominated regime in Eq.~\ref{eq:frag}). This is
simply due to irradiation setting a minimum disk temperature, so that the
disk needs to be hotter to cool.  With the higher internal energy and
longer cooling time, irradiated disks are harder to fragment.  From the
disk accretion point of view, even with the same $\alpha_{c}$, irradiated
disks have larger temperature so that the kinematic viscosity $\nu$
(which is $\propto T$) is larger and thus the critical accretion rate is higher.  
\footnote{The $\alpha$ description for GI assumes locality of the GI (see equations
\ref{eq:lambdac} and \ref{eq:tcool}), which has been studied by Cossins
\etal (2009).  They have shown that the pattern speed of spiral arms in
GI disks roughly equals the disk local Keplerian speed, so that the
anomalous term of the energy transport in GI disks is negligible.  On the
other hand, Cossins \etal (2009)  and Forgan \etal (2009) pointed out
that with higher and higher disk masses, the assumption of local energy
deposition gradually breaks down since lower mode spiral arms become
strong. Irradiated disks are normally massive, so our argument is not
exact and only acts as a guide.}. 

\subsection{Analytic models of clump evolution}

Although our simulations are two dimensional, we can construct simple
analytic models to study the clump evolution by assuming the clump is
spherical. In the following, we will calculate clump evolution using
models of hydrostatic self-gravitating gaseous spheres to represent the
first core of the clump, and follow the birthline calculations as in
protostar formation but with migration and accretion considered.  

The dynamical timescale for clump collapse is much shorter than its
Kelvin-Helmholtz (cooling) timescale (Nayakshin 2010a). Thus we can
assume the clump is in quasi-hydrostatic equilibrium (the first core)
before its central temperature reaches $\sim$1500 K when dust sublimates
and the opacity drops suddenly (\S 4.1).  Whether the first core is
convectively stable seems to depend on its density, and remains a matter
of debate (Wuchterl 1993; Nayakshin 2010a). In the following we will
discuss both cases.  We want to point out that our radiative core has a
polytropic index, n$_{e}$ (different from the gas adiabatic index), equal
to 1.5 and the convective core has a polytropic index equal to 2.5
(see Appendix C).  Boley \etal (2010) has found a polytrope with index 2.5
fits  Helled \& Bodenheimer (2010) data very well, which is not
surprising considering their core is convective.  Notice that for the
gaseous first cores we are discussing here, polytropic indices
running between 1.5 and 2.5 cover a wide parameter space. \footnote{A 
polytrope with index 1.5 can also represent the structure of a convective 
core for a monatomic ideal gas sphere with dust opacity having temperature 
dependence power a=1.5 (n$_{e}$=3-a).}

As derived from Eq.~\ref{eq:clumpL} in Appendix C, the luminosity is
given by
\begin{equation}
L_{c,r}=\frac{64\pi \sigma}{3\kappa_{0}}\left(\frac{G\mu}{\Re}\right)^{2.5}\frac{0.424^{1.5}M_{c}^{1.5}}{2.5 r_{c}^{-1.5}}\,.\label{eq:L}
\end{equation}
for a radiative core, and by
\begin{equation}
L_{c,c}=4\pi \sigma \left(20.48\kappa_{0}\right)^{-4/5}\left(\frac{G\mu}{\Re}\right)^{14/5}r_{c}^{4/5}M_{c}^{2}\,.\label{eq:LC}
\end{equation}
for a convective core (Eq. ~\ref{eq:Lconv}).

With our opacity this can be reduced to
\begin{equation}
L_{c,r}=2\times 10^{-3}{\rm L}_{\odot}\left(\frac{M_{c}}{0.01 {\rm M}_{\odot}} 
\frac{r_{c}}{10 {\rm AU}}\right)^{1.5}\label{eq:corel}\,.
\end{equation}
and (Eq. ~\ref{eq:Lconv2})
\begin{equation}
L_{c,c}=1.715\times 10^{-3}L_{\odot}\left(\frac{M_{c}}{0.01 M_{\odot}}\right)^{2} \left(\frac{r_{c}}{10 AU}\right)^{0.8}\,.
\end{equation}

Thus, the effective temperature is
\begin{equation}
T_{eff,r}=\left(\frac{L}{4\pi r_{c}^{2}\sigma}\right)^{0.25}=
26 {\rm K} \left(\frac{M_{c}}{0.01 {\rm M}_{\odot}}\right)^{0.375} 
\left( \frac{r_{c}}{10 {\rm AU}}\right)^{-0.125}\label{eq:coret}
\end{equation}
and
\begin{equation}
T_{eff,c}=\left(\frac{L}{4\pi r_{c}^{2}\sigma}\right)^{0.25}=
25 {\rm K} \left(\frac{M_{c}}{0.01 {\rm M}_{\odot}}\right)^{0.5} 
\left( \frac{r_{c}}{10 {\rm AU}}\right)^{-0.3}\label{eq:coretc}
\end{equation}

The clump central temperature is then (Eq. ~\ref{eq:tccox} and \ref{eq:tccox2})
\begin{equation}
T_{c,r}=138 K \left(\frac{ M_{c}}{0.01 {\rm M}_{\odot}}\right)
\left(\frac{r_{c}}{10 \, {\rm AU}}\right)^{-1}\,.\label{eq:Tc}
\end{equation}
\begin{equation}
T_{c,c}=179 K \left(\frac{ M_{c}}{0.01 {\rm M}_{\odot}}\right)
\left(\frac{r_{c}}{10 \, {\rm AU}}\right)^{-1}\,.\label{eq:Tcc}
\end{equation}
The clump loses energy through radiation and shrinks. Thus an energy
equation is required to derive the clump's evolution. The approach we
will take is quite similar to the  protostellar birthline calculations by
Hartmann \etal (1997), but instead of considering energy conservation,
we consider the conservation of the Jacobi-like energy integral
($\Gamma$) in a frame corotating with the clump.  We have
\begin{equation}
\Gamma=\int d^{3}x \rho \left(\frac{1}{2}v^{2}+u+w+\phi_{T}\right)
\end{equation}
where $\phi_{T} =- (3/2) \Omega^{2}x^{2}$ is the tidal expansion of the
effective potential about the clump, u is the internal energy per unit
mass, and w is the potential energy per unit mass.  Compared with the
total energy calculated by
\begin{equation}
E=\int d^{3}x\rho \left(\frac{1}{2}v^{2}+u+w\right)=K+U+W\,,
\end{equation}
an additional energy due to the tidal potential is present.  In the
picture of the clump accretion, each of these terms sensitively depends
on flow structure around the clump, which is uncertain.  First, due to
the tidal force,  the accretion flow from the outside of the Hill sphere
onto the clump does not conserve angular momentum with respect to the
clump unless a special flow geometry is met \footnote{In a shearing sheet
approximation, the flow needs to be symmetric/axi-symmetric in the R
direction to conserve angular momentum.}.  Thus, the structure within the
Hill sphere cannot be determined by rigorous  analytical arguments.
Second, even if the angular momentum is approximately conserved close to
the clump and far inside the Hill sphere, additional angular momentum
transport mechanism (e.g.  MRI) and the boundary layer physics may be
important. For example, if angular momentum within the Hill sphere is
weak, it is likely to form a rotationally supported giant clump.  On the
other hand, if it is like a mini-``protostellar system'', the clump is
compact and slowly rotating.  

For the most simple case, first we study the non-rotating hydrostatic
core with  a polytrope structure having the polytropic index $n_{e}$,
which has the following gravitational and internal energy 
\begin{equation}
E=U+W=-\left(1-\frac{1}{3(\gamma-1)}\right)\frac{3}{5-n_{e}}\frac{GM_{c}^{2}}{r_{c}}\,,
\end{equation}
With $\gamma=7/5$ for diatomic gas,
\begin{equation}
E=-\frac{0.5}{5-n_{e}}\frac{GM_{c}^{2}}{r_{c}}\,.
\end{equation} 

Consider $\Delta m$ amount of gas is added to the clump, with the total energy
\begin{equation}
\Delta E=-\varpi\frac{GM_{c}\Delta m}{r_{c}}\,
\end{equation}
where $\varpi$ represents the coefficient by adding all the factors from
gravitational energy,  thermal energy and the kinetic energy brought by
the accreted mass (Prialnik \& Livio 1985). For a cold accretion flow
which is in Keplerian rotation around the clump $\varpi$=0.5, while for
an accretion disk with boundary layer, $\varpi$=1. Then the star adjusts
to a new polytropic configuration with mass $M_{c}$+$\Delta m$ and radius
$r_{c}+\Delta r_{c}$. During the time interval $\Delta t$ of mass
addition, the conservation of the Jacobi-like energy L yields
\begin{eqnarray}
& & L_{c}\Delta t-\frac{0.5}{5-n_{e}}\frac{G(M_{c}+\Delta M_{c})^{2}}{(r_{c}+\Delta r_{c})} = \nonumber \\
& &  -\frac{0.5}{5-n_{e}}\frac{GM_{c}^{2}}{r_{c}}-\frac{GM_{c}\Delta m}{r_{c}}\varpi-\Delta \phi_{T}
\end{eqnarray}
This reduces to
\begin{equation}
L_{c}=-\frac{0.5}{5-n_{e}}\frac{GM_{c}^{2}}{r_{c}^{2}}\dot{r_{c}}+\left(\frac{1}{5-n_{e}}-\varpi\right)\frac{GM_{c}\dot{M_{c}}}{r_{c}}-\dot{\phi_{T}}\,.\label{eq:birth}
\end{equation}
At the Hill radius , $\phi_{T}$ is close to the gravitational potential
with respect to the clump.  Deep in the Hill sphere, $\phi_{T}$ is
significantly smaller than $u$. Thus if the clump is very small compared
with the Hill sphere, $\dot{\phi_{T}}$ can be ignored, as we will assume
in the following. 

As shown in Fig.~\ref{fig:f220}, the rotational energy becomes important
as the clump accretes a significant amount of disk mass, thus it is quite
uncertain what is the appropriate value for $\varpi$.

If the clump accretion is ignored, the cooling timescale is
\begin{equation}
\tau_{cool,r}=\frac{E}{L}=\frac{r_{c}}{\dot{r_{c}}}=102  
\left(\frac{r_{c}}{10 {\rm AU}}\right)^{-2.5}\left(\frac{M_{c}}{0.01 M_{\odot}}\right)^{0.5} 
{\rm yrs} \,,\label{eq:cooltimescale}
\end{equation}
\begin{equation}
\tau_{\rm cool,c}= 122  \left(\frac{r_{c}}{10 {\rm AU}}\right)^{-1.8} {\rm yrs}.
\end{equation}

In order to form a closed relationship so that we can evolve the clump's
properties with time, we have made several simplifications with attendant
large uncertainty. The first simplification is to treat the  structure of
the clump as a polytrope. Whether the clump is radiative or convective
changes the polytropic index significantly from $\sim$1.5 to 2.5. To
explore this uncertainty, we use both radiative and convective models.
The second uncertainty is the accretion physics/boundary layer around the
clump.  We will explore this by varying $\varpi$ from 0.5 to 1.  The
third simplification is assuming the clump is non-rotating, which is not
necessarily the case as shown in Fig.~\ref{fig:f220}.  The rotation of
the clump can change our analysis significantly, not only because the
clump changes shape due to rotation, but also the change of $\phi_{T}$ is
comparable to $u$ if the clump's radius is comparable to the Hill radius.
We do not attempt to explore this uncertainty since there is no
simple structure solution for a fast rotating clump.  Thus, this remains
a significant uncertainty in our analysis, which the reader should keep
in mind.  

Equations \ref{eq:dmclump3}, \ref{eq:L} or \ref{eq:LC}, \ref{eq:Tc} or
\ref{eq:Tcc}, \ref{eq:birth} form a closed model for clump evolution as
long as the initial clump properties ($M_{c}(t=0)$, $r_{c}(t=0)$) are
given. At time $t$, with known $M_{c}(t)$ and $r_{c}(t)$, the luminosity,
$L(t)$, can be derived from Eq.~\ref{eq:L}.  Then $r_{c}(t+\Delta t)$ can
be derived based on $\dot{r_{c}}$ from Eq.~\ref{eq:birth} with $L(t)$ and
$\dot{M_{c}}$ (Eq.~\ref{eq:dmclump3}). $M_{c}(t+\Delta t)$ can also be
derived with $\dot{M_{c}}$ from Eq.~\ref{eq:dmclump3}.  $T_{c}$ from
Eq.~\ref{eq:Tc} can be used to derive the opacity and the equation of
state.  At this point, all the quantities for the clump at time $t+\Delta
t$ have been updated.  Then with the migration velocity given by
Eq.~\ref{eq:mig3}, the clump's position can be updated. By comparing the
clump's core size with its Hill radius (which depends only on the clump's
mass and position in the disk), we can determine whether the clump can
continue accreting or will be tidally destroyed.

To compare with our simulations, we assume the initial clump at 150 AU
has mass 0.02 M$_{\odot}$ and the clump radius is half its Hill radius,
as suggested by the simulation results.  Following the above procedure,
the clump evolution is shown in Fig.~\ref{fig:clumpevolve1} with
n$_{e}$=1.5 or 2.5 and $\varpi$=0.5 or 1. As shown in the figure, the
clump evolution is not sensitive to the structure of the clump (n$_{e}$),
but very sensitive to accretion onto the clump ($\varpi$). This is due to
the high accretion rate onto the clump, which dominates the energy budget
in the energy equation above. In this sense, the interior structure of
the clump may affect the clump evolution significantly since a convective
core can transport angular momentum outwards more easily making the clump
rotate ridgedly.  Our simple theory predicts a higher central temperature
compared with our simulations shown in Figure \ref{fig:clumpmig}, which
could be due to our simple radiative cooling treatment in 2D simulations
or absence of clump rotation in the analytic models. 
  
As discussed in \S 5.2, the interplay between clump cooling, migration,
and accretion determines the clump's evolution and fate.  Cooling allows
the clump to shrink and inhibits tidal destruction, while inward
migration reduces the Hill radius and promotes tidal destruction. On the
other hand, accretion increases the Hill radius and inhibits tidal
destruction. Since these three timescales are comparable, as discussed in
\S 5.2, stochastic migration and accretion may lead to different clump
fates.  To examine the effect of varying parameters in our analytic
model, we calculated cases with 1/5 and 5 times the migration speed. As
shown clearly in Fig.~\ref{fig:clumpevolve2}, rapid migration leads to
tidal destruction around 30 AU. The clump properties here are similar to
our simulation results in Fig.~\ref{fig:clumptidal}.  On the other hand,
slower migration allows more time to cool, shrink and accrete.
Eventually, the clump becomes massive enough to open a gap during its
migration.  Clump properties from our simple model also agree with those
from our simulations of gap-opening clumps in Fig.~\ref{fig:clumpgap}.
  
Finally, we want to emphasize that this simple model is highly idealized,
and the behavior of $r_{c}$ depends on the polytropic index, the detailed
opacity, and the accretion physics around the clump. But it still
provides a useful tool for exploring different evolutionary scenarios,
and for comparison with hydrodynamic simulations.

\section{Discussion}

\subsection{Planets and multiple star systems}

Recent direct imaging observations have revealed companions around A type
stars (e.g. Fomalhaut, Kalas et al. 2008; HR 8799, Marois et al. 2008).
Although fragmentation from gravitational instability has been proposed
to explain these giant planets, our results pose two potential
challenges:

1.\emph{Mass challenge}: Including irradiation, the disk needs to be
quite massive to be gravitationally unstable ($M_{d}>0.3 M_{\odot}$).
Although the initial fragments can have masses around a few Jupiter
masses, they accrete at high rates $10^{-3}$--$10^{-1}$ M$_{\rm J}$
yr$^{-1}$, so that after only a few thousand years clumps may reach
their isolation masses, which are in the brown dwarf or low stellar mass
regime.
    
2. \emph{Migration challenge}: All clumps migrate inward from hundreds of
AU to tens of AU over timescales of $10^3$ yr, although the migration has
a stochastic component due to the interaction with the spiral arms. There
are at least two distinct fates for these migrating clumps.  The ones
reaching gap opening masses (corresponding to subsolar masses in
irradiated disks) slow their migration; less massive clumps migrate in
rapidly and are subject to tidal destruction. Generally, only the massive
clumps survive in our simulations, suggesting that GI-induced
fragmentation is best suited for forming massive brown dwarfs and close
binaries rather than gas-giant planets.

If binary systems are formed by GI fragmentation in disks, the spin axis
of these two stars should be parallel and perpendicular to the disk
plane. Recent observations on Kepler 16 systems have suggested the spin
axis of the binary and the planet around them are all closely aligned
(Winn et al.(2011)), which support the scenario that they are formed in
the disk.

The first cores produced by GI-induced fragmentation can be as large as
10 AU when the core is at 100 AU. Its luminosity can be as high as 0.1
L$_{\odot}$ if the core mass is 0.1 M$_{\odot}$ or 0.002 L$_{\odot}$ if
the core is 10 M$_{\rm J}$ (see Eq.~\ref{eq:corel}). With an effective
temperature $\sim 100$ K (Eq.~\ref{eq:coret}), these clumps can be easily
detected and resolved by ALMA, as illustrated by Fig.~\ref{fig:alma}, but
they may be rare because they only live for a few thousand years before
they collapse to the second core stage.
 
Boley \etal (2010) and Nayakshin (2010c) have proposed that dust can
settle and agglomerate within fragmented clumps, forming a solid core at
the center. Subsequent tidal disruption of the clump during inward
migration may then lead to the removal of the gas envelope, leaving a
surviving rocky body in the disk. Our simulations suggest the tidal
destruction of the envelope can happen and may not be rare (4 out of 13
cases) if the disk fragments. Whether the solid core can form is beyond
the scope of this paper (see Cha \& Nayakshin 2011).
 
Our simulations have several simplifications which require further
consideration.
First, our disks are subject to mass infall all the time. If the disk
fragments while there is no mass replenishment by infall, the clumps at
the outer edge of the disk may not necessarily migrate inwards; in other
words, small fragments formed near the end of infall may have the best
chance of survival at large radii. 

A second issue is that we have assumed axisymmetric, smooth infall, which
may not be the most typical case. Strong non-axisymmetric perturbations
in the infalling material can aid fragmentation (Burkert \& Bodenheimer
1993). We do not explore such large perturbations in order to develop
well-conditioned numerical simulations.

We emphasize that the two-dimensional nature of our simulations, while
enabling us to study clump formation and evolution over long timescales,
necessarily omits three-dimensional effects which could be important,
such as wave propagation in a stratified disk (Ogilvie \& Lubow 1999).
However, the fragmentation occurs at large radii, where we expect the
disk to be more nearly vertically isothermal due to the dominance of
irradiation heating.  Of more concern is the crude treatment of the
radiative cooling for clump evolution, which may differ significantly in
three dimensions. Finally, we have limited the calculations to a constant
value of $\gamma$, and have not considered possible instabilities related
to, for example, molecular hydrogen rotational thermalization or
dissociation. 

Considering our limitations, a comparison with existing 3-D simulations
is necessary.  Boley 2009 has done four simulations with similar mass
loading as our simulations, two of which (SIMB and SIMD) can be compared
with our simulations.  SIMB fragments with infall rate
$\dot{M}\sim10^{-4}\msunyr$ at 100 AU, which is consistent with our
results. And the clump in SIMD accretes from 6 M$_{J}$ to 11 M$_{J}$
supporting the clump accretion scenario.  Simulations by Kratter \etal
(2010a) have shown that the disk fragments when $\alpha>$1 and the clumps
are quite massive. However they used a piecewise polytropic equation of
state.  Harsono et al.(2011) suggested that infall has weakened the disk
fragmentation, in other words the critical $\alpha$ for fragmentation is
large. This seems to be consistent with our results. But caution has to
be made that we put infall manually by adding disk mass and no shear
between the infall and the disk has been set up.  We also used radiative
cooling instead of orbital cooling as Harsono \etal (2011). Hayfield et
al.(2011) have found in some cases, the clumps accrete at rates $\sim$
10$^{-5}\msunyr$ and can grow from several Jupiter masses to 40 M$_{J}$.
This is similar to what we have found in our simulations. But again,
our simple equation of state introduces uncertainties in the disk
dynamics and the thermal structure of the clumps.

Although theoretically the clump evolution and fate is complicated by the
disk infall history, and clump accretion/migration, observations may
provide us clues on  the formation mechanism for giant planets found at
large distances (e.g. Fomalhaut, HR 8799)  by measuring the metallicity
of these planets (Helled \& Bodenheimer 2010, 2011). 

\subsection{Dependence on stellar properties}

Figure \ref{fig:ba} shows the analytic estimates for the fragmentation
radii for disks around brown dwarfs and A type stars. We chose
$M_{*}$=0.08 M$_{\odot}$, $L_{*}$=0.01 L$_{\odot}$ for the brown dwarf,
and $M_{*}$=3 M$_{\odot}$, $L_{*}$=100 L$_{\odot}$ for the A type star.
The critical fragmentation $\alpha_{c}$ is 1 and minimum 10 K disk
temperature has been applied.

For viscous heating dominated disks, the fragmentation radius varies from
20 AU (brown dwarf) to 80 AU (A type star) simply due to $R_{f}\propto
M_{*}^{1/3}$ in Eq.~\ref{eq:fragr}. However, for irradiation dominated
disks, the fragmentation radius varies from 30 AU (brown dwarf) to 1000
AU (A type star) at 10$^{-5}\msunyr$. This sensitive radius dependence
for irradiation dominated disks is due to $R_{f}\propto L^{1/2}$ in
Eq.~\ref{eq:irr}, and the four orders of magnitude variation in
luminosity between the A type star and the brown dwarf. The lowest
fragmentation accretion rate is limited by the minimum disk temperature
which equals the envelope temperature. This analytic estimate is
confirmed by the R200\_3e-5L100 case where the disk did not fragment.

We want to emphasize that currently found far away planets ($>$50 AU)  by
direct imaging technique are around A type stars (e.g. HR 8799). Due 
to the stronger irradiation, the disk needs to be more massive to be gravitationally unstable and 
the clumps produced by GI can be more 
massive than our simulation produced. This poses a bigger challenge 
for GI to explain these giant planets.

These results apply to the formation of fragments for infall in a smooth,
non-structured pattern, which probably applies best to the formation of
objects of much lower mass than that of the central star.  Binary
fragmentation can be enhanced by non-axisymmetric infall (Burkert \&
Bodenheimer 1993; Kratter \etal 2010a), which are likely to dominate the
formation of companion stars from turbulent initial conditions.

\subsection{Disk evolution and outbursts}

GI-induced fragmentation has been proposed as a mechanism to explain
outbursting protostellar accretion phenomena, as in FU Orionis and Exor
objects (Vorobyov \& Basu 2005).  However, early calculations (Vorobyov
\& Basu 2005, 2006) used a polytropic equation of state which can
artificially enhance fragmentation (Boss 1997; Pickett \etal 2000; Gammie
2001).  Recent simulations (Voroboyov \& Basu 2010) with an improved
energy equation suggest that GI-induced fragmentation only occurs in a
fast rotating massive disk with a low background temperature; this is
consistent with our results that only infall to large disk radii at high
infall rates results in fragmentation.  Thus, whether outbursts can be
explained by protostellar disks fragmentation sensitively depends on the
initial conditions of prestellar cores, which need further observational
study. 

On the other hand, while Vorobyov \& Basu (2005, 2006) argued that
accretion of fragments explain FU Orionis outbursts of accretion from
disks, we showed (Zhu \etal 2008, 2009; see also Armitage \etal 2001,
Rice \etal 2010, Martin \& Lubow 2011) that the GI could trigger bursts
of accretion without clumping by thermally triggering the
magnetorotational instability (MRI) in the innermost disk.  We further
analyzed FU Ori's SED to show that the inner $\sim 0.5$~AU region of 
FU Ori during outburst is a steady accretion disk and it fits the
predictions of this picture of MRI triggering (Zhu \etal 2007, 2009). And
recently, this $\sim$0.5 AU disk has been confirmed by Keck
interferometric observations (Eisner \& Hillenbrand, 2011).  This means
that transport in the innermost disk is almost certainly dominated by
magnetic turbulence rather than gravitational instability, and the
symmetry of the rotationally-broadened line profiles (e.g., Zhu \etal
2009) yields no evidence for the strong departure from axisymmetry that
would be expected for pure clump accretion.

It is worth emphasizing that we find that some clumps can shear out (be
tidally destroyed) as they migrate to the innermost disk regions if they
don't open gaps and slow down their migration.  Thus, simply finding
fragmentation and inward migration is no guarantee that a clump will
actually accrete onto the central star as an entity; it is necessary to
follow the evolution of the mass inside of 1 AU, which no simulations
have done successfully yet, and consider the potential if not probable
effects of magnetic turbulence in addition to GI.  It may be, however,
that the tidal destruction of the clumps produced by GI-induced
fragmentation (Boley \etal 2010) or the episodic nature of the GI
accretion (Boley \& Durisen 2008) can serve as a mechanism to transport
mass to AU scales (compared with the steady accretion by GI as suggested
by Armitage \etal 2001 and Zhu \etal 2009) to trigger the MRI. 

\section{Conclusions}

We have presented two-dimensional hydrodynamic simulations of
self-gravitating protostellar disks subject to axisymmetric, steady mass
loading from an infalling envelope and irradiation from the central star
to explore the growth of gravitational instability (GI), fragmentation,
and the fragmented clump evolution.  Our results can be summarized as
follows: \begin{itemize} \item We confirm previous analytic estimates
that at high infall rates, there is a critical disk radius for infall
$\sim$ 50 AU, beyond which disk fragmentation occurs.  This critical
radius decreases with increasing mass accretion rate.

\item At low infall rates, irradiation suppresses fragmentation and
pushes the critical radius to larger values. For a solar-mass protostar,
this critical radius is 200 AU if the infall rate $\lesssim
10^{-5}\msunyr$.

\item With the gradual addition of the mass to the disk from the infall,
our simulations converge with the increasing numerical resolution for
both fragmentation and non-fragmentation cases. 

\item  The disk critical fragmentation $\alpha$ parameter seems to be
larger than that found by previous work; we suggest this is due to
differences in the infall model, the thermal history of the disk, and the
radiative energy trapping of the clumps.

\item For fragmenting disks, the clumps form at a mass which is close to
(in some cases slightly smaller) than the mass in the initial disk contained within the typical GI
length scale.  Rotational support becomes increasingly important as the
clump collapses and accretes high angular momentum material from the
disk.

\item The clumps accrete from the disk at a rate $\dot{M_{c}}\sim 4\Sigma
r_{H}^{2}\Omega$, which can be understood by assuming the accretion cross
section corresponds to 1-2 Hill radii.  In our irradiated disks, this is
10$^{-5}$ to 10$^{-4} \msunyr$, suggesting the clump can accrete to
sub-solar mass on the timescale of thousands of years.

\item The clumps migrate inward on the typical type I time scale ($\sim
2\times 10^3$~yr in our models) initially, but with a stochastic
component superposed because of interaction with the background spiral
waves.  However, the clump slows down when its mass is comparable to the
local disk mass, due to its inertia. A simple fit is given that agrees
well with the simulation results.

\item When a clump migrates inward its Hill radius decreases. Accretion by
the clump, however, counteracts the effect of the migration and slows
the shrinking of the Hill radius.

\item There are at least two distinct fates of the clumps, depending on
the migration speed. If the clump migrates slowly with enough time to
cool, shrink and accrete, it can become massive enough to open a gap
($M_{\rm c}>0.1 $M$_{\odot}$ in our models). This slows down its
migration (transition to type II migration) and may lead to the formation
of a binary companion (3 out of 13 clumps in our simulation).  On the
other hand, if the clump migrates inward quickly,  and is not massive
enough to open a gap, it is subject to tidal destruction (4 out of 13
clumps).  In the long run, however, the massive, gap opening clump 
should be the final fate for these disks, since new clumps can continue
forming and be quickly tidally destroyed until eventually a massive clump forms,
opens a gap and remains in the disc for a long period of time.

\item A simple analytic clump evolution model with clump accretion,
migration, and cooling included is presented and it confirms that the
different clump fates depend on the migration and accretion rates.

\item Both rapid clump migration and rapid gas accretion (leading to clumps much more massive than giant planets) pose challenges
to the scenario that giant planets are formed in situ by GI. And it is even
more challenging to explain currently found giant planets beyond 50 AU around A type
stars, since, under stronger irradiation, the fragmented clumps will be more massive
than those produced in our simulations. On the other hand, 
GI fragmentation can provide a
formation mechanism for close binary systems. 

\item The first cores produced by GI-induced fragmentation can be as
large as 10 AU when the core is at 100 AU. Its luminosity can be as high
as 0.1 L$_{\odot}$ if the core mass is 0.1 M$_{\odot}$, or 0.002
L$_{\odot}$ if the core mass is 10 M$_{\rm J}$. With an effective
temperature around 100 K, these clumps can be easily detected and
resolved by ALMA, although these first cores may only exist for thousands
of years before they collapse to form the second core.

\item Although accretion of discrete clumps does not agree with observational
constraints on FU Ori objects, tidal destruction can serve as a mechanism
to add a stochastic component to mass buildup in the inner disk,
potentially varying the resulting MRI-triggered outburst.

\end{itemize}

\acknowledgments
This work was supported in part by NASA grant NNX08AI39G from the Origins
of Solar Systems program, and in part by the University of Michigan. We thank
Aaron Boley for helpful discussion during the Gordon conference and Sergei Nayakshin for
giving us comments after reading our initial draft.
Particularly, we thank the referee for a very detailed and helpful report.
The simulation work reported here was initiated during the research programme
Dynamics of Discs and Planets hosted by the Isaac Newton Institute in 2009.
Thanks again to Jeremy Hallum for maintaining the compute cluster on which
these simulations were performed.

\appendix

\section{A: Opacity Table}\label{A}
The low temperature opacity is important for disk fragmentation. 
Thus, we have updated our low density opacity in Zhu \etal (2009), including water ice, 
carbonate and silicate grains. The opacity fits are given in Table 2.

The dust grain size distribution is chosen to be a power-law with slope 3.5 from 0.005 to 1 $\mu$m. Olivine
silicate to gas mass ratio is 0.0017, Pyroxene silicate to gas mass ratio is 0.0017, graphite to gas mass
ratio is 0.0041, and water ice to gas mass ratio is 0.0056. 
We use the dust destruction boundary as in D'Alessio \etal (1992).

\section{B: Boundary conditions}
We improve the boundary conditions of the publicly available version
of FARGOADSG so that the disk is in pressure equilibrium 
at the boundary. 
Take the inner boundary as an
example. Assuming the disk surface density distribution has the same
slope with radii in the ghost zone and the computational zone,
 we extrapolate the density ($\Sigma$) and internal energy
($E$) of the first two computational zones to the ghost zone as
\begin{equation}
log(\eta_{g,j}/\eta_{1,j})=\frac{log(\bar{\eta_{1}}/
\bar{\eta_{2}})}{log(R_{1}/R_{2})}(log(R_{g}/R_{1}))
\end{equation}
where $\eta_{i,j}$ is the density or the internal energy at the grid
with radial number i and azimuthal number j and $\bar{\eta_{1}}$ and
$\bar{\eta_{2}}$ are the azimuthal averaged value at the first and
second radii.

Then the azimuthal velocities of the ghost zone are calculated by
\begin{equation}
u_{\phi}^{2}=\frac{E(\gamma-1)}{\Sigma}\frac{\partial log
E}{\partial log R}+\frac{R\partial\Phi}{\partial R}
\end{equation}
where $R\partial\Phi/\partial R$ is just the GM/$R_{g}$ and
$\partial log E/\partial log R$ is given above. The radial velocity
of the ghost zone equals to that of the first computation zone only
if the radial velocity is inwards, otherwise it is 0.

With this treatment, we can reproduce the accretion rate of an axisymmetric
disk quite well without over-predicting the mass accretion rate as the normal free-flow boundary condition leads to.
 However, the extrapolation may lead to a negative density and energy value
in the ghost zone if the disk's density and energy fluctuate significantly close to the 
inner boundary.  This becomes more severe if a clump forms in the disk and migrates close
to the inner boundary. Thus for the fragmenting disks, we have also carried simulations
with normal outflow boundary condition (Stone 
\& Norman 1992).

\section{C: Radiative core luminosity-mass relation}
Assuming a radiative core (as our 2-D simulations implicitly assume in equation \ref{eq:cooling}),
with the uniform energy source model and ``radiative zero'' solution,   
the clump structure can be approximated by a polytrope P=K' T$^{n_{e}+1}$ (Chap. 23.3 of Cox 1968), where
 $n_{e}$=$-a+3$=1.5 and $a$ is the slope of the opacity on temperature 
 ($\kappa$=$\kappa_{0}$T$^{a}$).  Uniform energy source model should afford a fairly
 good approximation of gravitationally contracting stars (Chap. 17 of Cox 1968).  
 The structure of the polytropic sphere can be solved 
 by the Lane-Emden equation.  A radiative sphere with zero temperature boundary condition has luminosity
\begin{eqnarray}
L_{c}&=&\frac{64\pi \sigma}{3\kappa_{0}}(\frac{G}{\Re})^{-a+4}\frac{1}{n_{e}+1}[\frac{4\pi}{(n_{e}+1)^{n_{e}}\xi^{n_{e}+1}(-\theta_{n_{e}}')_{1}^{n_{e}-1}}]^{(-a+3)/n_{e}}\nonumber\\
&&\frac{\mu^{-a+4}M_{c}^{-a+3}}{r_{c}^{-a}}
\end{eqnarray}
where $\sigma$ is the Stefan-Boltzmann constant.
For dust opacity $a$=1.5 in Appendix A. The above equation is
\begin{equation}
L_{c}=\frac{64\pi \sigma}{3\kappa_{0}}(\frac{G\mu}{\Re})^{2.5}\frac{0.424^{1.5}M_{c}^{1.5}}{2.5 r_{c}^{-1.5}}\,.\label{eq:clumpL}
\end{equation}
With our opacity this can be reduced to
\begin{equation}
L_{c}=2\times 10^{-5}L_{\odot}\left(\frac{M_{c}}{M_{\odot}} \frac{r_{c}}{R_{\odot}}\right)^{1.5}=2\times 10^{-3}L_{\odot}\left(\frac{M_{c}}{0.01 M_{\odot}} \frac{r_{c}}{10 AU}\right)^{1.5}\,.\label{eq:radiativeL}
\end{equation}
And the clump central temperature, pressure and density are (notation as in Cox 1968 chap 23.1, and n=n$_{e}$=1.5, so $\xi_{1}$=3.65375 and $-\xi_{1}^{2}\theta'_{1}$=2.71406)
\begin{equation}
T_{c}=\frac{1}{(n_{e}+1)(-\xi \theta')_{1}}\frac{G}{\Re}\frac{\mu M_{c}}{r_{c}}=1.234\times 10^{7}  \frac{\mu M_{c}/M_{\odot}}{r_{c}/R_{\odot}} K=138\frac{ M_{c}/0.01 M_{\odot}}{r_{c}/10 AU} K \,,\label{eq:tccox}
\end{equation}
with $\mu$=2.4,
\begin{equation}
P_{c}=\frac{1}{4\pi (n_{e}+1)(\theta')^{2}_{1}}\frac{GM_{c}^{2}}{r_{c}^{4}}=0.08658\times 10^{17}  \frac{(M_{c}/M_{\odot})^{2}}{(r_{c}/R_{\odot})^{4}} dyne/cm^{2}\,,\label{eq:pccox}
\end{equation}
\begin{equation}
\rho_{c}=(1/3)(-\xi/\theta')_{1}\bar{\rho}=5.99071 \frac{M_{c}}{4/3\pi r_{c}^{3}}\,,
\end{equation}
From equation \ref{eq:tccox} we can derive
\begin{equation}
\frac{GM_{c}}{\Re r_{c} T_{c}}=\frac{1}{ \mu 0.537 }\,.\label{eq:gmort}
\end{equation}
With $\mu$=2.4, we can derive $GM_{c}/\Re r_{c}T_{c}$=0.7759.

On the other hand, for a completely convective core, $d ln T/d ln P=(\Gamma_{2}-1)/\Gamma_{2} $.
 With the ideal equation of state for diatomic gas (T is high enough that both the rotational and vibrational modes are active),  $\Gamma_{2}=\gamma=7/5$. Assuming $\Gamma_{2}$
 is constant throughout the core, integrating $d ln T/d ln P$ from the
 core surface to the center leads to  P= K' T$^{n_{e}+1}$ with $n_{e}$=1/($\gamma-1$)=2.5.  
 The clump structure is represented by a polytrope
P=K' T$^{3.5}$ which is steeper than our radiative case (in stellar cases with monatomic gas, 
n=1.5, which is the
same as our radiative cores). With n$_{e}$=2.5, we have  $\xi_{1}$=5.35528 and $-\xi_{1}^{2}\theta'_{1}$=2.18720. Thus the central temperature, pressure and density
are
\begin{equation}
T_{c}=1.603\times 10^{7}  \frac{\mu M_{c}/M_{\odot}}{r_{c}/R_{\odot}} K
=179\frac{M_{c}/0.01 M_{\odot}}{r_{c}/10 AU} K \,,\label{eq:tccox2}
\end{equation}
with $\mu$=2.4,
\begin{equation}
P_{c}=0.4935\times 10^{17}  \frac{(M_{c}/M_{\odot})^{2}}{(r_{c}/R_{\odot})^{4}} dyne/cm^{2}\,,\label{eq:pccox2}
\end{equation}
\begin{equation}
\rho_{c}=23.41 \frac{M_{c}}{4/3\pi r_{c}^{3}}\,,
\end{equation}

And the K' in the polytrope can be derived using Equation \ref{eq:tccox} and \ref{eq:pccox}
\begin{equation}
K'=\Re^{n_{e}+1}\left(\frac{n_{e}+1}{G}\right)^{n_{e}}\frac{\xi_{1}^{n_{e}+1}(-\theta')_{1}^{n_{e}-1}}{4\pi}\frac{1}{\mu^{n_{e}+1}M_{c}^{n_{e}-1}r_{c}^{3-n_{e}}}
\end{equation}
\begin{equation}
K'=\left(\frac{3.5}{G}\right)^{2.5}\Re^{3.5}\frac{5.3554^{3.5}0.07626^{1.5}}{4\pi}\frac{1}{\mu^{3.5}M_{c}^{1.5}r_{c}^{0.5}}\,,
\end{equation}
which is
\begin{equation}
K'=\frac{13.65\Re^{3.5}}{\mu^{3.5}M_{c}^{1.5}r_{c}^{0.5}G^{2.5}}\,.
\end{equation}
However the luminosity is determined by the boundary where the convective core becomes radiative.
For a fully convective core, we assume this occurs at the photosphere where the optical depth $\tau$=2/3. 
The pressure at the photosphere is
\begin{equation}
P_{eff}=\frac{\rho_{eff}\Re T_{eff}}{\mu}\sim\frac{2 g}{3 \kappa_{R}}=\frac{2 g}{3 \kappa_{0} T_{eff}^{a}}\,.
\end{equation}
Thus
\begin{equation}
\rho_{eff}=\frac{2\mu g}{3\kappa_{0}\Re T_{eff}^{a+1}}\,.
\end{equation}
Then with
\begin{equation}
\frac{P_{eff}}{T_{eff}^{7/2}}=\frac{\rho_{eff}\Re T_{eff}}{\mu T_{eff}^{7/2}}=K'\,,
\end{equation}
we can derive
\begin{equation}
T_{eff}=\left(20.48\kappa_{0}\right)^{-1/(a+3.5)}\left(\frac{G\mu}{\Re}\right)^{3.5/(a+3.5)}r_{c}^{-1.5/(a+3.5)}M_{c}^{2.5/(a+3.5)}\,.\label{eq:tceff2}
\end{equation}
And the luminosity is
\begin{equation}
L=4\pi \sigma \left(20.48\kappa_{0}\right)^{-4/(a+3.5)}\left(\frac{G\mu}{\Re}\right)^{14/(a+3.5)}r_{c}^{(2a+1)/(a+3.5)}M_{c}^{10/(a+3.5)}\,.
\end{equation}
With a=1.5,
\begin{equation}
L=4\pi \sigma \left(20.48\kappa_{0}\right)^{-4/5}\left(\frac{G\mu}{\Re}\right)^{14/5}r_{c}^{4/5}M_{c}^{2}\,.\label{eq:Lconv}
\end{equation}
With our opacity this is
\begin{equation}
L_{c}=1.715\times 10^{-3}L_{\odot}\left(\frac{M_{c}}{0.01 M_{\odot}}\right)^{2} \left(\frac{r_{c}}{10 AU}\right)^{0.8}\,.\label{eq:Lconv2}
\end{equation}
which is close to the estimate by Equation \ref{eq:radiativeL} at 0.01 $M_{\odot}$ and 10 AU size but with the different dependence on
$M_{c}$ and $r_{c}$.

\section{D: Analytic estimate for the critical fragmentation radius}
The parameter space for
GI disk fragmentation has been explored analytically by several authors (Rafikov
2005, 2007, 2010; Levin 2007; Kratter \etal 2008, 2010a;
Cossins \etal 2010; Zhu \etal 2010b). The effects
of the irradiation and opacity have been discussed in detail in
Rafikov (2009) and Cossins \etal (2010). Here we will give a short review on these analytic results, 
so that we can compare them with our numerical simulations. 

To calculate the fragmentation radius at different infall rates analytically,  we follow Rice (2005) in using the $\alpha$
 condition for fragmentation instead of using the cooling time condition
(Gammie 2001).
The $\alpha$ parameter is the scaling factor in the Shakura
\& Sunyaev (1973) treatment of viscosity, $\nu = \alpha c_{s}^{2}/\Omega$.

Rice (2005) has shown that for a variety of values of specific heat ratios ($\gamma$=5/3,7/5,2), for non-irradiated disks,
$\alpha > \alpha_{c}\sim 0.06$ is the condition for fragmentation and it is equivalent to cooling fragmentation
 condition by Gammie (2001).
For irradiated disks $\alpha_{c}$ decreases with stronger irradiation (Rice \etal 2011). 
If the energy balance of the GI disk is local,
\begin{equation}
\alpha\Omega c_{s}^{2} \Sigma \sim 2 \Lambda_{c}\,,\label{eq:lambdac}
\end{equation}
where $\Lambda_{c}$ represents the cooling of the excess generated energy due to GI heating beyond 
irradiation heating (defined in Eq. \ref{eq:cooling}); then
\begin{equation}
t_{cool}\equiv c_{s}^{2}\Sigma/\Lambda_{c}\sim\frac{1}{\alpha\Omega}\,.\label{eq:tcool}
\end{equation}
In this case if there is a critical fragmentation $t_{cool}$, it corresponds to a critical $\alpha$ (Rice \etal 2011).
Here, we will argue why $t_{cool}\equiv c_{s}^{2}\Sigma/\Lambda_{c}$ is the relevant cooling timescale 
for fragmentation in irradiated disks, instead of $c_{s}^{2}\Sigma/F$ where $F$ is the total outward flux, 
even though these two definitions degenerate in non-irradiated disks.
There are two heating sources in the disk: heating due to GI and heating due to the irradiation. The total 
outward flux $F$ from the disk surface also consists of two parts: the part corresponding to the net cooling
 ($\Lambda_{c}$) of the energy generated by GI and the part balancing the irradiation. Consider a gas 
 parcel in the spiral arm collapses due to its self-gravity, this gas parcel gets heated up by the compression. 
 Wether this collapse can continue or not depends on how fast the compressed heat
 can be cooled (as $\Lambda_{c}$ in Eq. \ref{eq:cooling} 
 describes). So that the net cooling ability  ($\Lambda_{c}$) determines the fragmentation instead of the
  total outward flux from the disk. In the extreme case for a locally isothermal disk, $\Lambda_{c}$ is infinite, 
  so that $t_{cool}$ defined in equation \ref{eq:tcool} is 0, and the disk fragments.

With our definition  of $t_{cool}$, $\alpha$ and $t_{cool}$ are directly related and no assumption about the 
irradiation is needed. In the following derivation, we just use the critical $\alpha_{c}$,
 as the criterion for the fragmentation in our analytic estimate.

Since $\alpha=\dot{M}G/2c_{s}^{3}$ for a Q=1.5 steady
accretion disk with accretion rate $\dot{M}$ and $c_{s}$, the disk can fragment if
\begin{equation}
\dot{M}G/2c_{s}^{3}>\alpha_{c}\,.\label{eq:frag}
\end{equation}
Since the disk temperature which determines $c_{s}$ decreases with radii, there is a critical
radius ($R_{f}$) beyond which equation \ref{eq:frag} is satisfied and the disk will fragment for a given $\dot{M}$. 
This critical radius has been derived by several authors before (Levin 2007; Rafikov 2005, 2010; Clarke 2009; Cossins \etal
2010; Kratter \etal 2010a; Zhu \etal 2010b). In order to explore its properties and compare with our numerical 
simulations, we summarize previous derivations and results below briefly.

\begin{figure}
\includegraphics[width=0.85\textwidth]{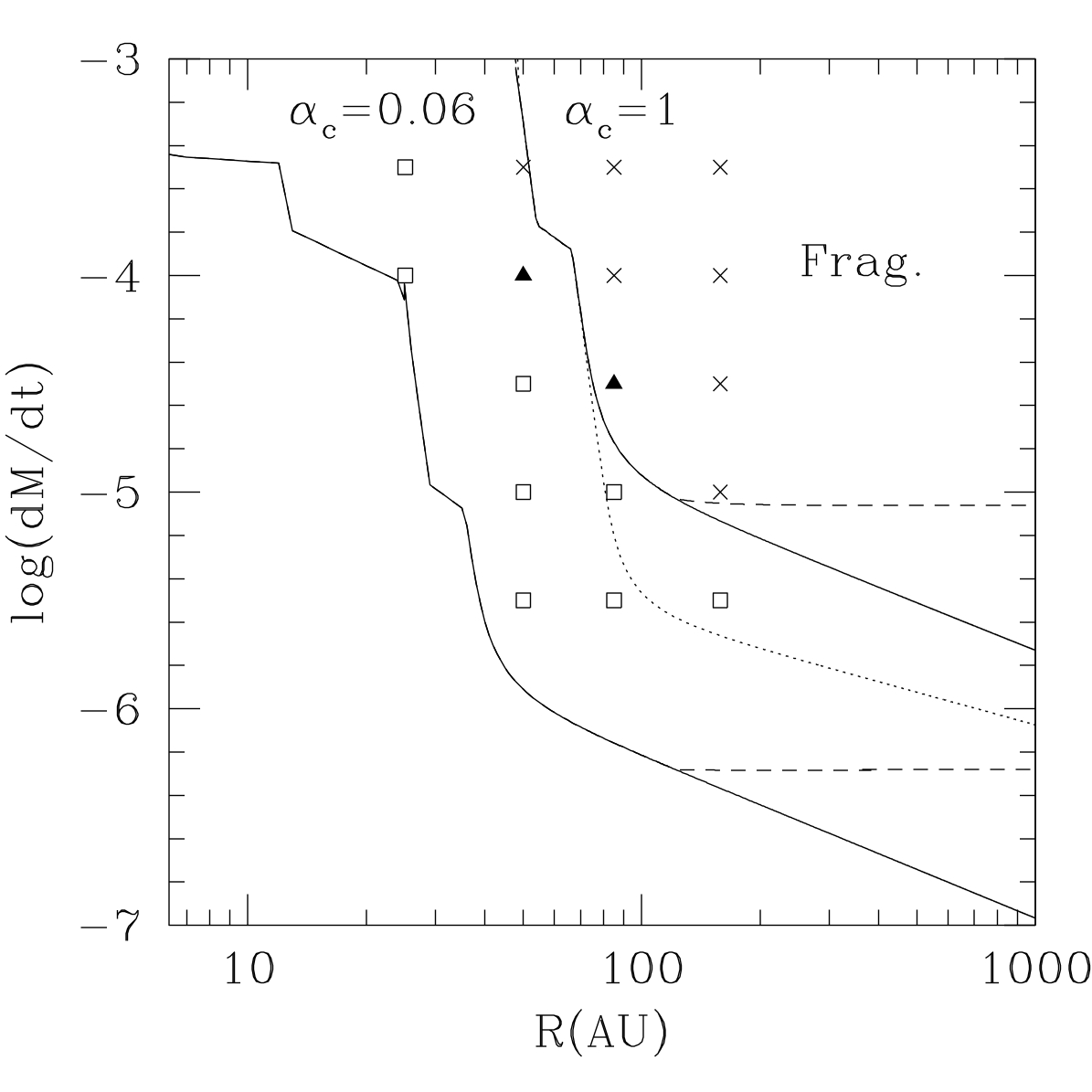} \caption{The fragmentation radii for 
gravitationally unstable disks with different infall rates (in unit of M$_{\odot}$/yr). 
Analytical estimates are shown with the critical fragmentation 
viscosity parameter $\alpha_{c}$=0.06 (left solid curve)
and $\alpha_{c}$=1 (right solid curve). The dotted curve represents the 
fragmentation radii without irradiation  ($\alpha_{c}$=1 is assumed).
The dashed curves represent the same estimates but with the minimum
irradiation temperature set to 20 K.  Crosses label the disks
fragmenting with multiple clumps, while the solid triangle label the disk which only has 
one or two clumps (marginally fragmentation cases). 
Open squares label the disks which fail to fragment and can accrete steadily.} 
\label{fig:frag}
\end{figure}

\begin{figure}
\includegraphics[width=0.85\textwidth]{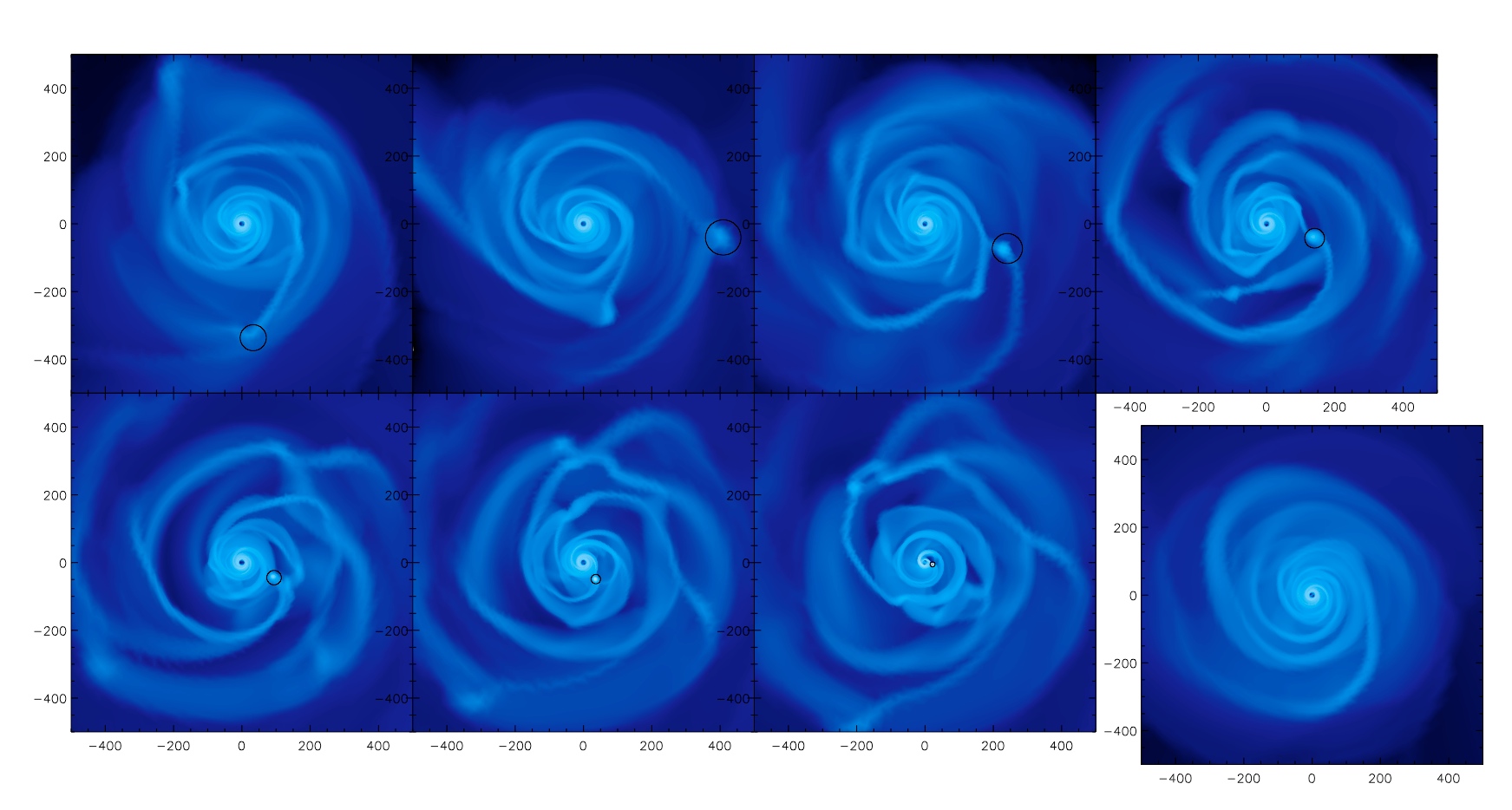} \caption{Clump formation and migration for the 
R100\_1e-5I (decreasing irradiation) case compared with the R100\_1e-5 case 
(lower right panel). For the R100\_1e-5I case, the upper 
left panel shows the clump formation and then each consecutive panel shows the clump
one orbital period later (5.78, 5.89, 6.27, 6.44, 6.53, 6.57, 6.61$\times$10$^{4}$ years 
after the start of the simulation). After 5 orbits the clump moves 
from 300 AU to 10 AU. The black circle labels the Hill radius of the clump. } 
\label{fig:figclump2}
\end{figure}

\begin{figure}
\includegraphics[width=0.85\textwidth]{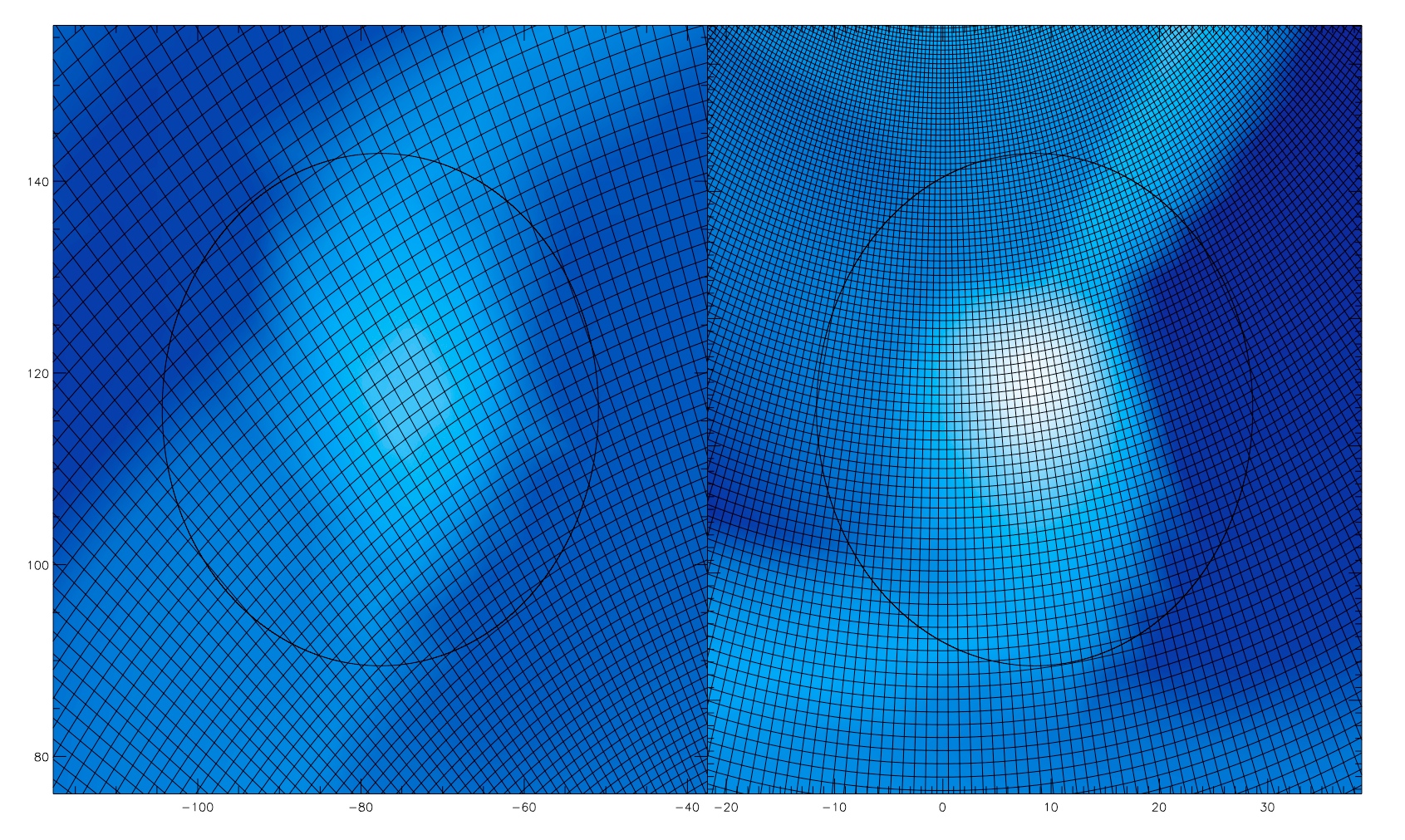} \caption{The surface density of the clump (run R100\_3e-5 ) 
which just forms (2.2$\times$10$^{4}$ years after the start of the simulation, left panel) 
and fully developed (2.4$\times$10$^{4}$ years after the start of the simulation, 
right panel). The black circle labels the Hill radius of the clump and the numerical 
grids have been plotted on top of the density distribution.  } \label{fig:fig2grid}
\end{figure}

\begin{figure}
\includegraphics[width=0.85\textwidth]{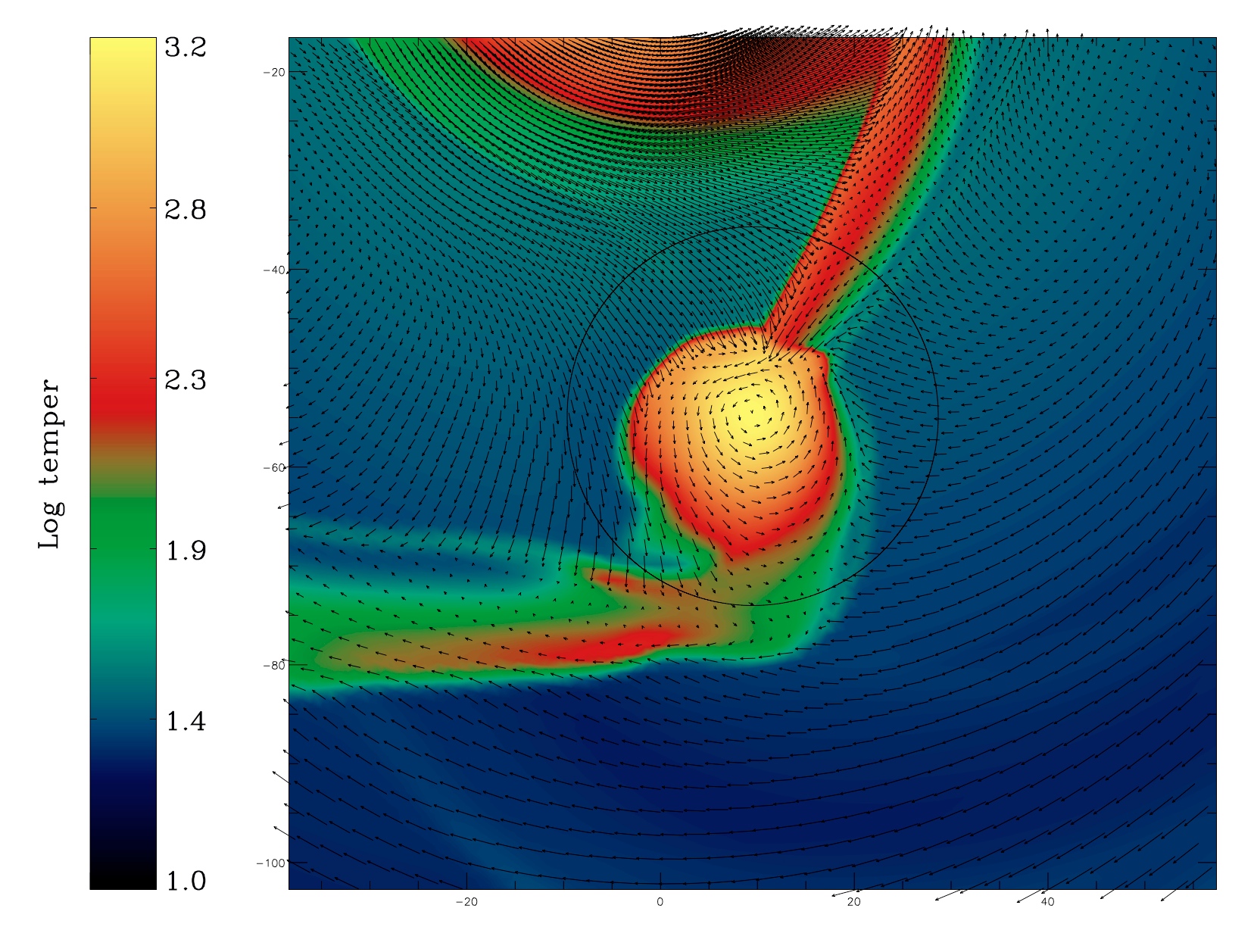}
\caption{The midplane temperature distribution of the clump as shown in the right panel of 
Fig.~\ref{fig:fig2grid} plotted with velocity vectors in the frame corotating with the 
clump center.}
\label{fig:relT}
\end{figure}

\begin{figure}
\includegraphics[width=0.5\textwidth]{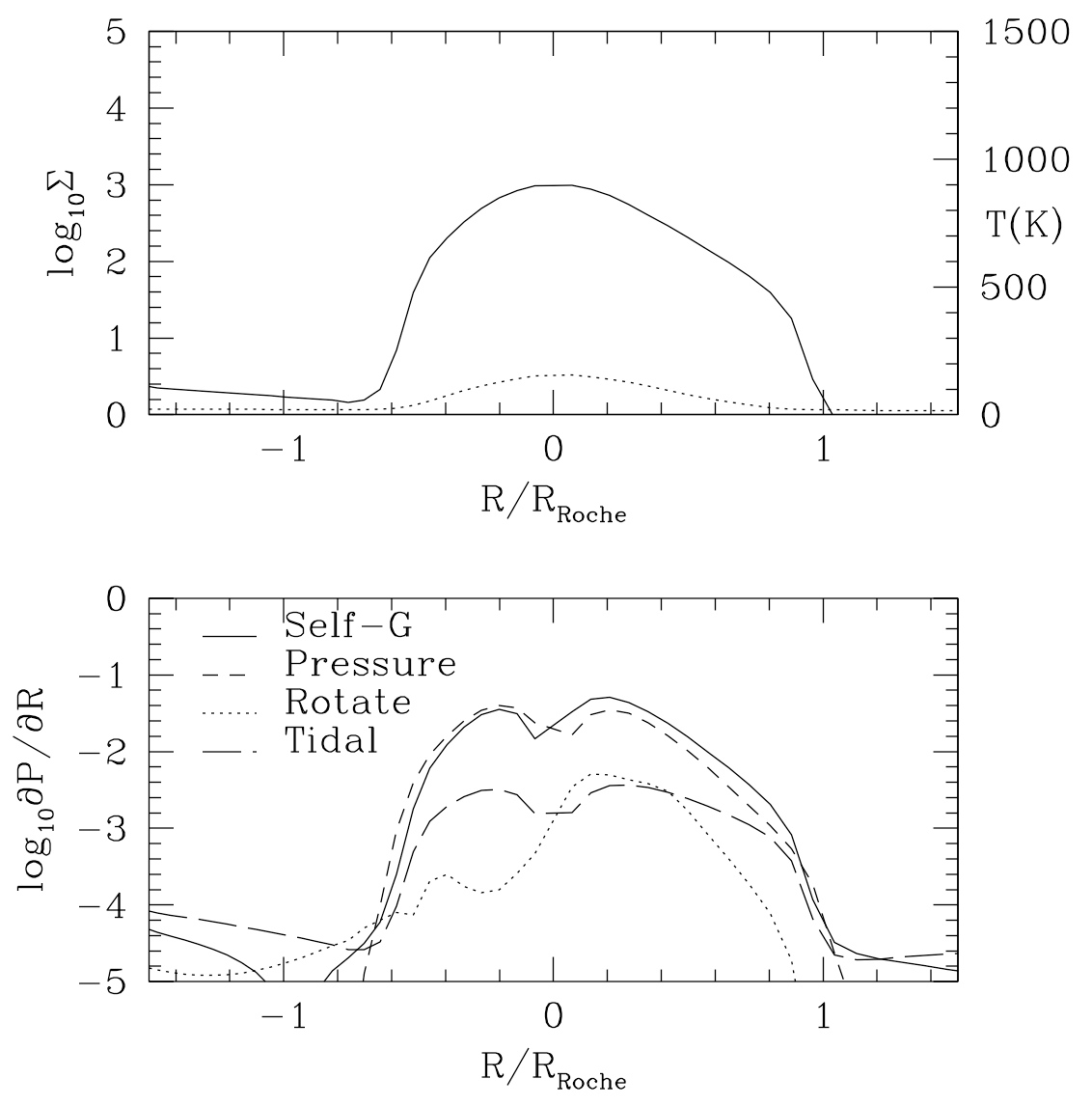} \hfil
\includegraphics[width=0.5\textwidth]{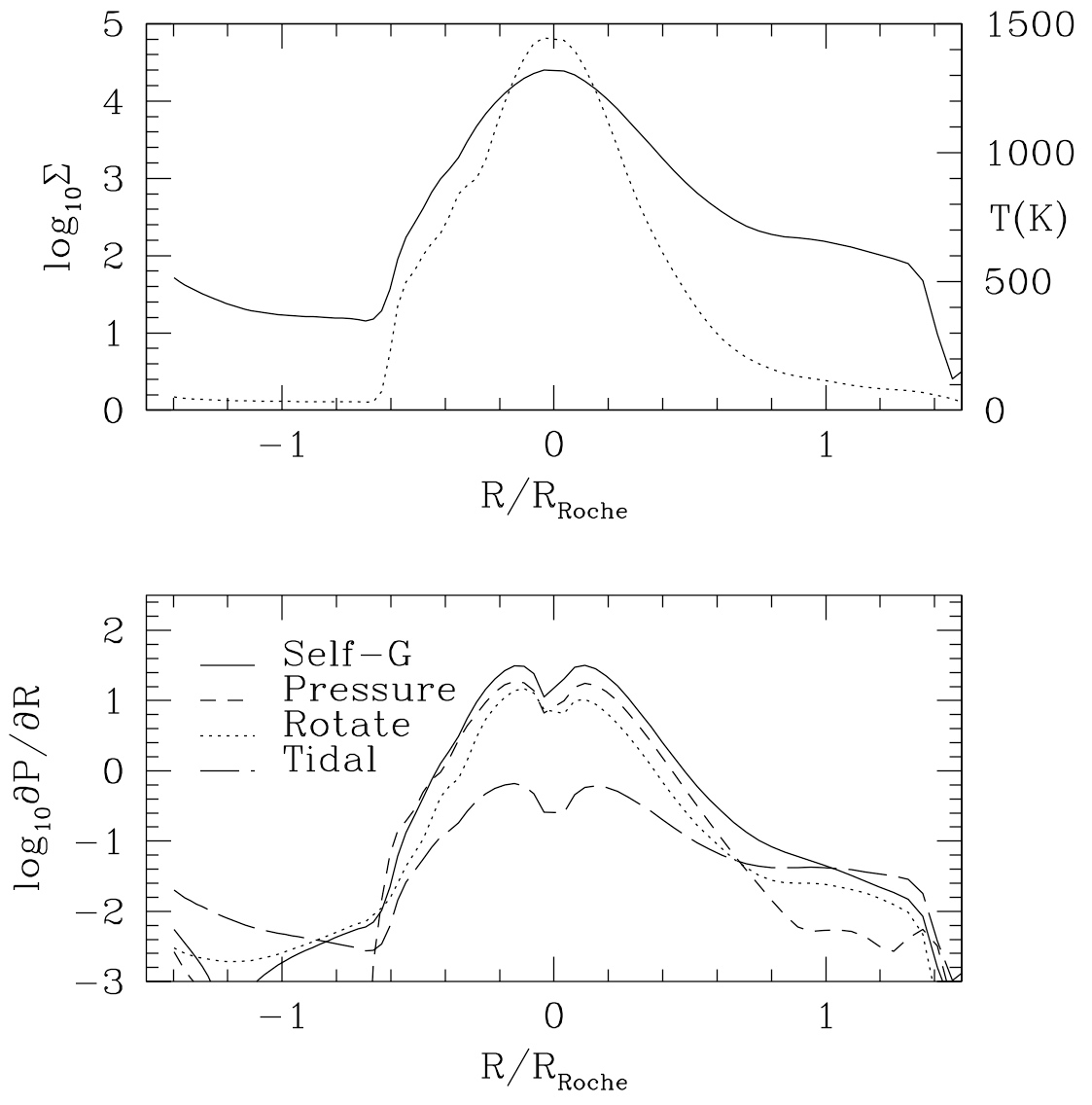} \\
\caption{The clump's surface density (dotted curve) and temperature (solid curve) profile 
along the radial cut across the clump center (upper panels) at the two different times 
corresponding to the left and right panel of Fig.~\ref{fig:fig2grid}. Lower panels show 
various forces along this cut. In the left panels the rotational support 
is negligible, while in the right panels the rotational support is as important as 
the thermal support. }\label{fig:f220}
\end{figure}

\begin{figure}
\includegraphics[width=0.85\textwidth]{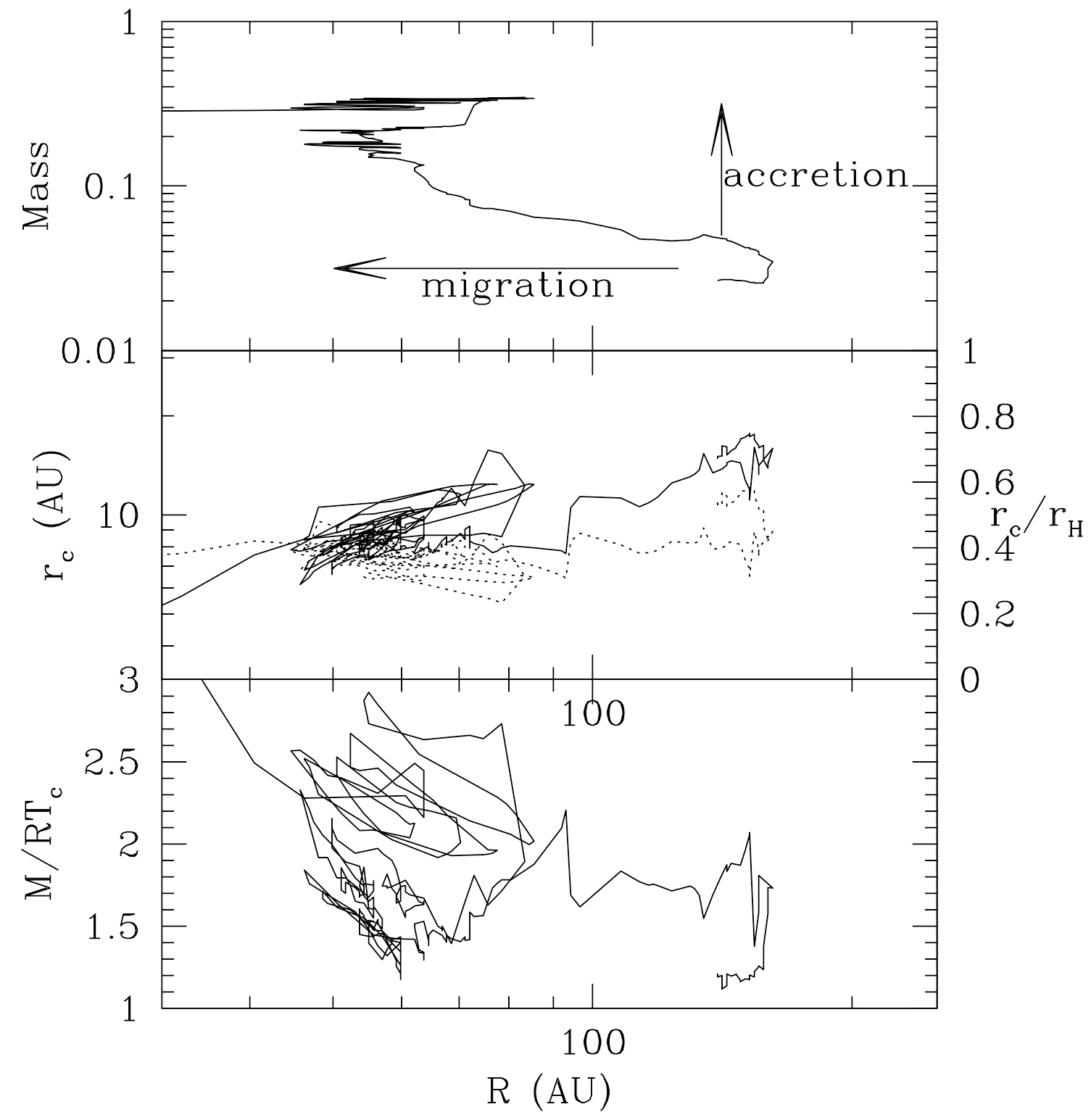} 
\caption{Clump's mass (in unit of solar mass), core radius (solid curve), $r_{c}$/$r_{H}$ (dotted curve), and  
$GM_{c}/\Re T_{c}$ with the clump's position in the disk for the R100\_3e-5 case.}
\label{fig:clump}
\end{figure}

\begin{figure}
\includegraphics[width=0.5\textwidth]{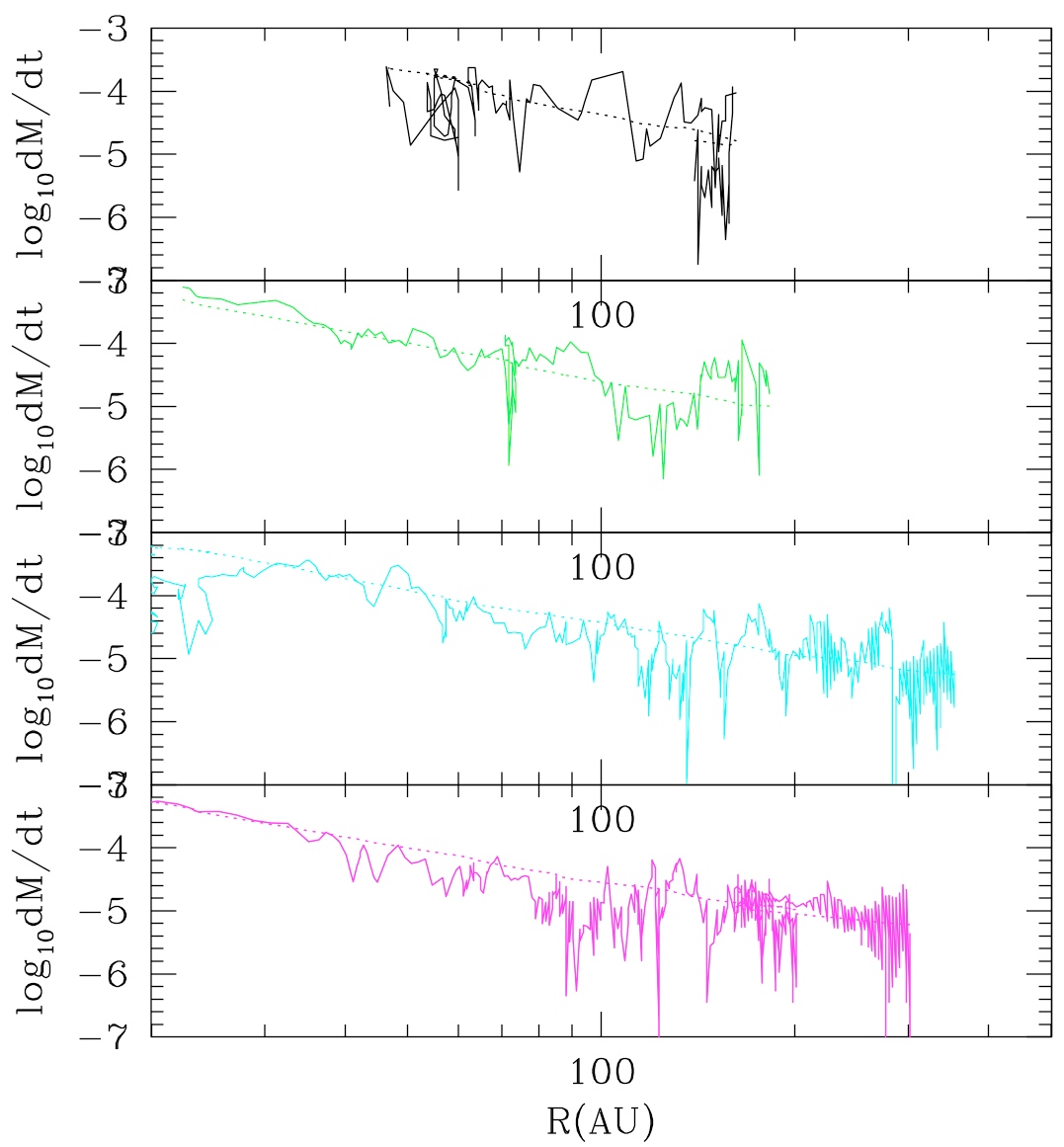} \hfil
\includegraphics[width=0.5\textwidth]{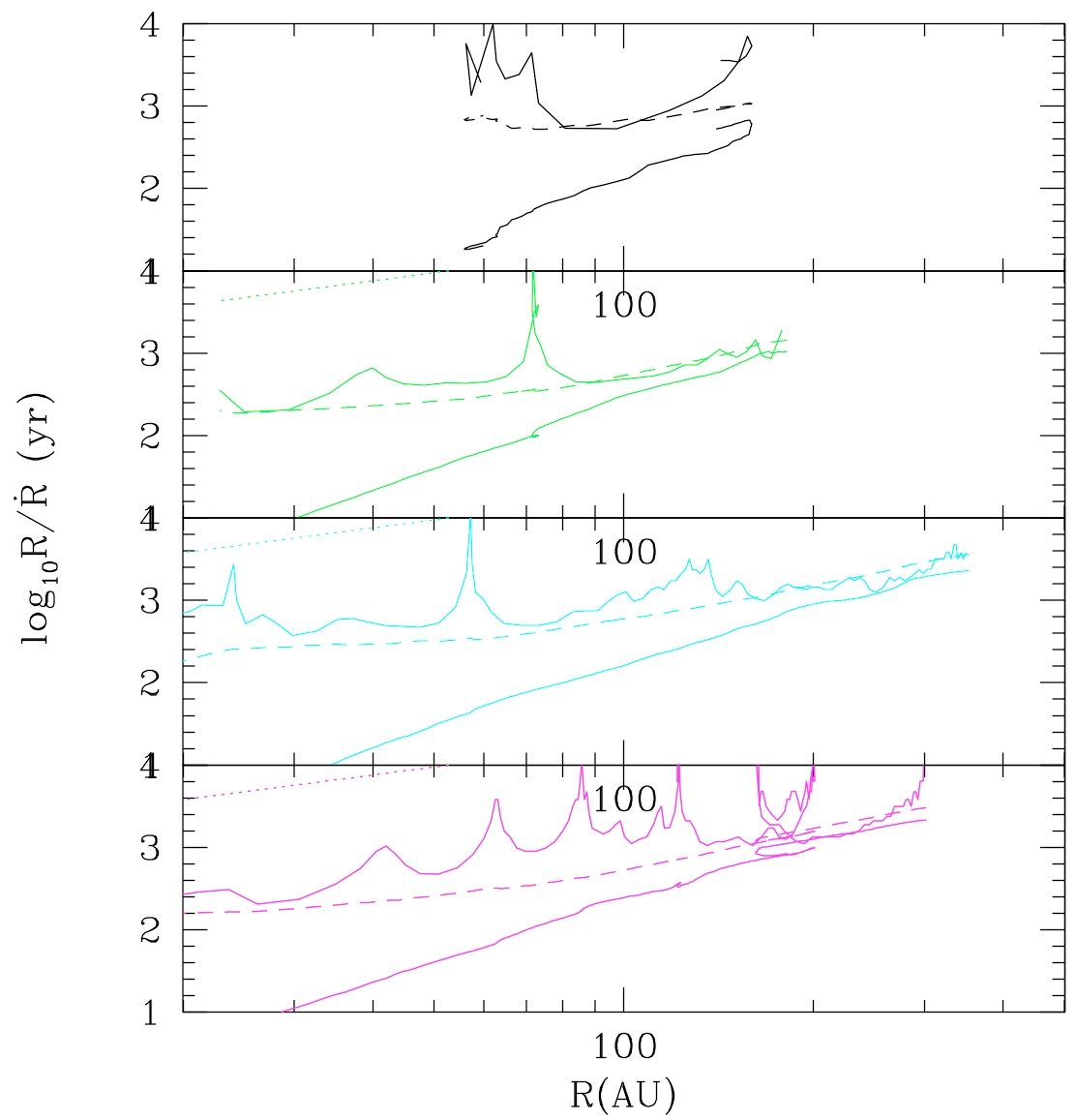} \\
\caption{Left: The clump mass accretion rate (in unit of M$_{\odot}$/yr) when the clump migrates inwards in the disk.
Reading from the top, we show cases R100\_3e-5 (clump C), R100\_1e-5I (clump H), 
R200\_1e-5 (clump D), R200\_3e-6noirr (clump K);
Different colors correspond to the cases described in Table 1. 
The dotted lines show the analytic estimates based on Eq.~\ref{eq:dmclump3}. 
Right: The migration timescale in years $R/\dot{R}$ at each radius for the same cases.  
The lower solid curves show the type I migration timescales estimated with 
Eq.~\ref{eq:mig}, while the upper dotted curves show the type II migration 
timescales with $\alpha=1$. The dashed curves show our fit using Eq.~\ref{eq:mig3}.
}\label{fig:clumpdm}
\end{figure}

\begin{figure}
\includegraphics[width=0.8\textwidth]{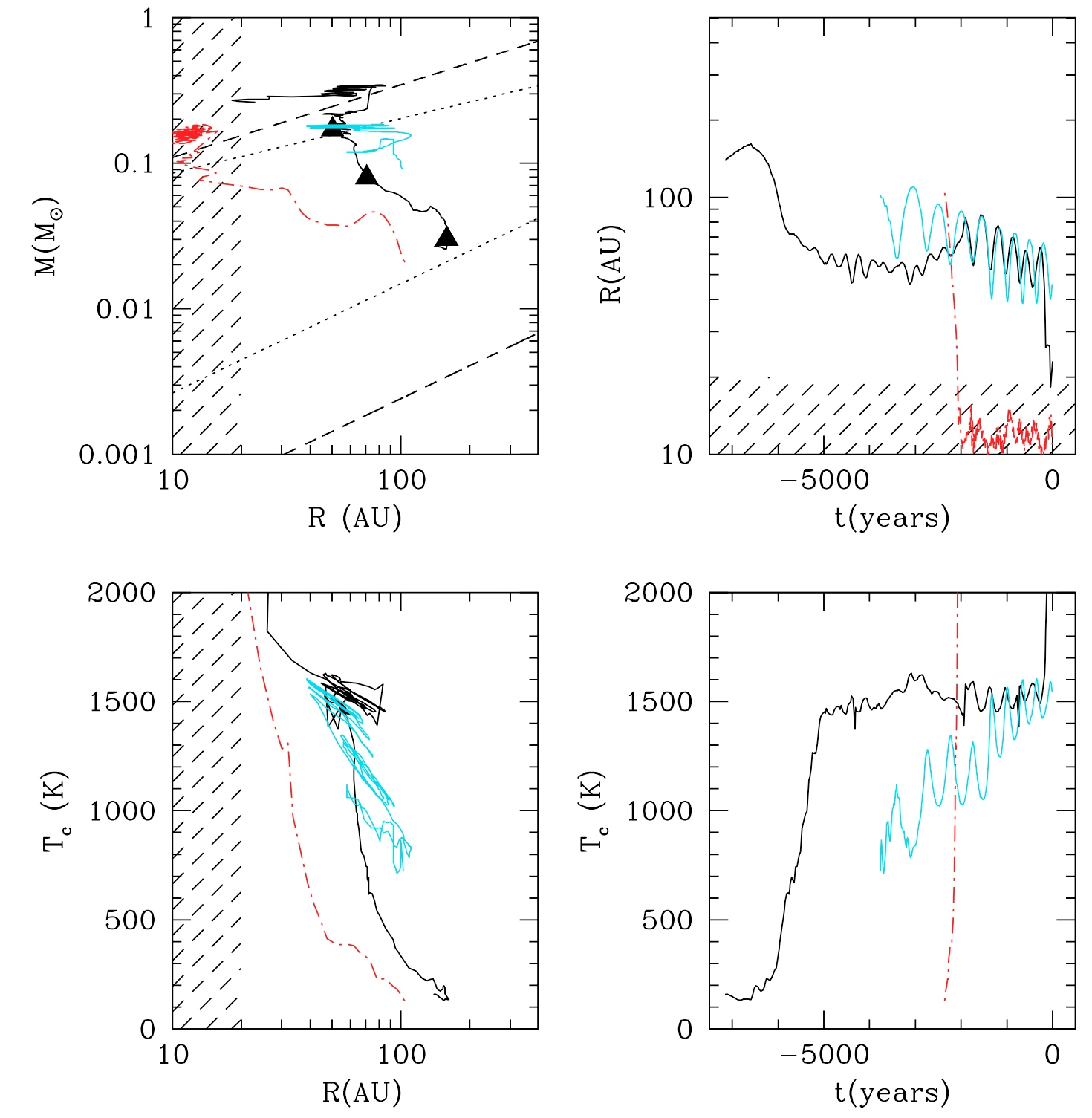}\hfill 
\caption{Left: Clump masses and central temperatures as a function of position for cases 
when a gap is opened and migration stops (different colors correspond to different cases 
in Table 1, and the clump names are given in Table 2). The dotted 
lines represent the isolation mass (Eq. \ref{eq:miso}) and minimum clump masses 
(Eq. \ref{eq:mclump}), while the dashed lines represent the gap opening masses 
(Eq. \ref{eq:mgap}). When the clump is close to the inner boundary, its evolution 
is significantly affected by the inner boundary. Thus we shade the region
where the clump evolution may not be reliable. The three triangles label when
the clump mass is 0.03, 0.08 and 0.17 M$_{\odot}$. The disk surface densities at 
these three times are shown in Fig. \ref{fig:surf}. 
Right:The clump radial position in the disk (distance to the star) and the central temperature
evolution with time for these cases. The black curve is for R100\_3e-5 case, the cyan curve is for
R200\_1e-5 case, and the red curve is for R65\_1e-4 case.}
 \label{fig:clumpgap}
\end{figure}

\begin{figure}
\includegraphics[width=0.8\textwidth]{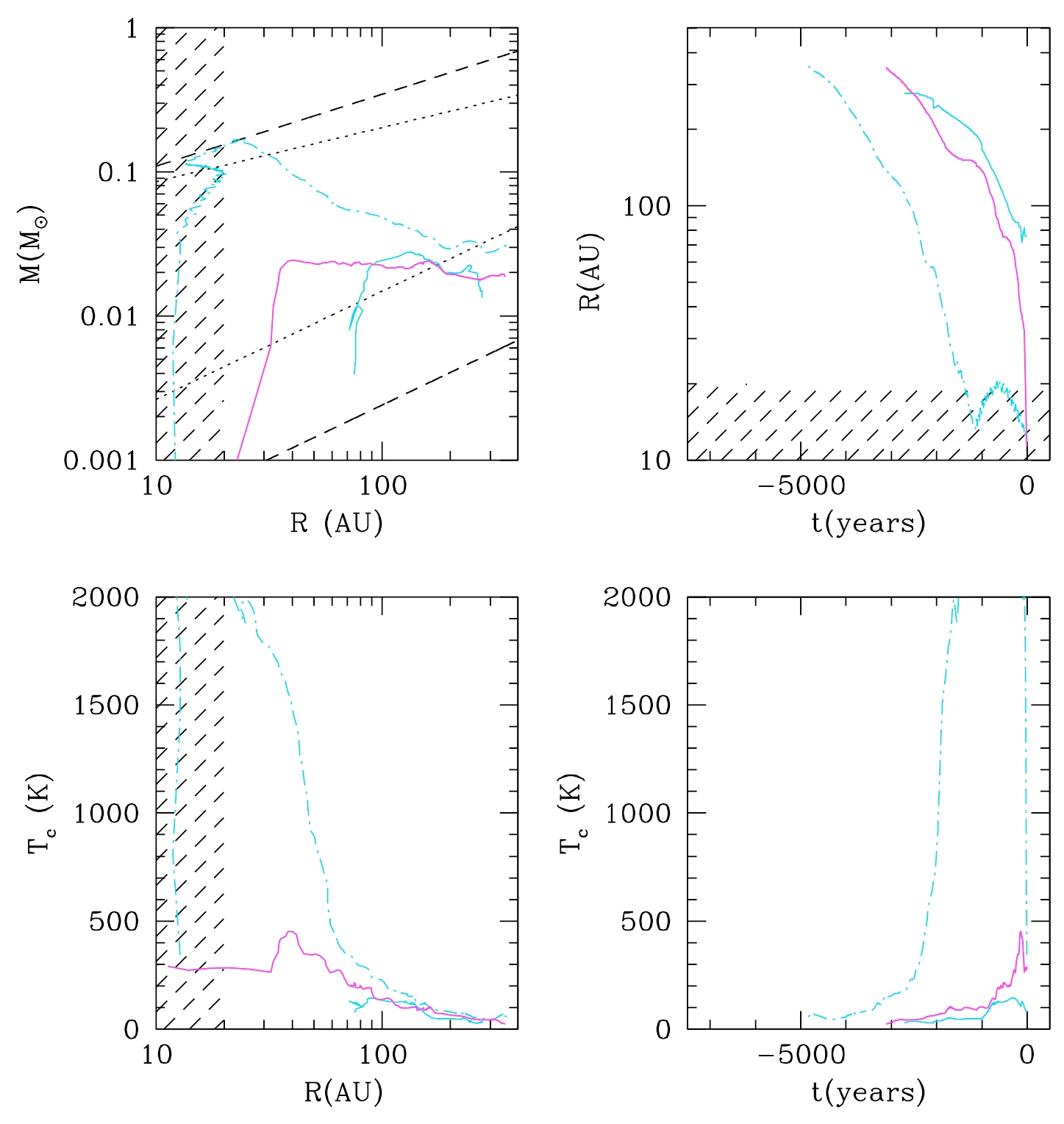}\hfill 
\caption{The same as Fig. \ref{fig:clumpgap} but for cases when tidal destruction occurs.
These clumps correspond to the clumps labeled in Table 1 and 2.   The cyan curves are for
R200\_1e-5 case, and the magenta curve is for R200\_3e-6noirr case. } \label{fig:clumptidal}
\end{figure}

\begin{figure}
\includegraphics[width=0.8\textwidth]{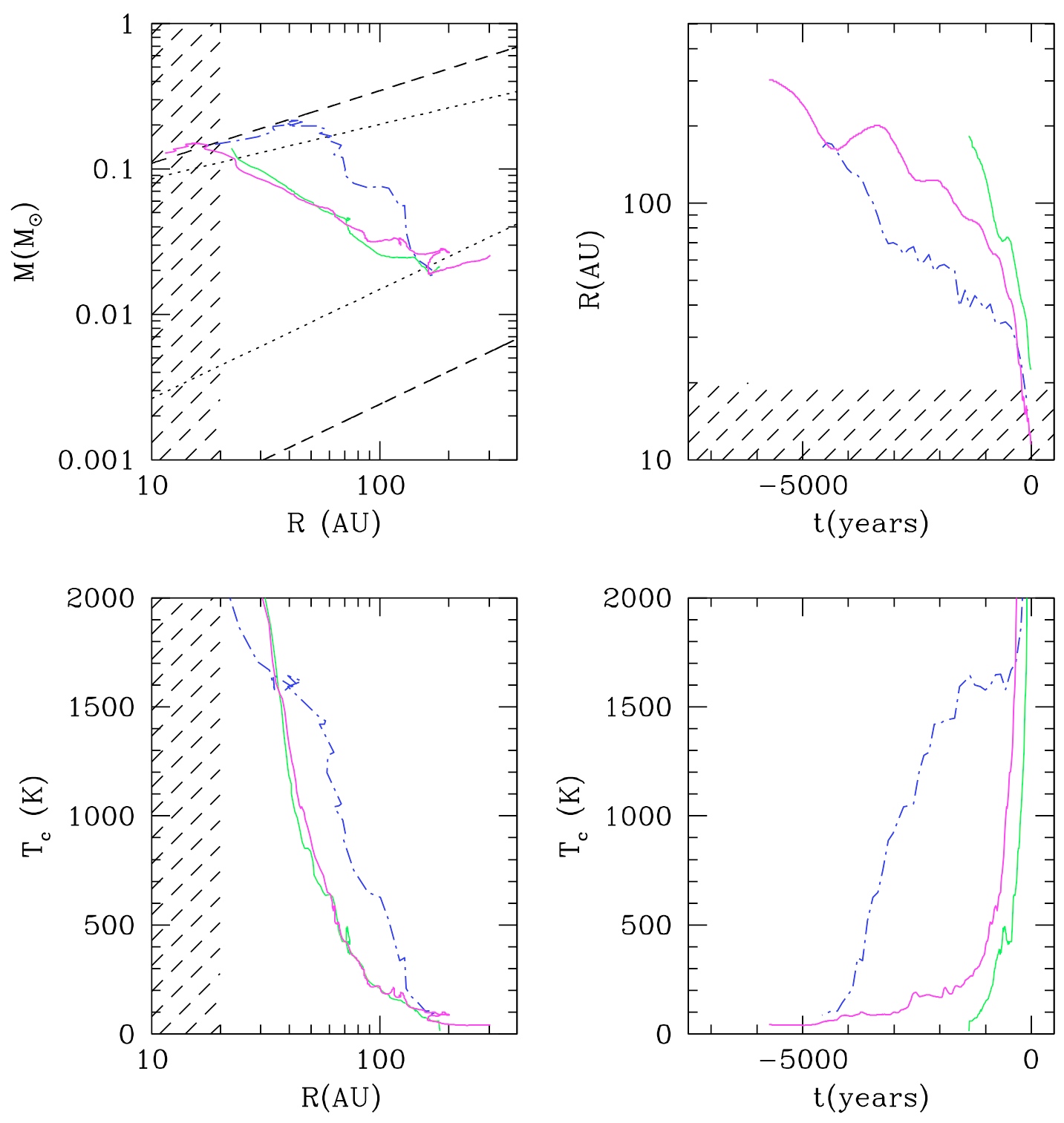}\hfill 
\caption{The same as Fig. \ref{fig:clumpgap} but for cases when the clumps 
migrate to the inner boundary. 
These clumps correspond to the clumps labeled in Table 1 and 2. The magenta curve is for R200\_3e-6noirr case,
the green curve is for R100\_1e-5I case, and the blue curve is for R100\_1e-4 case.
 } \label{fig:clumpmig}
\end{figure}

\begin{figure}
\includegraphics[width=0.8\textwidth]{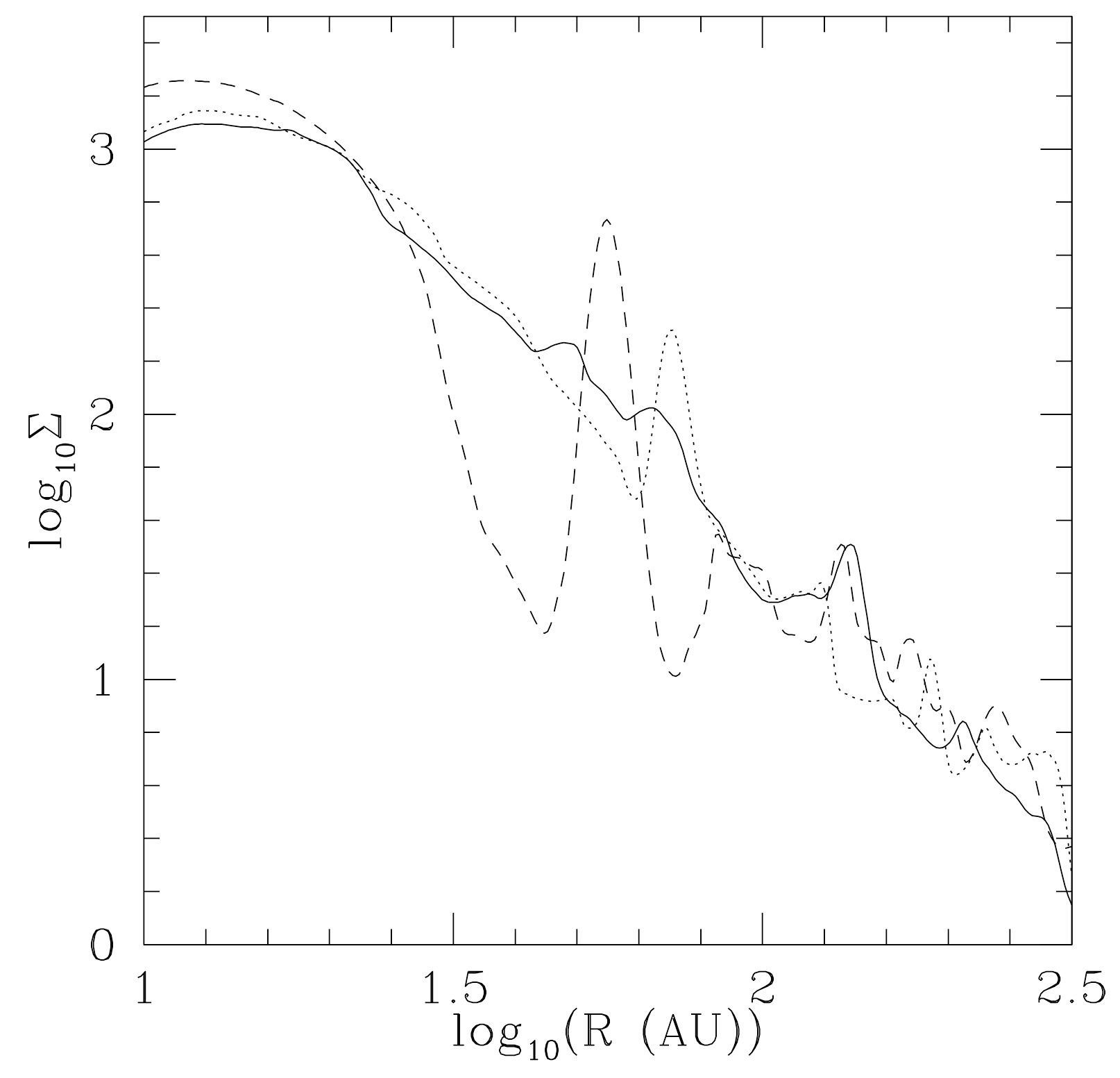}\hfill 
\caption{The disk azimuthal averaged surface density for the R100\_3em5 case when 
the clump mass is 0.03 M$_{\odot}$ (solid curve), 0.08 M$_{\odot}$ (dotted curve), 
and 0.17 M$_{\odot}$ (dashed curve). As clearly shown, the clump stops its migration 
(corresponding to mass 0.17 M$_{\odot}$ in Fig.~\ref{fig:clumpgap}) when a gap forms. } 
\label{fig:surf}
\end{figure}

\begin{figure}
\includegraphics[width=0.8\textwidth]{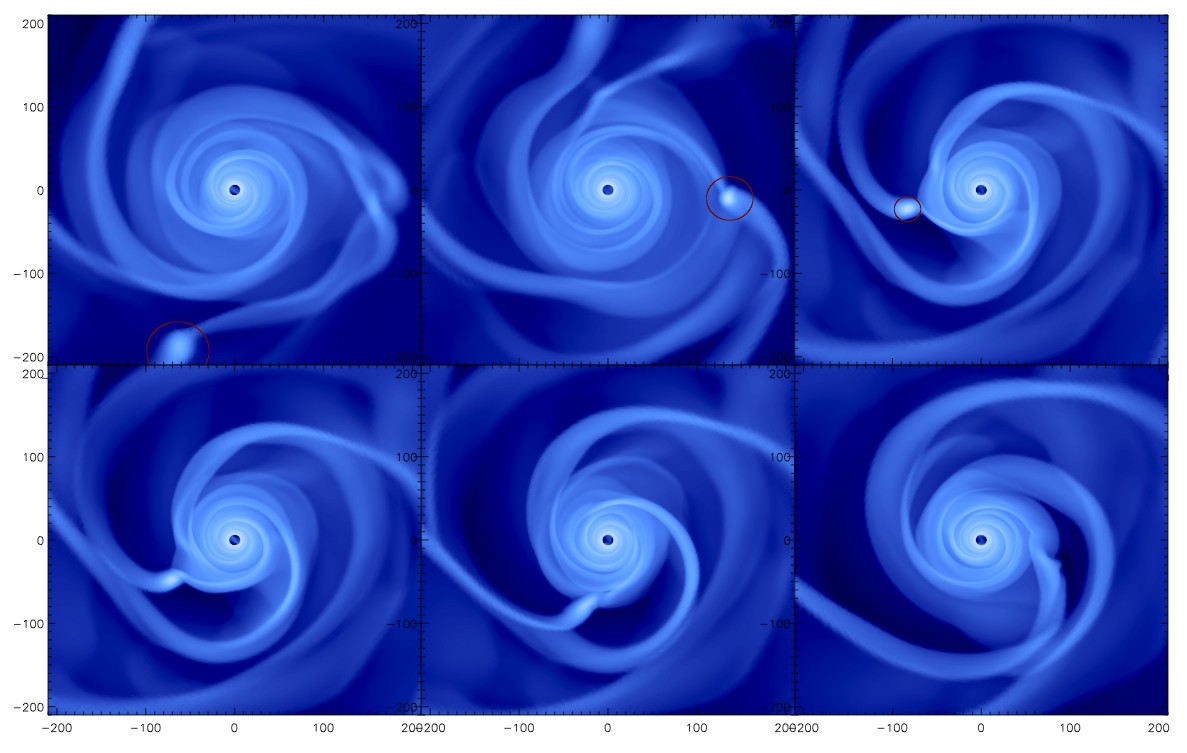} 
\caption{Clump formation, migration, and tidal destruction for the R200\_1e-5
case. From left to right and top to bottom, setting the first image at time $t=0$, 
the time slots for subsequent images are 546 yrs, 1067 yrs, 
1103 yrs, 1174 yrs, and 1328 yrs.}\label{fig:figtidal}
\end{figure}

\begin{figure}
\includegraphics[width=0.8\textwidth]{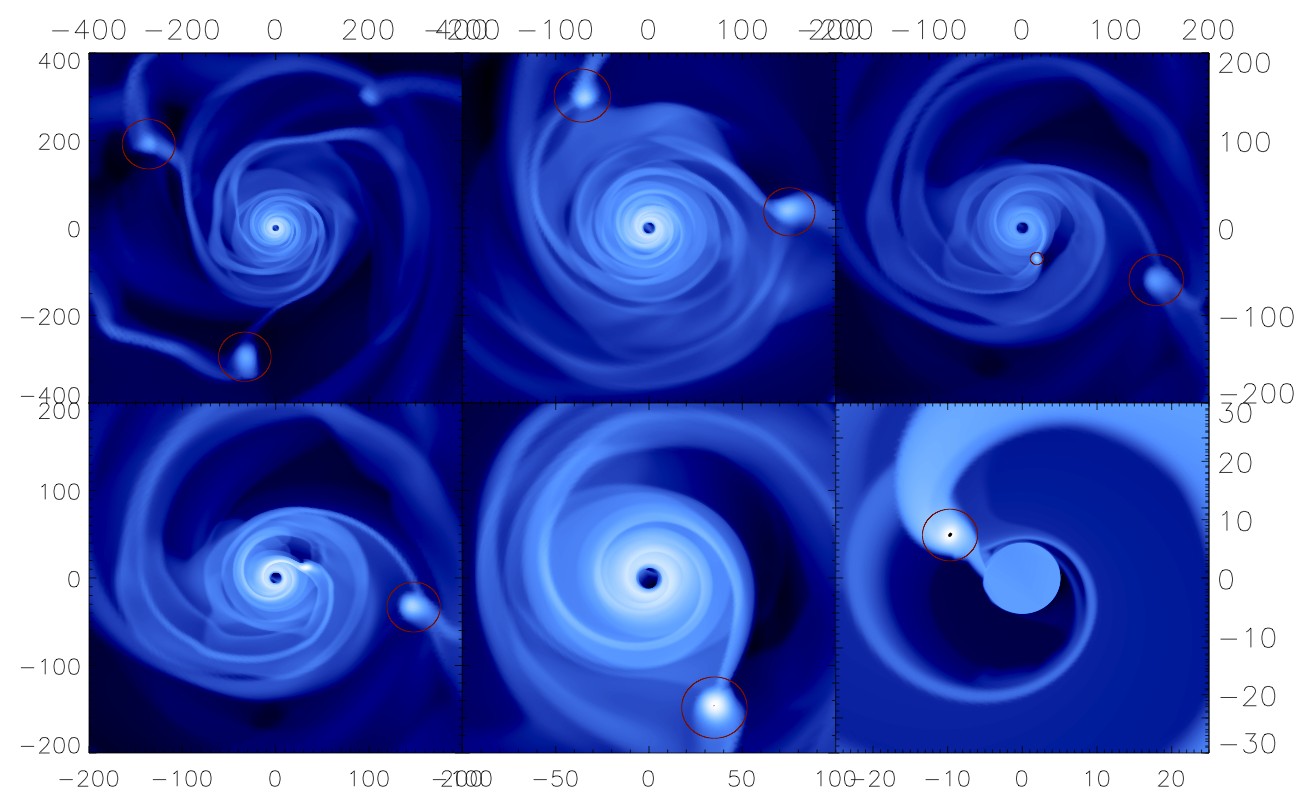} 
\caption{Two clumps forming at the same time but with different fates for the
R200\_3e-6noirr case. From left to right and top to bottom, setting the first 
image at time $t=0$, the time slots for subsequent images are
1660 yrs, 3036 yrs, 3084 yrs, 4898 yrs, and 6072 yrs.}\label{fig:figtidal2}
\end{figure}

\begin{figure}
\includegraphics[width=0.8\textwidth]{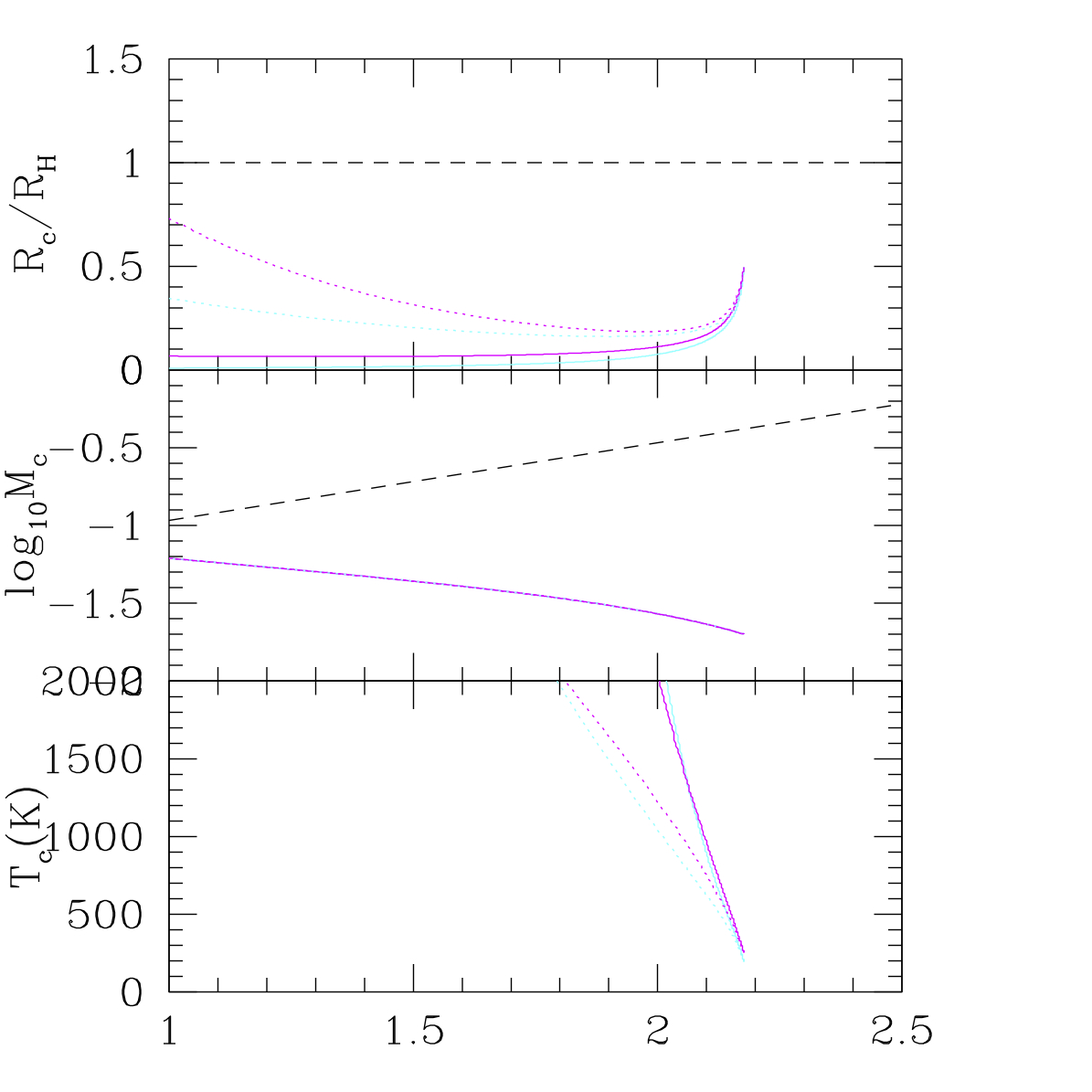} 
\caption{The values of r$_{c}$/r$_{H}$, mass (in unit of M$_{\odot}$), 
and central temperature for a clump as a function of its position in the disk
for different clump parameters. The purple curves are for the convective cores with $n_{e}$=2.5,
while the cyan curves are for the radiative cores with $n_{e}$=1.5. The
solid curves are calculated with $\varpi$=1, while the dotted curves are
calculated with $\varpi$=0.5. The dashed line in the top panel shows
when the tidal destruction happens, and the dashed line in the middle plane
shows when the gap opening occurs in the disk.
  }\label{fig:clumpevolve1}
\end{figure}

\begin{figure}
\includegraphics[width=0.8\textwidth]{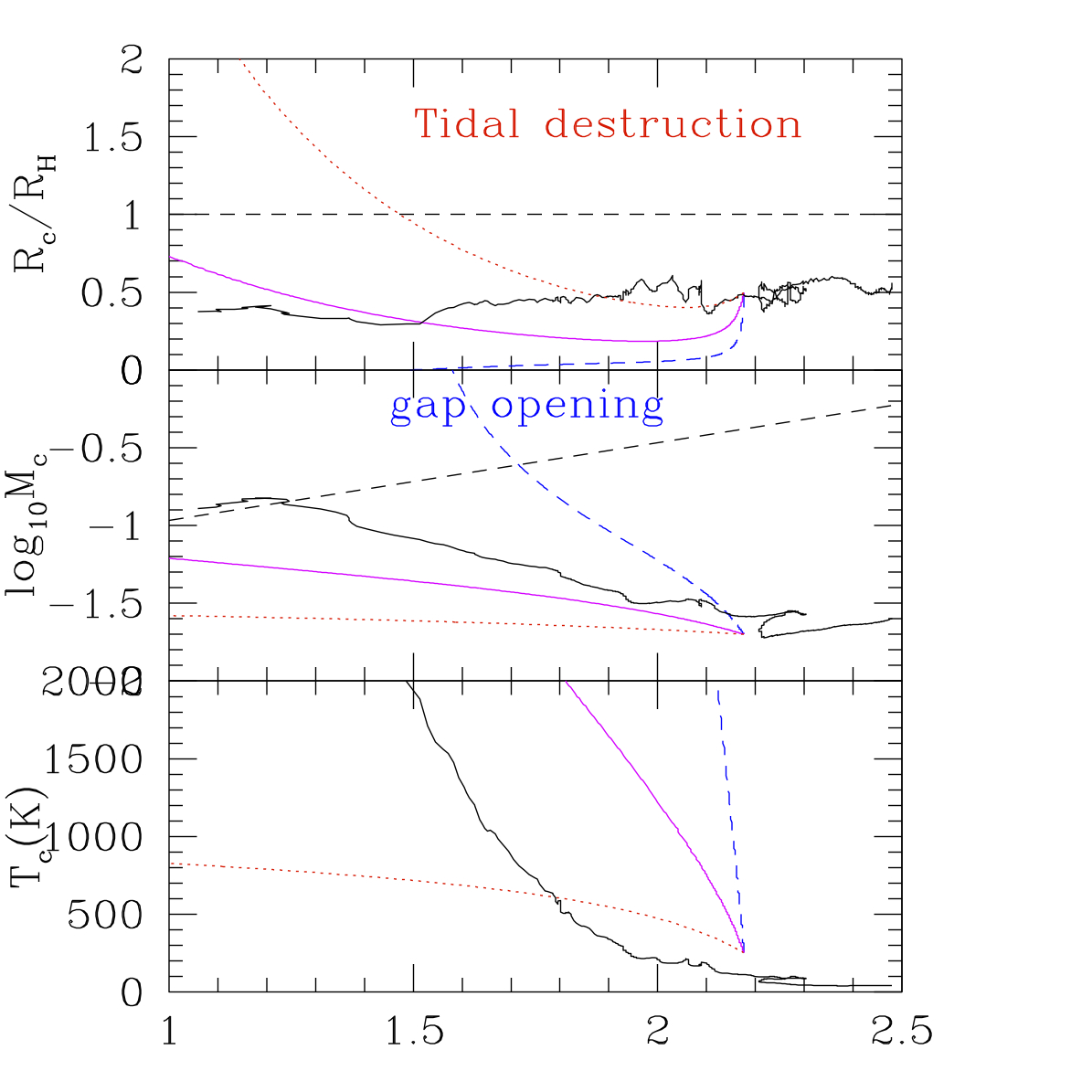} 
\caption{The values of r$_{c}$/r$_{H}$, mass (in unit of M$_{\odot}$), 
and central temperature for a clump as a function of its position in the disk. 
The clump's evolution is from right to left, while it migrates 
inwards. The solid black curve is from our simulation with clump K 
(shown in Fig.~\ref{fig:clumpdm} and  \ref{fig:clumpmig}). The solid purple 
curve is from our simple analytic model with a convective core assumption and
$\varpi$=0.5.
 The dotted red curve is from our 
analytic model but with 5 times the nominal migration speed  (Eq.~\ref{eq:mig3}), 
and the clump is 
tidally destroyed, which can be compared with Fig.~\ref{fig:clumptidal}. The 
dashed blue curve is from our analytic model but with 1/5 of the nominal migration
speed, and it open gaps in the disk around 30 AU, which can be compared 
with our simulations in Fig.~\ref{fig:clumpgap}. The dashed line in the top panel shows
when the tidal destruction happens, and the dashed line in the middle plane
shows when the gap opening occurs in the disk.
}\label{fig:clumpevolve2}
\end{figure}

\begin{figure}
\includegraphics[width=0.8\textwidth]{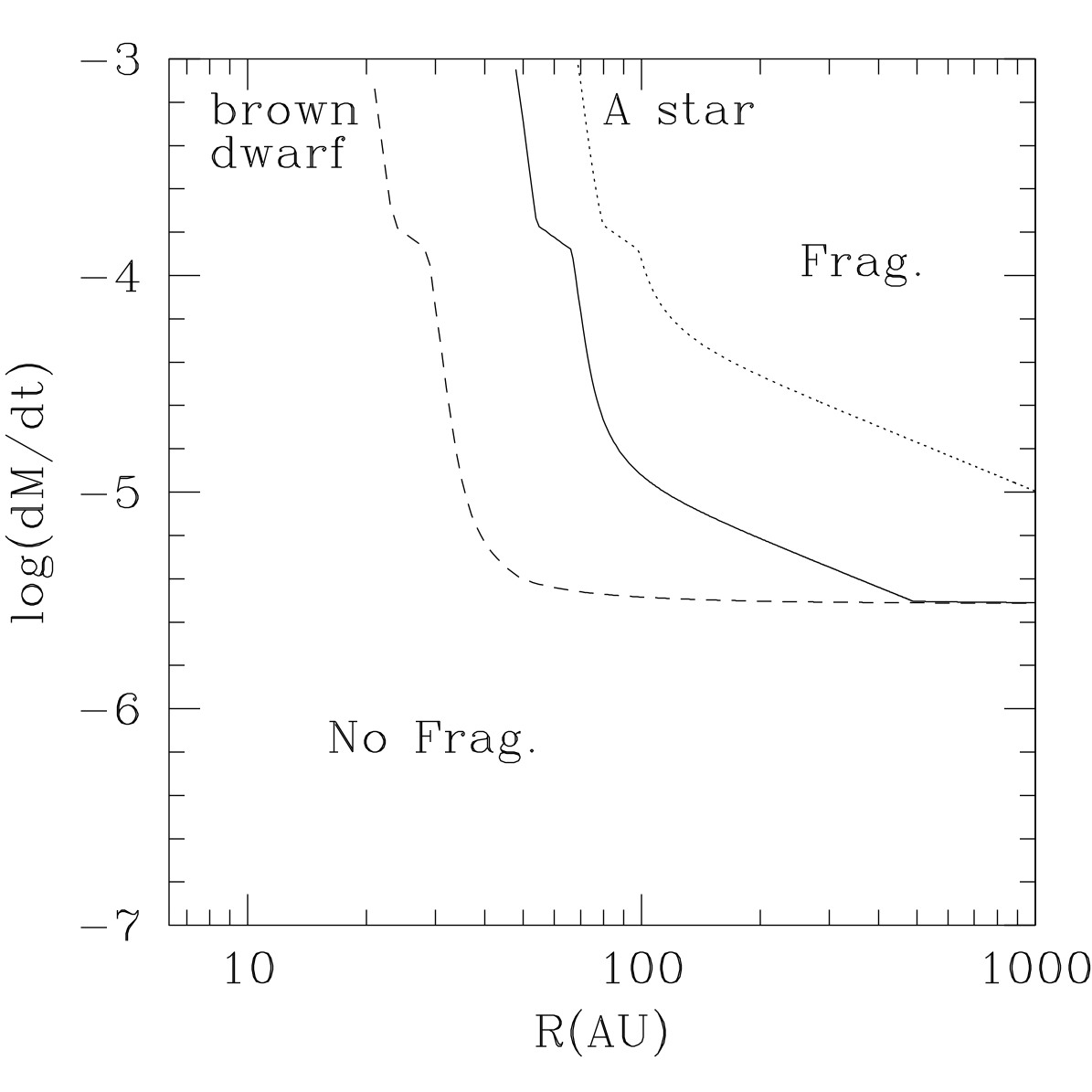} 
\caption{Similar to Fig. \ref{fig:frag}, the fragmentation radi for a typical brown dwarf 
(0.08 M$_{\odot}$, 0.01 L$_{\odot}$) and an A-type star (3 M$_{\odot}$,100 L$_{\odot}$) 
compared with our standard case (1 M$_{\odot}$, 1 L$_{\odot}$, solid curve). 
A $T=20$ K minimum disk temperature is assumed and $\alpha_{c}$=1 has been used. } 
\label{fig:ba}
\end{figure}

\begin{figure}
\includegraphics[width=0.8\textwidth]{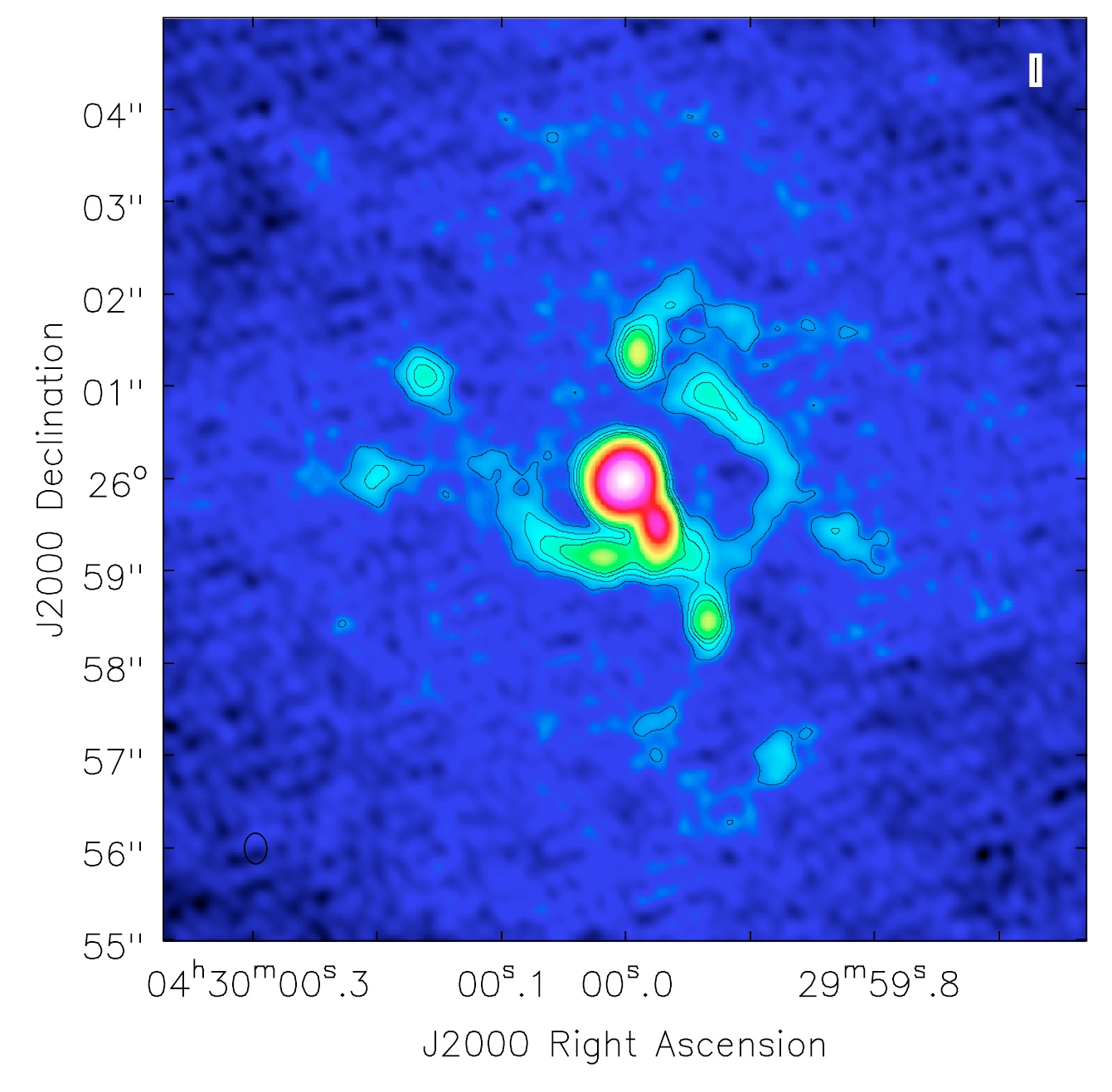} \caption{The synthetic ALMA image (with CASA) at 3mm 
for the R100\_3e-5 case, assuming this disk is in Ophiuchus. The disk effective temperature 
at 3mm is approximated with Equation 4 and the central star is assumed to have 5000 K effective temperature. 
The integration is only 
1 minute with Full ALMA (resolution 0.1''). Besides the massive clump close to
the central star, several other clumps at $\sim 100$ AU scales are also apparent.  The tidally induced
spiral arms by the clumps have 5 $\sigma$ detection. } 
\label{fig:alma}
\end{figure}

 Before R$_{f}$ is determined, the relationship between $\dot{M}$ and the  disk sound speed $c_{s}$ at disk 
 radius R needs to be derived first.
$\dot{M}$, T$_{eff}$, and R for a steady accretion disk with zero torque
inner boundary are related with
\begin{equation}
T_{eff}^{4}=\frac{3GM_{*}\dot{M}}{8\pi R^{3}\sigma}(1-\frac{R_{*}}{R})\,.\label{eq:teffr}
\end{equation}
T$_{eff}$ and T$_{c}$ are related by
\begin{equation}
T_{eff}^{4}=\frac{8}{3}(T_{c}^{4}-T_{ext}^{4})\frac{\tau}{1+\tau^{2}}\,,\label{eq:teffl}
\end{equation}
where $\tau=\Sigma\kappa(T_{c},\rho_{c})/2$ and
$\rho_{c}=\Sigma/2H$. Substituting $\Sigma$ with T$_{c}$ by assuming $Q\equiv c_{s}\Omega$/$\pi$G$\Sigma\sim$1.5,
T$_{c}$ can be solved by equating Equation\ref{eq:teffr} and \ref{eq:teffl} with the Newton-Raphson method at a given $\dot{M}$.
Finally, with the derived $\dot{M}$-T$_{c}$, the critical fragmentation radii (R$_{f}$)
can be solved from Equation \ref{eq:frag} for any given
$\alpha_{c}$ (The results are shown as curves in Figure \ref{fig:frag}).

The analytical solution of the optically thick limit
and the irradiation dominated limit are summarized below:

With the fragmentation condition assumption $\dot{M}G/2c_{s}^{3}>\alpha_{c}$, and $T_{c}$=$T_{c,0}$(R/R$_{0}$)$^{\gamma}$, 
the disk will fragment if
\begin{equation}
\frac{R}{R_{0}}>\frac{R_{f}}{R_{0}}=\left(\frac{\dot{M}G}{2\alpha_{c}c_{s,0}^{3}}\right)^{2/3\gamma}\,.\label{eq:frag1}
\end{equation}
where c$_{s,0}$ is the sound speed at T$_{c,0}$.
In the special case where $\gamma$=0 ($T_{c}$=$T_{c,0}$, e.g. constant irradiation temperature at every radius), 
Eq. \ref{eq:frag1} is not valid, and the disk will fragment everywhere if and only if $\dot{M}>2\alpha_{c}c_{s}^{3}/G$, 
independent on the disk radius (horizontal part of the curve in Figure \ref{fig:frag} at large radii). However,
we want to point out that $\alpha_{c}$ depends on the irradiation strength (Rice \etal 2011) and in this paper
we assume $\alpha_{c}$ is a constant. Thus the key
 parameter to define the fragmentation radii (R$_{f}$) is the temperature slope $\gamma$ which will be derived as following.

In the viscous heating dominated limit, T$_{c}^{4}$=3/16 T$_{eff}^4$ $\Sigma \kappa$, assuming the disk is 
optically thick which will be justified later. Considering the Rossland mean opacity 
$\kappa=\kappa_{r} (T_{c}/T_{r})^{a}$ (T$_{r}$ is some arbitrary temperature to scale the 
opacity, and c$_{s,r}$ is its corresponding sound speed), T$_{eff}^{4}$=3GM$_{*}$$\dot{M}$/8$\pi$R$^{3}$$\sigma$,
 and the Toomre Q parameter of gravitationally unstable disk (Q=c$_{s}$$\Omega$/$\pi$G$\Sigma$) is close 
 to some critical value Q$_{c}$$\sim$1.5, we have
\begin{equation}
T_{c}=T_{r}\left[\frac{3}{16}\frac{3GM_{*}\dot{M}}{8\pi R_{d,0}^{3}\sigma T_{r}^{4}}\frac{c_{s,r}\Omega_{0}\kappa_{r}}{\pi G Q_{c}}\right]^{2/(7-2a)}\left(\frac{R}{R_{d,0}}\right)^{-9/(7-2a)}\,.\label{eq:Tcq1}
\end{equation}
Please notice that, R$_{d,0}$ here could be any value since it is actually canceled out. We keep 
it here to seperate R from the square bracket, but try to maintain the other dimensionless parameters. 

With the dust opacity given in Appendix A,  $a$=1.5, the midplane temperature has slope
 $\gamma$=-9/(7-2$a$)=-2.25 (T$_{c}\propto R^{-2.25}$). If we plug Eq. \ref{eq:Tcq1} into
  $\dot{M}G/2c_{s}^{3}>\alpha_{c}$, we can derive the disk will fragment if
\begin{equation}
\frac{R}{R_{0}}>\left(\frac{\dot{M}G}{2c_{s,r}^{3}\alpha_{c}}\right)^{-(14-4a)/27}\left[\frac{3}{16}\frac{3GM_{*}\dot{M}}{8\pi R_{d,0}^{3}\sigma T_{r}^{4}}\frac{c_{s,r}\Omega_{0}\kappa_{r}}{\pi G Q_{c}}\right]^{2/9}\,.
\end{equation}
Thus, with the opacity given above, we have 
\begin{equation}
R_{f}\propto\alpha_{c}^{0.3}\dot{M}^{-0.07}M_{*}^{1/3}\kappa_{r}^{2/9}\,.\label{eq:fragr}
\end{equation}
R$_{f}$ has a very weak dependence on $\dot{M}$.
This has clearly been shown in Figure \ref{fig:frag} where R$_{f}$ is almost vertical at $\sim$ 50 AU as long as
 the disk accretion rate is high and is viscously heating dominated. The effect of the opacity is also 
 apparent: if the opacity depletes by a factor of 1000 ($\kappa_{r}$ is 0.001 of the nominal value), 
 R$_{f}$ decreases by a factor of 5 ($\sim$10 AU). Finally we will justify why we assume optically 
 thick  limit for the viscous heating dominated cases. If we plug the opacity law $\kappa=\kappa_{r} (T_{c}/T_{r})^{a}$,
  and the disk midplane temperature $T_{c}$=$T_{c,0}$(R/R$_{d,0}$)$^{\gamma}$ into Toomre Q 
  parameter we can derive the disk is optically thick if R$<$R$_{t}$, where R$_{t}$ is
\begin{equation}
\left(\frac{R_{t}}{R_{0}}\right)^{3/2-\gamma(a+1/2)}=\frac{c_{s,0}\Omega_{0}\kappa_{0}}{\pi G Q}
\end{equation}
where 0 denotes the corresponding  physical quantities at radius R$_{0}$. Assuming T$_{0}$=220 K at 
R$_{0}$=1 AU, and $\gamma$=-0.5, R$_{t}$ is 80 AU, which is larger than the fragmentation radius
 of the viscous dominated cases $\sim$50 AU. Thus, it is justified that the disk is optically thick at the fragmentation radius.

For the central star irradiation dominated case, T$_{c}$=T$_{irr}$. Thus with $T_{irr}^{4}=f(R) L/(4 \pi R^{2} \sigma)$, we have $\gamma$=-1/2 so that
\begin{equation}
R_{f}=\left( \frac{\dot{M}G}{2\alpha_{c}}\right)^{-4/3}\left(\frac{k}{\mu
m_{h}}\right)^{2}\left(\frac{f(R) L}{4\pi\sigma}\right)^{1/2}\,.\label{eq:irr}
\end{equation}
Considering L/L$_{\odot}$$\sim$(M$_{*}$/M$_{\odot}$)$^{2}$, we can derive
\begin{equation}
 R_{f}\propto(\dot{M}/\alpha_{c})^{-4/3}M_{*} \,,\label{eq:fragirr}
\end{equation}
thus, $R_{f}$ sensitively depends on
  $\dot{M}$ as shown in Figure \ref{fig:frag}. In this limit, R$_{f}$ does not depend on the opacity. If
$T_{irr}$=const. which is similar to the envelope irradiation
case, the disk will fragment everywhere as long as  $\dot{M}>2c_{s,irr}^{3}\alpha_{c}/G$ (where c$_{s,irr}$ is the sound
speed at T$_{irr}$, this $\dot{M}$ is represented as the horizontal 
dotted line in Figure \ref{fig:frag}), below which the disk will not fragment at any radius.

\FloatBarrier
\clearpage

\begin{table}
\begin{center}
\caption{Models \label{tab1}}
\begin{tabular}{cccccccccc}

\tableline\tableline
Case name    & Resolution & $R_{\rm in}$ & $R_{\rm out}$ & Infall $R_{a}$ & Infall $R_{b}$ & $\dot{M_{\rm in}}$  & Time\tablenotemark{a}      & Fragments ? & Color in Fig.~\ref{fig:clumpdm}, \ref{fig:clumpgap}, \ref{fig:clumptidal}, \& \ref{fig:clumpmig}\\
             & & AU       & AU        & AU              & AU        &  M$_{\odot}$ yr$^{-1}$ & 10$^{4}$yr    &  &   and comments\\
\tableline
R40\_3e-4 & 488$\times$512 & 2.5 & 1000 & 25 & 40 & 3$\times$10$^{-4}$ & 0.48 & No &  \\
R40\_1e-4 & 488$\times$512 & 2.5 & 1000 & 25 & 40 & 10$^{-4}$ & 0.4 & No  & \\
R65\_3e-4 & 432$\times$512 & 5 & 1000 & 50 & 65 & 3$\times$10$^{-4}$ & 0.1 & Yes & \\
R65\_1e-4 & 432$\times$512 & 5 & 1000 & 50 & 65 & 10$^{-4}$ & 1.24 & Yes/M\tablenotemark{b} & red, clump A \\
R65\_3e-5 & 432$\times$512 & 5 & 1000 & 50 & 65 & 3$\times$10$^{-5}$ & 2.04 & No & \\
R65\_1e-5 & 432$\times$512 & 5 & 1000 & 50 & 65 & 10$^{-5}$ &4 & No & \\
R65\_3e-6 & 432$\times$512 & 5 & 1000 & 50 & 65 & 3$\times$10$^{-6}$ &4 & No & \\
R100\_3e-4 & 432$\times$512 & 5 & 1000 & 85 & 100 & 3$\times$10$^{-4}$ & 0.2 & Yes & \\
R100\_1e-4 & 432$\times$512 & 5 & 1000 & 85 & 100 & 10$^{-4}$ & 0.67 & Yes & blue, clump B \\
R100\_3e-5 & 432$\times$512 & 5 & 1000 & 85 & 100 & 3$\times$10$^{-5}$ & 2.4 & Yes/M & black, clump C \\
R100\_1e-5 & 432$\times$512 & 5 & 1000 & 85 & 100 & 10$^{-5}$ & 4 & No & \\
R100\_3e-6 & 432$\times$512 & 5 & 1000 & 85 & 100 & 3$\times$10$^{-6}$ & 4 & No & \\
R200\_3e-4 & 408$\times$512 & 10 & 1500 & 175 & 200 & 3$\times$10$^{-4}$ & 0.28 & Yes & \\
R200\_1e-4 & 408$\times$512 & 10 & 1500 & 175 & 200 & 10$^{-4}$ & 0.47 & Yes  & \\
R200\_3e-5 & 408$\times$512 & 10 & 1500 & 175 & 200 & 3$\times$10$^{-5}$ & 1.2 & Yes &  \\
R200\_1e-5 & 408$\times$512 & 10 & 1500 & 175 & 200 & 10$^{-5}$ & 3.3 & Yes/M & cyan, clump D,E,F,G \\
R200\_3e-6 & 408$\times$512 & 10 & 1500 & 175 & 200 & 3$\times$10$^{-6}$ & 20 & No & \\
 \tableline
Special cases&&&&&&&&&\\
\tableline
R100\_1e-5I \tablenotemark{c}& 432$\times$512 & 5 & 1000 & 85 & 100 & 10$^{-5}$ & 6.6 & Yes/M  & green, clump H \\
R200\_3e-5L100 \tablenotemark{d}  & 408$\times$512 & 10 & 1500 & 175 & 200 & 3$\times$10$^{-5}$ & 4 & No &  \\
R200\_3e-6noirr \tablenotemark{e} & 408$\times$512 & 10 & 1500 & 175 & 200 & 3$\times$10$^{-6}$ & 8 & Yes/M & Magenta, clump I,J,K \\
R100\_1e-5DG1p57 \tablenotemark{f} & 916$\times$1024 & 10 & 1500 & 85 & 100 & 10$^{-5}$ & 4 & No & \\
R200\_1e-5G1p57 \tablenotemark{f} & 816$\times$1024 & 10 & 1500 & 175 & 200 & 10$^{-5}$ & - & Yes \\
R200\_1e-5S0p06 \tablenotemark{g} & 816$\times$1024 & 10 & 1500 & 175 & 200 & 10$^{-5}$ & - & Yes \\
\tableline
Resolution studies &&&&&&&&&\\
\tableline
R65\_1e-5D & 816$\times$1024 & 10 & 1500 & 50 & 65 & 10$^{-5}$ & 3.6 & No & \\
R65\_3e-5D & 816$\times$1024 & 10 & 1500 & 50 & 65 & 3$\times$10$^{-5}$ & 1.7 & No & \\
R100\_1e-5D & 916$\times$1024 & 10 & 1500 & 85 & 100 & 10$^{-5}$ & 3.96 & No & \\
R100\_1e-5D2 &  704$\times$1024 & 20 & 1500 & 85 & 100 & 10$^{-5}$ & 4 & No & \\
R100\_1e-5Q2 &  1408$\times$2048 & 20 & 1500 & 85 & 100 & 10$^{-5}$ & 2.65 & No & \\
R200\_1e-5 & 816$\times$1024 & 10 & 1500 & 175 & 200 & 10$^{-5}$ & 3.3 & Yes \\
\tableline

\tableline
\end{tabular}
\tablenotetext{a}{If fragmentation occurs, this gives time when the first clump 
migrates to the inner boundary.}
\tablenotetext{b}{These are the marginal fragmentation cases where only one or 
two clumps form in the disk.}
\tablenotetext{c}{Decreasing irradiation after 5$\times$10$^{4}$ yr.}
\tablenotetext{d}{Luminosity is 100 L$_{\odot}$ from the central star.}
\tablenotetext{e}{No irradiation included.}
\tablenotetext{f}{The same as the corresponding case in the resolution studies but with $\gamma$=1.57 instead of 1.4.}
\tablenotetext{f}{The smoothing length is 0.06 disk scale height at 5 AU, compared with other cases with 0.6 disk scale height smoothing length at 5 AU.}
\end{center}
\end{table}

\begin{table}
\begin{center}
\caption{Clump Fates \label{tabfate}}
\begin{tabular}{ccc}

\tableline\tableline
Gap Opening & Tidal Destruction & Across inner boundary\\
\tableline
\tablenotemark{a}A, C, G & E, F, I, J & A\tablenotemark{b}, B, D, E\tablenotemark{c}, H, K  \\
\tableline
3 & 4 & 6\\
\tableline
\end{tabular}
\tablenotetext{a}{The clump label corresponds to the clump label in the last column of Table 1.}
\tablenotetext{b}{This clump opens a gap within twice the disk inner radius. 
The gap opening may be due to the inner boundary condition. Thus we count this case 
as both crossing the inner boundary and gap opening.}
\tablenotetext{c}{This clump is tidally destroyed within twice the disk inner radius. 
This tidal destruction may be due to the inner boundary condition. 
Thus we count this case as both crossing the inner boundary and as being tidally destroyed. }
\end{center}
\end{table}

\clearpage
\begin{table}
\begin{center}
\caption{Fit to Zhu \etal (2007, 2008) opacity but with opacity from water ice, graphite, and silicate dust included \label{fitopa}}
\begin{tabular}{clcl}
\tableline\tableline
$\log_{10}T$& $\log_{10}\kappa$ & comments \\
\tableline
$< 0.019 \log_{10}P +2.14$ & $1.5 \log_{10}T - 2.48$
& water ice opacity \\
$< 0.019 \log_{10}P +2.21$ & $-3.53 \log_{10}T +0.095 \log_{10} P+8.29$
& water ice evaporation \\
$< 2.79$ & $1.5 \log_{10}T - 2.84$
& metal grain opacity \\
$< 2.97$ & $-5.83 \log_{10}T + 17.7$
& graphite corrosion \\
$< 0.027 \log_{10}P+3.16 $ & $2.13 \log_{10}T - 5.94$&
grain opacity \\
$< 0.0281  \log_{10}P + 3.19 $&$ -42.98 \log_{10}T + 1.312 \log_{10}P +
135.1$ & silicate grain evaporation \\
$< 0.03 \log_{10}P + 3.28$ & $4.063 \log_{10}T -15.013$ & water vapor
\\
$< 0.00832 \log_{10}P + 3.41$ &$ -18.48 \log_{10}T + 0.676 \log_{10}P
+58.93$ &  \\
$< 0.015 \log_{10}P + 3.7$ & $2.905 \log_{10}T + 0.498 \log_{10}P
-13.995$ & molecular opacities \\
$< 0.04 \log_{10}P + 3.91 $&$ 10.19 \log_{10}T + 0.382\log_{10}P
-40.936$ & H scattering \\
$< 0.28 \log_{10}P + 3.69$ & $-3.36 \log_{10}T + 0.928 \log_{10}P
+12.026$ & bound-free,free-free \\
else \tablenotemark{a} & $-0.48$ & electron scattering \\
\tableline
\end{tabular}
\tablenotetext{a}{With one additional condition to set the boundary:
if $\log_{10}\kappa <$3.586 $\log_{10}T$ -16.85 and $\log_{10}T$ $<$
4, $\log_{10}\kappa=$ 3.586 $\log_{10}T$-16.85}
\tablenotetext{*}{Both C and fortran subroutine can be downloaded at http://http://www.astro.princeton.edu/$\sim$zhzhu/opacity.html}
\end{center}
\end{table}

\clearpage

\clearpage

\clearpage

\end{document}